\documentclass[iop]{emulateapj2}
\usepackage{epstopdf}
\newcommand{\kms}{km~s$^{-1}$}

\newcommand{\cmms}{cm$^2$~s$^{-1}$}
\newcommand{\msun}{M$_{\sun}$}
\newcommand{\rjup}{R$_{Jup}$}
\newcommand{\ldl}{$\lambda/{\Delta}{\lambda}$}

\newcommand{\lbol}{$\log_{10}{L_{bol}/L_{\sun}}$}
\newcommand{\teff}{T$_{eff}$}
\newcommand{\logg}{$\log{g}$}
\newcommand{\fsed}{$f_{sed}$}
\newcommand{\kzz}{$K_{zz}$}
\newcommand{\lii}{\ion{Li}{1}}
\newcommand{\ki}{\ion{K}{1}}

\newcommand{\meth}{CH$_4$}

\newcommand{\wat}{H$_2$O}
\newcommand{\ammon}{NH$_3$}
\newcommand{\name}{Ross~458C}

\slugcomment{Submitted to ApJ 16 Aug 2010; Accepted for publication 28 Sep 2010}

\shorttitle{Clouds in Ross~458C}
\shortauthors{Burgasser et al.}

\begin{document}

\title{Clouds in the Coldest Brown Dwarfs: FIRE Spectroscopy of Ross~458C\footnote{This paper includes data gathered with the 6.5-m Magellan Telescopes located at Las Campanas Observatory, Chile.}}

\author{
Adam J.\ Burgasser\altaffilmark{1,2,3},
Robert A.\ Simcoe\altaffilmark{2,4},
John J.\ Bochanski\altaffilmark{2},
Didier Saumon\altaffilmark{5}, 
Eric E.\ Mamajek\altaffilmark{6},
Michael C.\ Cushing\altaffilmark{7},
Mark S.\ Marley\altaffilmark{8},
Craig McMurtry\altaffilmark{6},
Judith L.\ Pipher\altaffilmark{6},
and
William J.\ Forrest\altaffilmark{6}
}

\altaffiltext{1}{Center for Astrophysics and Space Science, University of California San Diego, La Jolla, CA 92093, USA; aburgasser@ucsd.edu}
\altaffiltext{2}{Massachusetts Institute of Technology, Kavli Institute for Astrophysics and Space Research, Building 37, Room 664B, 77 Massachusetts Avenue, Cambridge, MA 02139, USA}
\altaffiltext{3}{Hellman Fellow}
\altaffiltext{4}{Alfred P. Sloan Foundation Research Fellow}
\altaffiltext{5}{Los Alamos National Laboratory, P.O. Box 1663, MS F663, Los Alamos, NM 87545, USA}
\altaffiltext{6}{Dept.\ of Physics and Astronomy, University of Rochester, Rochester, NY USA 14627}
\altaffiltext{7}{Jet Propulsion Laboratory, California Institute of Technology, Pasadena, CA, USA}
\altaffiltext{8}{NASA Ames Research Center, Mail Stop 245-3, Moffett Field, CA 94035, USA}

\begin{abstract}
Condensate clouds are a salient feature of L dwarf atmospheres, but have been assumed to play little role in shaping the spectra of the coldest T-type brown dwarfs. Here we report evidence of condensate opacity in the near-infrared spectrum of the brown dwarf candidate Ross~458C, obtained with the Folded-Port Infrared Echellette (FIRE) spectrograph at the Magellan Telescopes.  These data verify the low-temperature nature of this source, indicating a T8 spectral classification, $\log_{10}{L_{bol}/L_{\sun}}$ = $-5.62{\pm}0.03$, T$_{eff}$ = 650$\pm$25~K, and a mass at or below the deuterium burning limit.   The data also reveal enhanced emission at $K$-band associated with youth (low surface gravity) and supersolar metallicity, reflecting the properties of the Ross~458 system (age = 150--800~Myr, [Fe/H] = +0.2 to +0.3). We present fits of FIRE data for Ross~458C, the T9 dwarf ULAS~J133553.45+113005.2, and the blue T7.5 dwarf SDSS~J141624.08+134826.7B, to cloudless and cloudy spectral models from Saumon \& Marley. For Ross~458C we confirm a low surface gravity and supersolar metallicity, while the temperature differs depending on the presence (635$^{+25}_{-35}$~K) or absence (760$^{+70}_{-45}$~K) of cloud extinction.  ULAS~J1335+1130 and SDSS~J1416+1348B have similar temperatures (595$^{+25}_{-45}$~K), but distinct surface gravities ($\log{g}$ = 4.0--4.5~cgs versus 5.0--5.5~cgs) and metallicities ([M/H] $\approx$ +0.2 versus -0.2). In all three cases, cloudy models provide better fits to the spectral data, significantly so for Ross~458C.  These results indicate that clouds are an important opacity source in the spectra of young cold T dwarfs, and should be considered when characterizing the spectra of planetary-mass objects in young clusters and directly-imaged exoplanets.  The characteristics of Ross~458C suggest it could itself be regarded as a planet, albeit one whose cosmogony does not conform with current planet formation theories. 
\end{abstract}


\section{Introduction}

Mineral condensate clouds are a unique and prominent constituent of the atmospheres of the 
coldest brown dwarfs, the L dwarfs and T dwarfs (\citealt{2005ARA&A..43..195K} and references therein).  For the L dwarfs, condensates are a primary contributor to shaping spectral energy distributions (SEDs).  Their formation depletes the atmosphere of TiO and VO gases, transforming optical SEDs and defining the L dwarf
spectral sequence \citep{1997ApJ...480L..39J,1999ApJ...519..802K,2002ApJ...577..974L}.
Condensate opacity contributes to muted infrared {\wat} bands and the characteristic red
near-infrared colors of L dwarfs ($J-K$ = 1.5...2.5; \citealt{2000ApJ...542..464C, 2004AJ....127.3553K}).  Evidence of silicate grain absorption has also been found in the mid-infrared
spectra of L dwarfs \citep{2006ApJ...648..614C, 2006A&A...451L...9H, 2008ApJ...686..528L}.  For the T dwarfs, condensates appear to play a minor role, as 
evidenced by strong molecular gas absorption features and typically blue near-infrared colors
($J-K$ = $-$0.5...0.5;  \citealt{2002ApJ...564..421B, 2006ApJ...637.1067B, 2002ApJ...564..466G, 2005ApJ...623.1115C}).  This shift is believed to be the result of clouds sinking
below the visible photosphere \citep{2001ApJ...556..872A, 2003ApJ...586.1320C, 2005ApJ...621.1033T} or breaking apart \citep{2002ApJ...571L.151B,marley10}, although the details of this transition remain poorly understood \citep{2006ApJ...640.1063B,2007ApJ...659..655B, 2008ApJ...689.1327S}.

Even without considering condensates, modeling the highly structured SEDs of T dwarfs remains a challenge.   The abundances and opacities of prominent gas
species -- H$_2$ (collision induced absorption; \citealt{1969ApJ...156..989L}), {\wat}, {\meth}, {\ammon} and neutral alkalis (pressure-broadened wings; \citealt{2003ApJ...583..985B}) --
are highly sensitive to photospheric gas conditions.  Temperature and pressure variations,
elemental abundances, and vertical mixing all have observable effects on T dwarf spectra
(e.g., \citealt{2004AJ....127.3553K, 2006ApJ...639.1095B, 2007ApJ...656.1136S, 2009ApJ...702..154S}).
In principle, one could disentangle the contributions of these atmospheric properties 
on observed spectra and derive bulk characteristics---mass, age and metallicity---for individual sources (e.g., \citealt{2006ApJ...639.1095B, 2009ApJ...695.1517L}).  However, a variety of effects drive
systematic deviations between model and observed spectra.  These include
incomplete opacities for key absorbers such as {\meth}, {\ammon} and \ion{K}{1} (e.g., \citealt{2003A&A...411L.473A, 2008ApJ...678.1372C, 2008ApJS..174..504F});
non-equilibrium chemistry arising from vertical mixing (e.g., \citealt{2006ApJ...647..552S,2007ApJ...656.1136S});
metallicity effects on chemical pathways (e.g., \citealt{2002Icar..155..393L, 2003ApJ...592.1186B, 2010ApJ...716.1060V});
and the influence of condensate clouds in regions of minimum gas opacity 
(e.g., \citealt{1999ApJ...520L.119T, 2001ApJ...556..872A, 2002ApJ...573..394B}).  These processes
lead to persistent uncertainties in inferred atmospheric and physical
parameters for individual sources (cf., \citealt{2010MNRAS.404.1952B, 2010AJ....139.2448B}).

Benchmark T dwarfs,
with independent determinations of age, mass and metallicity, are important calibrators
for brown dwarf atmospheric studies \citep{2006MNRAS.368.1281P}.  
Such benchmarks include widely-separated T dwarf
companions to nearby stars, several examples of which have been identified 
over the past 15 years (e.g., \citealt{1995Natur.378..463N, 2000ApJ...531L..57B, 2007ApJ...654..570L, 2009MNRAS.395.1237B}). Most recently, \citet{2010MNRAS.405.1140G} and \citet{2010A&A...515A..92S} reported the discovery
of a faint ($J$ = 16.67$\pm$0.01), co-moving candidate T dwarf companion to the Ross 458 system
in the UKIRT Infrared Deep Sky Survey (UKIDSS; \citealt{2007MNRAS.379.1599L}).
The source, ULAS~J130041.72+122114.7 (hereafter {\name}), is separated by 102$\arcsec$ from
the M0.5~Ve + M7~Ve Ross 458AB pair, implying a projected separation of over 1100~AU
at a distance of 11.7$\pm$0.2~pc \citep{2007A&A...474..653V}.  
The late-type
nature of {\name} was inferred from its optical and near-infrared colors ($z-J$ = 3.55$\pm$0.19; $J-K$ = -0.21$\pm$0.06), faint absolute magnitude ($M_J$ = 16.40$\pm$0.04) and large flux contrast in filters sampling the 1.6~$\micron$ {\meth} band (cf., \citealt{2005AJ....130.2326T}).  From these photometric measurements, \citet{2010A&A...515A..92S} and \citet{2010MNRAS.405.1140G} estimated spectral types of T7$\pm$1 and T8-9 for {\name}, respectively.  However, no spectrum was reported in
either study.  

If {\name} is a brown dwarf, it is a potentially interesting
benchmark for several reasons.
Kinematics of the Ross~458 system coincide with the Hyades supercluster
\citep{1960MNRAS.120..540E, 2001A&A...379..976M}, an 
extended population of relatively young and metal-rich stars that includes
the 625$\pm$50~Myr Hyades open cluster \citep{1998A&A...331...81P}.  The primary exhibits conspicuous activity in H$\alpha$ and \ion{Ca}{2} lines \citep{2010AJ....139..504B}, and is an X-ray source \citep{2007AcA....57..149K, 2009A&A...499..129L}.  {\name} could therefore be a young and/or metal-rich brown dwarf benchmark, useful for constraining both evolutionary and atmospheric models (cf. \citealt{2004ApJ...615..958Z, 2009ApJ...692..729D}).

In this article, we present the first spectroscopic observations of {\name}, obtained in the near-infrared with the newly-commissioned Folded Port Infrared Echellette (FIRE; \citealt{2008SPIE.7014E..27S}).  These data
confirm the late-type T dwarf nature of the source, and indicate a
low surface gravity and supersolar metallicity consistent with characteristics of the primary.
Furthermore, model fits indicate that cloud opacity plays a prominent role in shaping
its near-infrared spectrum.  
In Section~2 we constrain the age and metallicity of the Ross~458 system
based on the photometric, spectral, kinematic and rotational properties of its M0.5~Ve primary.
In Section~3  we describe our FIRE observations of {\name} and two other late-type T dwarfs, from which we derive spectral types and basic physical characteristics.  
In Section~4 we fit
the spectral data for these three sources to cloudless and cloudy atmospheric models from \citet[hereafter SM08]{2008ApJ...689.1327S},
and compare the inferred atmospheric and physical parameters to prior studies.
In Section~5 we discuss the role
of condensates in shaping the near-infrared SEDs of young, cold T dwarfs, and their
influence on T dwarf temperature scales.
In Section~6 we summarize our results.

\section{Characterizing the Ross~458 System}

\subsection{Examining Association with the Hyades Open Cluster}

Under the assumption of coevality, some physical properties of {\name} can be inferred from its stellar companions, specifically age and metallicity.
As such, it is necessary to examine the association of this system with the Hyades supercluster, 
a kinematic stream composed of several dozen nearby stars with similar space
motions \citep{1960MNRAS.120..540E, 2001A&A...379..976M}.  
While co-moving associations typically have similar ages and compositions,
the ages of the Hyades supercluster stars span the star-formation history of the disk
(0.4--2.0~Gyr; \citealt{2005A&A...430..165F, 2008A&A...490..135A}), so 
stream membership is not a particularly useful age indicator. 
Within the stream is the eponymous Hyades open cluster, the nearest open
cluster to the Sun, whichi constitutes a coeval group with an
age of 625$\pm$50 Myr \citep{1998A&A...331...81P}.  Because of its
proximity, members of the Hyades cluster halo are distributed broadly across the sky.  It is therefore not unreasonable for Ross~458 to be associated with this structure.

To examine this possibility, we calculated the systemic space velocity\footnote{We adopted a heliocentric, right-handed rectangular coordinate system with U velocity locally pointing away from the Galactic center, V velocity pointing in the direction of Galactic rotation 
($\ell$, $b$ = 90$^{\circ}$, 0$^{\circ}$) and W velocity pointing towards the north Galactic pole ($b$ = +90$^{\circ}$).} of the Ross~458 system. We used the long-baseline proper motion from Tycho-2 
($\mu_{\alpha}$ = $-$640.1$\pm$1.5 mas~yr$^{-1}$; 
$\mu_{\delta}$ = $-$25.1$\pm$1.4 mas~yr$^{-1}$; \citealt{2000A&A...355L..27H}),
the radial velocity from \citet[$V_r$ = -11.2$\pm$0.1~{\kms}]{2002ApJS..141..503N}, 
and the revised Hipparcos parallax ($\pi$
= 85.5$\pm$1.5 mas; \citealt{2007A&A...474..653V}). From these we find heliocentric
(U, V, W) = (30.6, -18.8, -9.9)$\pm$(0.5, 0.3, 0.2)~{\kms}.
The velocity of the Hyades cluster in the same coordinate system \citep{2001A&A...367..111D} is (U, V, W) = (42.25, -19.06, -1.45)$\pm$(0.14, 0.13, 0.21)~{\kms}, a difference of (11.7, -0.3, 8.5)$\pm$(0.5, 0.3, 0.3)~{\kms} (see also \citealt{1998A&A...331...81P}). This is much larger than the 1D velocity dispersion of Hyades cluster members, $\sim$0.35~{\kms} \citep{2003A&A...401..565M},
amounting to a $\sim$20$\sigma$ discrepancy.  Hence, Ross 458 is not currently kinematically associated with the Hyades open cluster.

Has Ross~458 ever been associated with the Hyades cluster?
Its current separation from the cluster center is 53 pc. 
Using the epicyclic orbit approximation to model the
past orbits of Ross 458 and the Hyades (adopting the Oort parameters
of \citealt{1997MNRAS.291..683F} and Local Standard of Rest from \citealt{1998MNRAS.298..387D}), 
we find that Ross 458 was  never closer to the Hyades than $\sim$50~pc during the
past 100~Myr, or about 5 tidal radii.  
Hence, there is no evidence of past or present physical association
between Ross~458 and the Hyades open cluster, and association with the
Hyades supercluster provides no meaningful constraint on the age of this system.

\subsection{The Age of the Ross~458 System}

Despite the wide range of ages associated with the Hyades kinematic stream there
is considerable evidence that the Ross~458 system itself is quite young.
The M0.5~Ve primary\footnote{The M dwarf components of Ross~458 are separated by 
0$\farcs$5 \citep{2004A&A...425..997B}, implying that reported optical spectroscopy of this system probably includes both components. However, Ross~458B is $\approx$8~mag fainter at $V$ (Section~2.3), implying that its contribution to combined light optical spectra is negligible.}
\citep{1995AJ....110.1838R}  is an ``ultra-fast rotator'' with a period between
1.54 and 2.89 days \citep{2003A&A...397..147P,2007AcA....57..149K}. 
Ross~458A had the highest measured
projected rotational velocity ($v\sin{i}$ = 9.7$\pm$0.5 {\kms}) 
among 69 nearby M0-M3 dwarfs in the California
Planet Search program, in which only $\sim$3\%\, of early-M dwarfs had any
detectable rotation ($v\sin{i}$ $\gtrsim$ 3~{\kms};  \citealt{2010AJ....139..504B}).
\citet{2004A&A...425..997B} have argued for an age less than 1~Gyr on the basis 
of this rotation.

In addition, Ross~458A exhibits strong magnetic activity.  
The presence of H$\alpha$ emission indicates an upper activity age limit of 400--800~Myr \citep{2008AJ....135..785W}.  This limit is roughly consistent with the analysis of 
\citet{2007ApJS..168..297T}, who deduce a 68\% upper limit of 440~Myr based
on theoretical modeling of the primary's high-resolution spectrum (however, this does not appear to be a ``well-defined'' age according to that study).  
Ross~458 is also an X-ray source, detected in three R\"{o}ntgen Satellite (ROSAT) Position Sensitive Proportional Counter (PSPC) pointings and
in the All-Sky Survey \citep{2000IAUC.7432....3V}.  Combining these four
observations yields a mean ROSAT PSPC flux in
the 0.1-2.4 keV band of 4.35 ct\,s$^{-1}$, and mean hardness ratios of
HR1 = -0.13 and HR2 = -0.09. 
The mean X-ray flux, coupled with the energy conversion factors of \citet{1995ApJ...450..401F} and the
Hipparcos parallax, implies a soft X-ray luminosity of
$L_X$ = 5$\times$10$^{29}$ erg\,s$^{-1}$ and an
X-ray to bolometric luminosity ratio of $\log{L_X/L_{bol}}$ = -2.7. 
Such a high ratio defines Ross~458 as a``saturated'' X-ray emitter, 
as expected for stars of similar color
and rotation periods of $\lesssim$4 days
\citep{2003A&A...397..147P}. The combination of color and rotation
period for Ross~458A also classifies it as a C-sequence (``convective'') rotator in the
scheme of \citet{2003ApJ...586..464B}. The characteristic spin-down
timescale for such stars is $\sim$300 Myr.   From a
cross-comparison of age-dated cluster and field populations of
low-mass, solar-type stars, \citet{2003ApJ...586..464B} concludes that the
fraction of C-sequence ultrafast rotators reduces to zero by age
$\sim$800 Myr.  Hence, this age can be considered a robust upper limit 
for the Ross~458 system that is consistent
with activity estimates.

\begin{figure}[t]
\epsscale{1.0}
\includegraphics[width=0.5\textwidth]{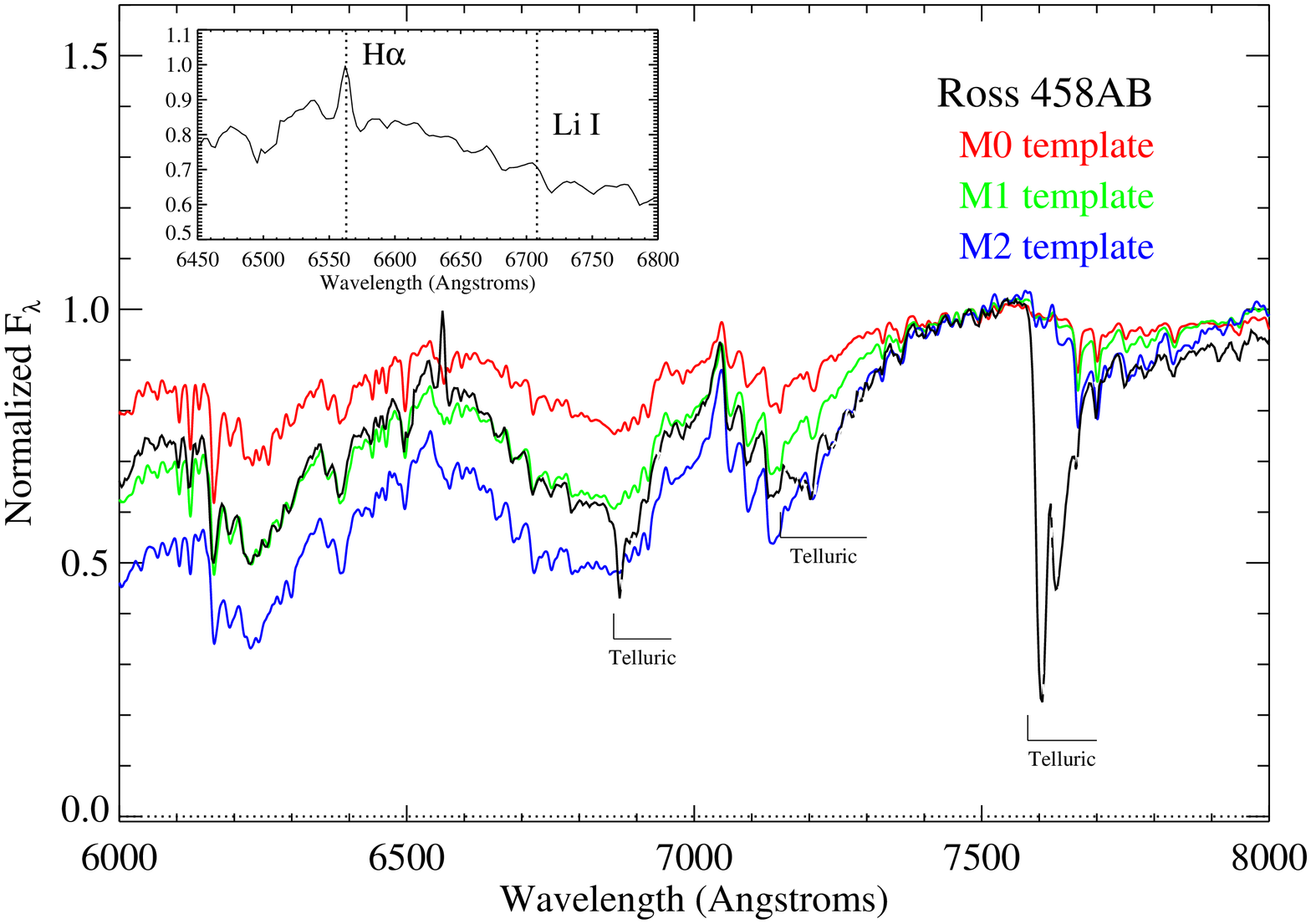}
\caption{Red optical spectrum of Ross~458AB (black line) compared to M0 (red line), M1 (green line) and M2 (blue line) spectral templates from \citet{2007AJ....133..531B}.  All spectra are normalized at 7500~{\AA}, and templates are smoothed to the resolution of the observational data ({\ldl} $\approx$ 800) using a gaussian kernel.  Regions of strong telluric absorption
in the observational data are indicated.   Inset box details the 6450--6800~{\AA} region highlighting the prominent H$\alpha$ emission and absence of {\lii} absorption.
\label{fig_gj494ab}}
\end{figure}

A lower limit on the age of this system can be inferred from the optical spectrum of
the primary.  Figure~\ref{fig_gj494ab} displays {\ldl} $\approx$ 800 optical spectroscopy for the combined AB components obtained by J.\ D.\ Kirkpatrick with the CTIO-1.5m RC Spectrograph on 22 May 1996\footnote{These data are available at DwarfArchives, \url{http://dwarfarchives.org}.}.  The spectrum is compared to M0, M1 and M2 spectral templates from
 \citet{2007AJ....133..531B}, the M1 showing the strongest agreement with the data.  
The prominent 6563~{\AA} H$\alpha$ line is clearly seen in these data, but there is no indication of {\lii} absorption at 6708~{\AA}, implying an age older than 30-50~Myr \citep{1993ApJ...404L..17M, 1996ApJ...459L..91C,1997ApJ...482..442B}.  Furthermore, there is no indication low surface gravity features, such as the weakened 7700~{\AA} {\ki} doublet lines observed of cooler young M dwarfs up to the age of the Pleiades (e.g., \citealt{1996ApJ...469..706M, 2008ApJ...689.1295K}).  These spectroscopic constraints place a lower limit on the age of this system of $\sim$150~Myr.

\subsection{The Metallicity of the Ross~458 System}

The metallicity of Ross~458A can be determined from its position
on the $V-K$/$M_K$ color magnitude diagram (CMD), a combination that
effectively segregates mass and metallicity variations in field M dwarfs
\citep{2005A&A...442..635B,2009ApJ...699..933J,2010arXiv1006.2850S}.
For better accuracy, we disentangled the flux of Ross~458A from its 
closely-separated M7 companion.  Using $\Delta{K}$ = 4.41 from \citet[a 0.1~mag uncertainty is assumed]{2004A&A...425..997B}, we
determined $M_{K_A}$ = 5.26$\pm$0.04 and $M_{K_B}$ = 9.67$\pm$0.11.  We then used unresolved Hipparcos/2MASS
photometry ($V-K$ = 4.180$\pm$0.016) and the mean $V-K$/$M_K$ relation of 
\citet{2010arXiv1006.2850S} to estimate $\Delta{V}$ = 7.93$\pm$0.14 and
hence $(V-K)_A$ = 4.16$\pm$0.02.  The $M_K$/metallicity calibration of
\citet{2009ApJ...699..933J} then yields [Fe/H] = +0.31$\pm$0.05, while the $V-K$/metallicity
calibration of \citet{2010arXiv1006.2850S} yields [Fe/H] = +0.20$\pm$0.05.  Hence, both relations indicate a supersolar metallicity for the Ross~458 system.  However, these quantitative estimates should be regarded with care given the young age of the system. The
primary may be systematically brighter/redder than the control samples used by \citet{2009ApJ...699..933J} and \citet{2010arXiv1006.2850S}, resulting in a systematically higher metallicity.

In summary, the Ross~458 system has a probable age of 150--800~Myr based on its rotation, strong activity indicators and spectroscopic features.  It also appears to be metal-rich, with [Fe/H] = +0.2 to +0.3, albeit with possible age biases.  We will compare these values to results from spectral modeling of the T dwarf companion in Section~4.

\section{Near-Infrared Spectroscopy}

\subsection{The FIRE spectrograph}

\begin{figure*}[t]
%\epsscale{1.0}
\includegraphics[width=0.9\textwidth]{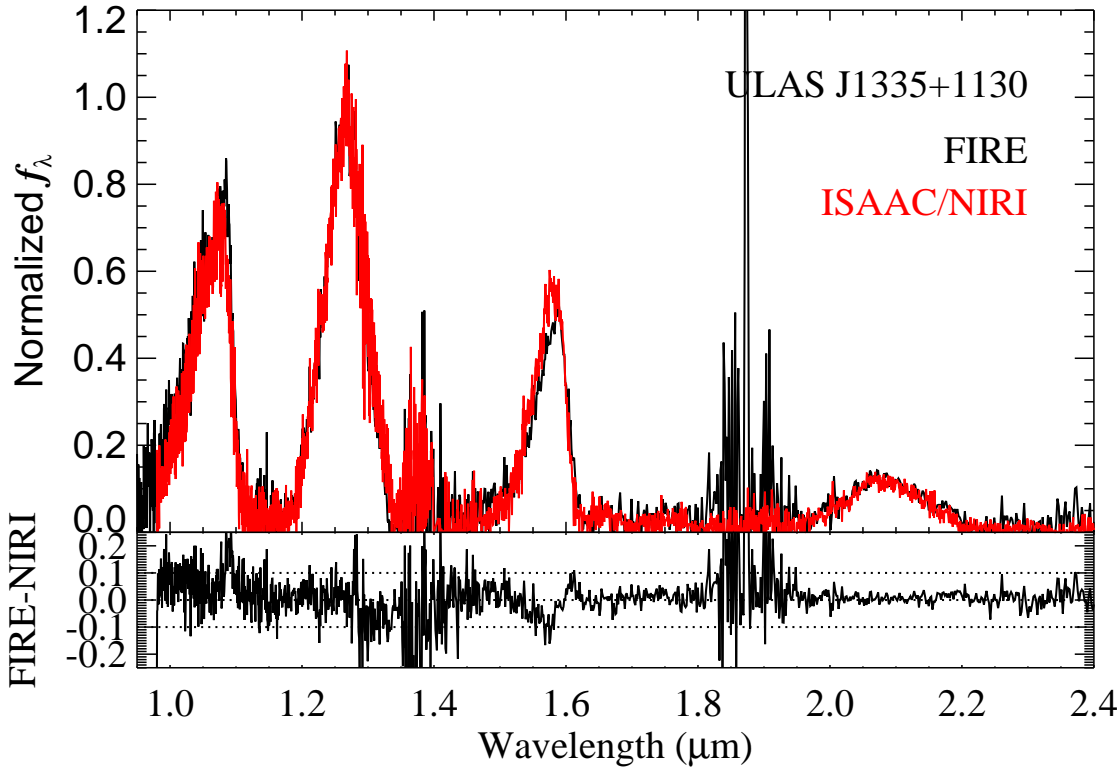}
\caption{Spectral observations of ULAS~J1335+1130 based on Magellan/FIRE (this study) and
Gemini-North/NIRI plus VLT/ISAAC \citep[red line]{2008MNRAS.391..320B}.  Both spectra are normalized at the 1.27~$\micron$ flux peak.   The difference between these spectra (bottom panel) is less than 10\% over most of the near-infrared band.
\label{fig_colorcheck}}
\end{figure*}

FIRE is a cross-dispersed, near-infrared spectrometer recently installed at
the auxiliary Nasmyth focus of the Magellan 6.5-meter Baade Telescope.
It is a single-object spectrograph, designed primarily to deliver moderate-resolution ({\ldl} $\approx$ 6000 for a 0$\farcs$6-wide slit) spectra over the 0.85--2.5~$\micron$ range in a single exposure.  
The 21 orders dispersed by a combination of prisms and reflection grating are imaged onto a 2048$\times$2048, HAWAII-2RG (H2RG) focal plane array.  A mirror can also be rotated in place of the reflection grating to obtain a single-order, high-throughput spectrum at
lower resolution ({\ldl} = 250-350; see below).   A second H2RG camera images the
entrance slit through an MKO\footnote{Mauna Kea Observatory filter system; see \citet{2002PASP..114..180T} and \citet{2002PASP..114..169S}.} $J$-band filter for source acquisition and manual guiding.
Further details on this instrument can be found in \citet{2008SPIE.7014E..27S, simcoe_fire} and on the instrument webpage.\footnote{\url{http://www.firespectrograph.org}}

\subsection{Observations and Data Reduction}

Data presented here were obtained during FIRE's commissioning run
during 31 March through 7 April 2010 (UT).
Observations were obtained for three T dwarfs: {\name}, the T7.5 ULAS~J141623.94+134836.3 (aka SDSS~J141624.08+134826.7B and hereafter SDSS~J1416+1348B; \citealt{2010MNRAS.404.1952B, 2010A&A...510L...8S, 2010AJ....139.2448B})
and the T9  ULAS~J133553.45+113005.2 (hereafter ULAS~J1335+1130; \citealt{2008MNRAS.391..320B}).
A summary of the observations is given in Table~\ref{tab_obs}.
Conditions ranged from clear (3 and 7 April) to patchy cirrus (6 April), with seeing of 0$\farcs$6--0$\farcs$8 at $J$-band.  
For all sources, we employed FIRE's prism-dispersed mode
with the 0$\arcsec$6 wide slit aligned to the parallactic angle (rotator angle 89$\fdg$5).
The resolution for this setting ranges from {\ldl} $\approx$ 350 in the 1.0--1.4~$\micron$ $Y$- and $J$-band region to {\ldl} $\approx$ 250 in the 1.4--2.5~$\micron$ $H$- and $K$-band region, as determined from gaussian fits to 
emission lines in the arc lamp spectra and 
telluric emission lines in the science images.
The spectrograph detector was read out using the 4-amplifier mode at ``high gain'' 
(1.2~counts/e$^-$).
Each source was observed in four 150~s exposures using an 
ABBA dither pattern, nodding 9$\arcsec$ along the 30$\arcsec$ slit,
followed by observation of a nearby A0~V star 
at similar airmass for
telluric correction and flux calibration. Exposures of arc lamps 
(HeNeAr on 3 April; NeAr on 6 and 7 April) and quartz lamps, reflected off 
of the Baade secondary screen, were obtained with the target and A0~V
stellar observations for wavelength and pixel response calibration.

\begin{deluxetable*}{llccccl}
\tabletypesize{\small}
\tablecaption{Magellan/FIRE Observing Log. \label{tab_obs}}
\tablewidth{0pt}
\tablehead{
\colhead{Source} &
\colhead{SpT} &
\colhead{UKIDSS $J$} &
\colhead{Date (UT) } &
\colhead{Airmass} &
\colhead{$t_{int}$\tablenotemark{a}} &
\colhead{A0~V Calibrator}  \\
}
\startdata
Ross 458C & T8 &  16.69$\pm$0.01 &   3 Apr 2010 & 1.34--1.36 & 642   & HD~114381 \\
ULAS~J133553.45+113005.2 &  T9 & 17.90$\pm$0.01 &  7 Apr 2010 &  1.38--1.42 & 642 & HD~114381 \\
ULAS~J141623.94+134836.3\tablenotemark{b} & T7.5p & 17.35$\pm$0.02 &  6 Apr 2010 & 1.40--1.44 & 642 & HD~124773 \\
\enddata
\tablenotetext{a}{Total integration time divided into four ABBA dithered pointings.  Total integration includes 10.61~s frame readout.}
\tablenotetext{b}{aka SDSS~J141624.08+134826.7B.}
\end{deluxetable*}

Data were reduced using a combination of standard IRAF\footnote{Image Reduction and Analysis Facility (IRAF; \citealt{1986SPIE..627..733T}) is distributed by the National Optical Astronomy Observatory, which is operated by the Association of Universities for Research in Astronomy, Inc., under cooperative agreement with the National Science Foundation.} and custom IDL\footnote{Interactive Data Language.} routines.  A pixel response frame
was first constructed by combining two quartz lamp images taken with the lamp at high (2.5~V) and low (1.5~V) voltage settings.  This was necessary to compensate for the substantial thermal emission 
from the lamp screen which tended to saturate the detector at the longest wavelengths.  The two images were stitched together at column 1300 (roughly 1.4~$\micron$), then normalized by a median combination of all columns along the dispersion region, which eliminated the intensity step function at the stitch and isolated pixel-to-pixel response variations.  The science images were divided by this pixel response frame, and bad pixels interpolated by averaging over nearest neighbors.  The images were then pairwise-subtracted to remove sky background.  Spectra were extracted using the IRAF {\em apall} task.  Calibrator stars were extracted first, then used as traces for the T dwarf and arclamp spectra.
A wavelength solution was generated from a fifth-order spline fit to 60 lines in the arclamp spectra using the IRAF task {\it identify}, producing a mean dispersion of 8.5~{\AA} pixel$^{-1}$ and scatter of 9~{\AA} ($\approx$1 pixel).  Note that the actual dispersion varies considerably along the chip, from 4~{\AA}~pixel$^{-1}$ at 1.0~$\micron$ to 20~{\AA}~pixel$^{-1}$ at 2.4~$\micron$. 
The individual spectra were combined using the {\it xcombspec} routine from SpeXtool \citep{2004PASP..116..362C}, using a robust weighted mean after normalizing in the 1.2--1.4~$\micron$ region.   The SpeXtool task {\it xtellcor{\_}general} was used to compute telluric correction and flux calibration from the A0~V star, following the prescription of 
\citet{2003PASP..115..389V} and assuming a line profile width of 18~{\AA}.

To verify the flux calibration of our data, we computed MKO spectrophotometric colors from all three sources from the FIRE spectra following \citet{2005ApJ...623.1115C}, computing 100 iterations with each spectrum varied by its noise spectrum to determine uncertainties.  
For ULAS~J1335+1130 and SDSS~J1416+1348B, these colors match UKIDSS photometry to within the uncertainties, indicating robust relative flux calibration.
For Ross~458C, the MKO colors are intermediate between the UKIDSS and ground-based GROND \citep{2008PASP..120..405G} photometry reported in \citet{2010MNRAS.405.1140G}, which themselves differ significantly (UKIDSS $J-K$ = 0.51$\pm$0.06;  GROND $J-K$ = -0.21$\pm$0.06).  This discrepancy may be due to differences in the filter basspands between UKIDSS, MKO and GROND filter passbands, or may reflect intrinsic variability in the source.
In Figure~\ref{fig_colorcheck}, we compare our reduced FIRE spectrum of ULAS~J1335+1130 to that obtained
by \citet{2008MNRAS.391..320B} using the Gemini-North Near Infrared Imager and Spectrometer (NIRI; \citealt{2003PASP..115.1388H}) and Paranal Very Large Telescope Infrared Spectrometer and Array Camera (ISAAC; \citealt{1998Msngr..94....7M}).  
The Burningham et al.\  spectrum was stitched together from four orders individually flux-calibrated using UKIDSS $YJHK$ photometry.  Our spectrum is in excellent agreement with the NIRI and ISAAC data, with deviations of less than 10\% outside the telluric 
absorption regions.  Importantly, the FIRE data reproduce the relative $YJHK$ flux peaks, which are essential for spectral model fits (Section~4).

\subsection{Results}

\begin{figure*}
\epsscale{0.8}
\includegraphics[width=0.9\textwidth]{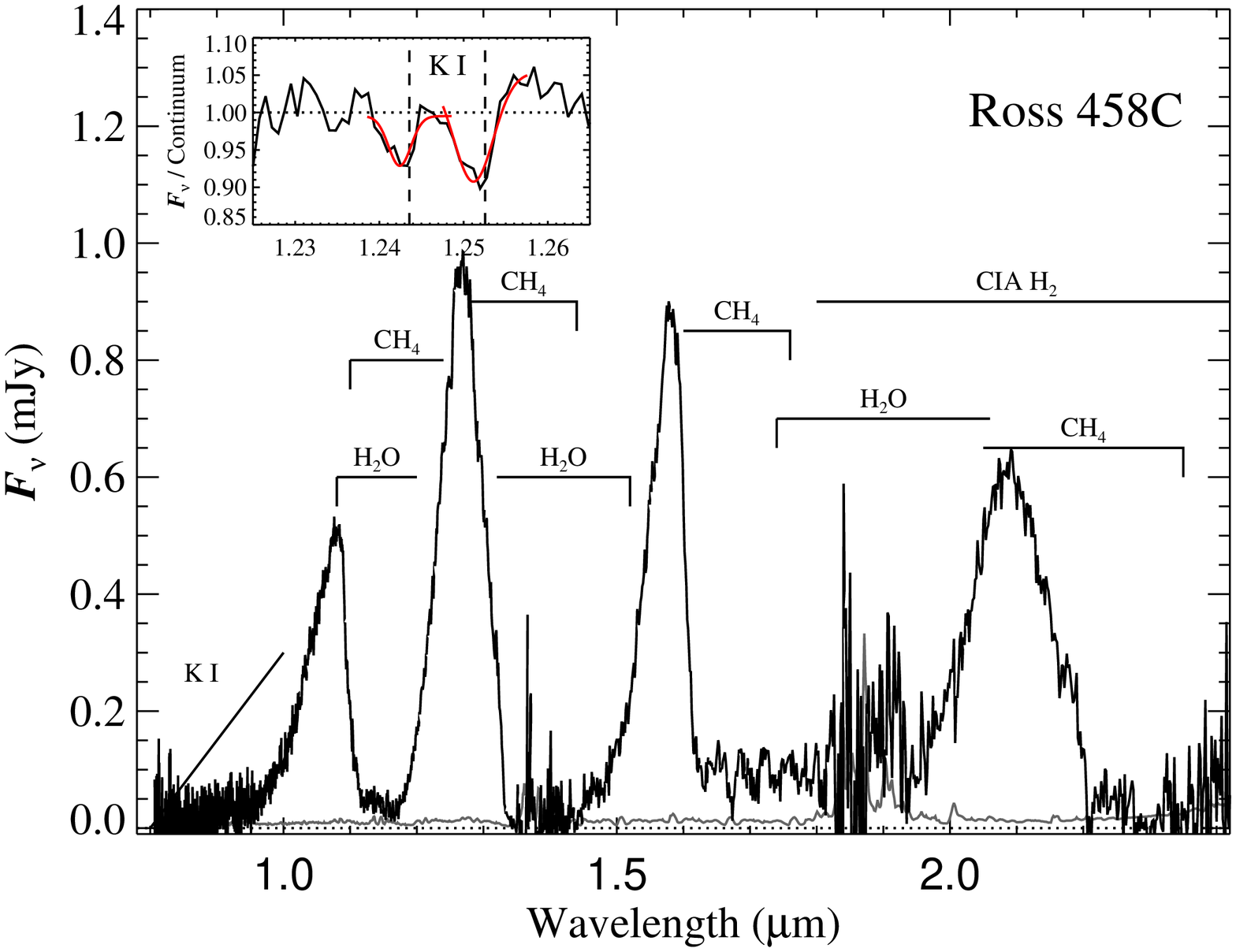}
\caption{FIRE prism spectrum of {\name} in absolute flux units ($F_{\nu}$).
The uncertainty spectrum is indicated by the grey line.  
Prominent {\wat} and {\meth} absorption features are labeled, as well as regions influenced by the pressure-broadened \ion{K}{1} doublet wing ($\lambda \lesssim 1$~$\micron$) and collision-induced H$_2$ opacity ($\lambda \gtrsim 1.75$~$\micron$).  The inset box shows
continuum-normalized flux between 1.23 and 1.26~$\micron$, highlighting
the weak 1.24/1.25~$\micron$ {\ki} doublet present in the spectrum of this source (red lines show gaussian fits).
\label{fig_nirspec}}
\end{figure*}

\begin{figure*}
\epsscale{1.0}
\includegraphics[width=0.9\textwidth]{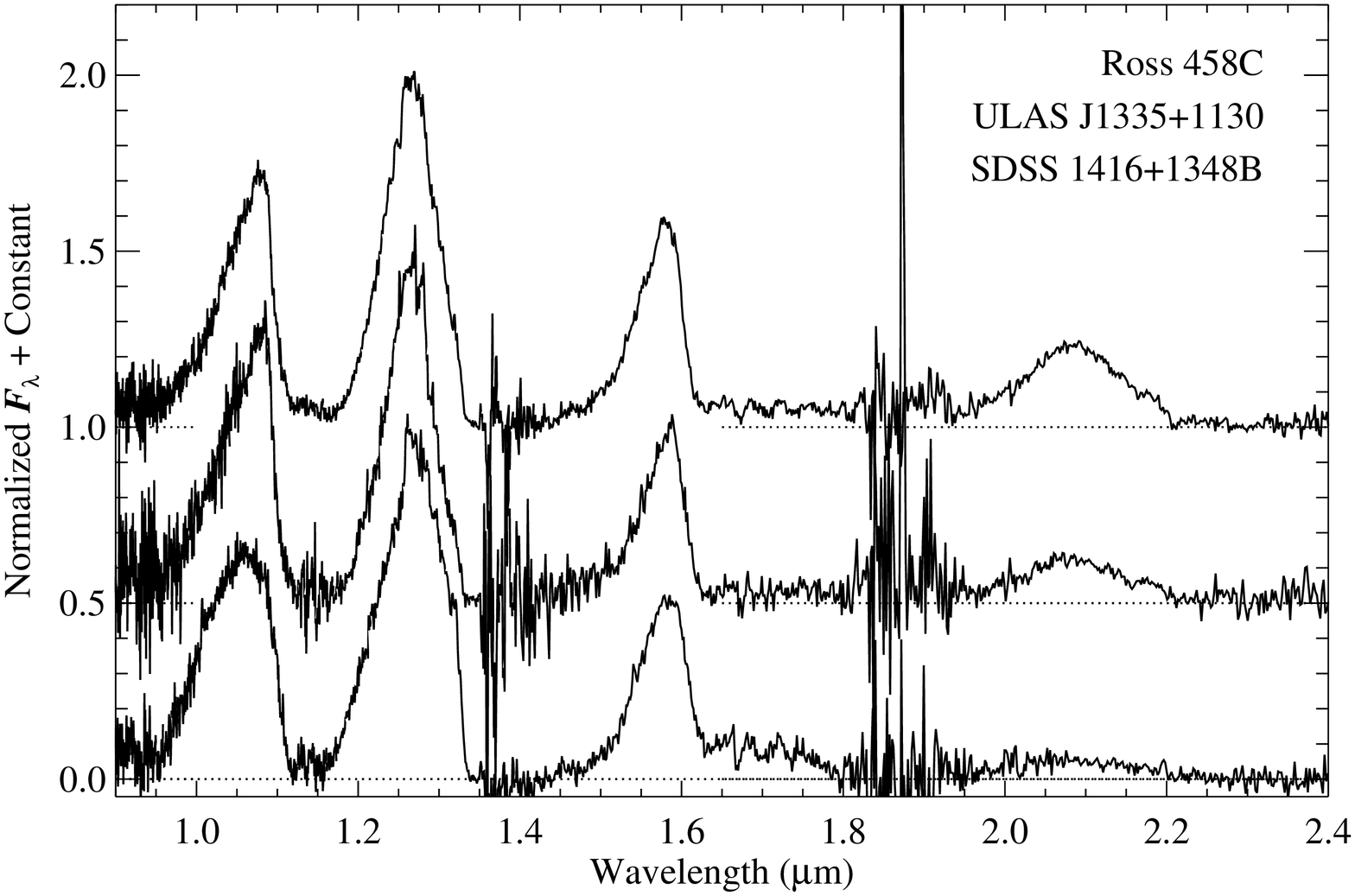}
\caption{FIRE prism spectra of (top to bottom) {\name}, ULAS~J1335+1130  and SDSS~J1416+1348B.
Data are normalized at 1.27~$\micron$ and offset by constants (dotted lines).
Note the variations in the $Y$- and $K$-band flux peaks.
\label{fig_comp}}
\end{figure*}

Figure~\ref{fig_nirspec} displays the reduced spectrum of {\name}, scaled to its absolute flux
($F_{\nu}$ units at 10~pc) based on UKIDSS $J$-band photometry and the Hipparcos parallax
of the Ross~458 system.  
Signal-to-noise (S/N) in the spectrum ranges from 50 to 80 in each of the $YJHK$ flux peaks.
{\name} exhibits the classic spectral signatures of a T dwarf, with strong {\wat} and {\meth} absorption bands at 1.15, 1.4, 1.6, 1.8 and 2.2~$\micron$; a steep red optical slope shaped by the pressure-broadened
0.77~$\micron$ \ion{K}{1} wings; and a relatively
blue spectral energy distribution from 1.2 to 2.1~$\micron$,
due in part to collision-induced H$_2$ absorption \citep{1969ApJ...156..989L, 1994ApJ...424..333S, 2002A&A...390..779B}.
As anticipated from its extreme {\meth}-filter colors, 
the 1.6~$\micron$ {\meth} band of {\name} is quite deep, while 
the $J$- and $H$-band flux peaks are narrow, 
all indicative of a very late-type, low-temperature T dwarf.  Indeed, as shown in Figure~\ref{fig_comp}, these features in the spectrum of {\name} are intermediate between those of SDSS~J1416+1348B and ULAS~J1335+1130, indicating an intermediate spectral type.

Our spectrum of {\name} also reveals the presence 
of the 1.24/1.25~$\micron$ \ion{K}{1} doublet, highlighted in the inset box in Figure~\ref{fig_nirspec}.
These lines are known to fade in the latest-type T dwarfs \citep{2002ApJ...564..421B,2003ApJ...596..561M}, likely a consequence
of KCl(g) formation at temperatures below 1000~K \citep{1999ApJ...519..793L}.
We measured pseudoequivalent widths pEW = 2.7$\pm$0.2~{\AA}
for each doublet line, comparable to measurements for other T7--T8 dwarfs
\citep{2002ApJ...564..421B,2003ApJ...596..561M}.
We do not see these lines in our spectra of ULAS~J1335+1130 or SDSS~J1416+1348B
to limits of 0.5 and 0.3~{\AA}, respectively.   For the latter source, this may be a consequence
of its subsolar metallicity in analogy to the metal poor T6 dwarf 2MASS~J09373487+2931409 (\citealt{2002ApJ...564..421B, 2007ApJ...658.1217M}; see Section~4.2)

Another feature of note in the spectrum of {\name} is its fairly pronounced 
2.1~$\micron$ $K$-band peak.  The relative brightness of this peak is greater than those 
of both ULAS~J1335+1130 and SDSS~J1416+1348B, in accord with its reddest $J-K$ color.  
The $K$-band peak is normally suppressed by H$_2$ absorption in late-type T dwarfs. This absorption is weakened in brown dwarfs with low surface gravities (i.e., young, low-mass) and/or higher metallicities, as both effects result in a lower
photospheric pressure and weaker collision-induced absorption.  For {\name}, both of these properties likely contribute based on the characteristics of the system (Section~2).

\subsection{Spectral Classification}

\begin{deluxetable*}{cccccccc}
\tabletypesize{\footnotesize}
\tablecaption{Spectral Indices for Observed T Dwarfs. \label{tab_indices}}
\tablewidth{0pt}
\tablehead{
 & \multicolumn{2}{c}{Ross 458C} & \multicolumn{2}{c}{ULAS~J1335+1130} & \multicolumn{2}{c}{SDSS~J1416+1348B} & \\
 \cline{2-3} 
 \cline{4-5}
 \cline{6-7}
\colhead{Index} &
\colhead{Value\tablenotemark{a}} &
\colhead{SpT} &
\colhead{Value\tablenotemark{a}} &
\colhead{SpT} &
\colhead{Value\tablenotemark{a}} &
\colhead{SpT} &
\colhead{Ref.}  \\
}
\startdata
{\wat}-J & 0.051$\pm$0.002 & T8.0 &  0.028$\pm$0.003 & T8.5  & 0.044$\pm$0.003 & T8.0 &   1  \\
{\meth}-J & 0.154$\pm$0.001 & T8.5 &  0.109$\pm$0.002 & T9.0 &  0.262$\pm$0.002 & T7.0 &  1  \\
$W_J$ & 0.272$\pm$0.002 & T9.0 &  0.253$\pm$0.002 & T9.0 & 0.360$\pm$0.002 & T7.0 &  2,3  \\
{\wat}-H & 0.193$\pm$0.003 & T7.5 & 0.192$\pm$0.007 & T7.5 & 0.148$\pm$0.004 & T9.0 &   1  \\
{\meth}-H & 0.112$\pm$0.003 & T8.0  &  0.068$\pm$0.005 & T9.0 & 0.192$\pm$0.003 & T7.0 &   1  \\
{\ammon}-H & 0.632$\pm$0.004 & \nodata & 0.537$\pm$0.007 & \nodata & 0.610$\pm$0.005 & \nodata &   4  \\
%{\wat}-K & 0.360$\pm$0.006 & T7.0 &  0.349$\pm$0.016 & T7.0 &  0.365$\pm$0.014 & T6.5 &  1  \\
{\meth}-K & 0.078$\pm$0.005 & T7.5 &  0.073$\pm$0.013 & T7.5 & 0.191$\pm$0.032 & T5.5 &   1  \\
K/J & 0.167$\pm$0.001 & \nodata & 0.130$\pm$0.002 & \nodata & 0.044$\pm$0.001 & \nodata &  1  \\
$J-H$\tablenotemark{b} & -0.14$\pm$0.01 & \nodata & -0.34$\pm$0.02 & \nodata & -0.25$\pm$0.01 & \nodata &  \\
$H-K$\tablenotemark{b} & 0.38$\pm$0.02 & \nodata & 0.06$\pm$0.05 & \nodata & -1.02$\pm$0.09 & \nodata &  \\
$J-K$\tablenotemark{b} & 0.24$\pm$0.02 & \nodata & -0.28$\pm$0.04 & \nodata & -1.25$\pm$0.09 & \nodata &  \\
\enddata
\tablenotetext{a}{Values for each source were measured for 1000 realizations of the spectrum (100 realizations for spectrophotometric colors), each with a normal distribution of random values scaled by the noise spectrum added to the original fluxes.  The reported values are the means and standard deviations of these measurements.}
\tablenotetext{b}{Spectrophotometric colors computed according to \citet{2005ApJ...623.1115C}.} 
\tablerefs{(1) \citet{2006ApJ...637.1067B}; (2) \citet{2007MNRAS.381.1400W}; (3) \citet{2008MNRAS.391..320B}; (4) \citet{2008A&A...484..469D}.}
\end{deluxetable*}

To classify the T dwarfs, we used the near-infrared spectral indices and index-spectral type relations
defined in \citet{2006ApJ...637.1067B, 2007MNRAS.381.1400W}; and \citet{2008MNRAS.391..320B}; these are listed in Table~\ref{tab_indices}.  
Our index measurements for ULAS~J1335+1130 and SDSS~J1416+1348B are in rough agreement with previous
results \citep{2008MNRAS.391..320B, 2010MNRAS.404.1952B, 2010AJ....139.2448B}, with minor
variances attributable to differences in spectral resolution and S/N between data sets.  
We derive mean classifications of T8, T8.5 and T7.5p for {\name}, ULAS~J1335+1130 and SDSS~J1416+1348B, respectively.  Uncertainties of 0.5~subtypes are assumed for the first two sources, while the last classification is deemed ``peculiar'' due to the larger scatter in indices ($\pm$1.2 subtypes).   The last two classifications formally agree with previously published results \citep{2008MNRAS.391..320B, 2010MNRAS.404.1952B, 2010AJ....139.2448B}.  Our classification of
{\name} splits the difference between the photometric estimates of \citet{2010A&A...515A..92S}
and \citet{2010MNRAS.405.1140G}, and is consistent with the relative ordering of the spectra in Figure~\ref{fig_comp}.
The absolute $J$ magnitude of {\name} is also comparable to other T8 dwarfs \citep{2004AJ....127.2948V}.

\subsection{Physical Properties of {\name}}

With a spectral type and known distance, it is possible to assess some 
of the basic physical properties of {\name}.  A bolometric luminosity was
determined using the absolute UKIDSS
$JHK$-band magnitudes and 
bolometric corrections ($BC$) from \citet{2010arXiv1008.2200L}.  Adopting\footnote{Error estimates quoted here take into account uncertainties in the spectral classification ($\pm$0.5 subtype), photometry, parallax of Ross~458 and the spectral type/$BC$ relations (0.14, 0.07 and 0.08 mag for $J$, $H$ and $K$, respectively).  Systematic errors---e.g., from gravity or metallicity effects---are not accounted for.} $BC_{J}$ = 2.62$\pm$0.14, $BC_H$ = 2.19$\pm$0.07 and $BC_K$ = 2.05$\pm$0.09 for a T8 dwarf, we derive estimates of {\lbol} = -5.69$\pm$0.06, {\lbol} = -5.65$\pm$0.04 and {\lbol} = -5.55$\pm$0.05 based on $J$-, $H$- and $K$-band fluxes, or a weighted mean of {\lbol} = -5.62$\pm$0.03.
This luminosity is similar to that of the
T8 spectral standard 2MASS J04151954-0935066 (hereafter 2MASS~J0415-0935; {\lbol} = -5.67$\pm$0.02; \citealt{2007ApJ...656.1136S}).  The {\teff} of {\name}
can be inferred from its luminosity by adopting a radius from the evolutionary models of SM08.
For an age of 150--800~Myr (Section~2.2), these models
predict a radius of 0.12--0.13~R$_{\sun}$, {\teff} = 650$\pm$25~K and a  
mass in the range 0.006--0.011~{\msun}.
The inferred temperature is cooler than that of 2MASS~J0415-0935 ({\teff} = 750$\pm$25~K; \citealt{2007ApJ...656.1136S}) due to the inflated radius, while the inferred mass is below the deuterium burning minimum mass limit (0.014~{\msun}; \citealt{2000ARA&A..38..337C, 2010arXiv1008.5150S}).

\section{Spectral Model Fits}

\subsection{Spectral Models and Method}

We compared the FIRE spectra of these three T dwarfs to a suite of 
atmosphere models from \citet{2002ApJ...568..335M} and SM08, following the prescriptions detailed in \citet{2008ApJ...678.1372C,2008ApJ...689L..53B} and \citet{2009ApJ...706.1114B}.  
The one-dimensional models include all significant gas species considered in \citet{2008ApJS..174..504F} as well as CO$_2$ (cf. \citealt{2010arXiv1008.3732Y}), albeit with
line lists for {\meth} and {\ammon} that are known to be incomplete (e.g., \citealt{2006ApJ...647..552S}). Solar abundances from \citet{2003ApJ...591.1220L} were assumed. 
Non-equilibrium chemistry of CO, CH$_4$, H$_2$O and NH$_3$ (the relevant molecules influencing the near-infrared spectra of T dwarfs) by vertical transport was accounted for using
an eddy diffusion parameterization, {\kzz} \citep{1996ApJ...472L..37F, 1999ApJ...519L..85G, 2006ApJ...647..552S, 2007ApJ...669.1248H}.
Equilibrium condensate formation was included in the chemistry\footnote{The chemical equilibrium calculation accounts for all important condensed species \citep{2006asup.book....1L, 2010ApJ...716.1060V}.  For the cloud opacity we included only silicates, iron, and Al$_2$O$_3$ (alumina or corundum).  Other condensates are expected to form above the silicate cloud at low effective temperature, but of these only Na$_2$S is likely to have any significant column opacity \citep{2000ASPC..212..152M}.  In a future work we will consider the possible contribution of this species.}, with grain size and vertical distributions parameterized by a sedimentation rate {\fsed}
\citep{2001ApJ...556..872A,2002ApJ...568..335M}.
While contemporaneous studies of late-type T dwarfs have generally ignored the opacity from condensate grains 
(e.g., \citealt{2007ApJ...667..537L, 2009ApJ...695.1517L, 2007ApJ...656.1136S, 2008ApJ...689L..53B}), we considered two sets of SM08 models with (``cloudy'') and without (``cloudless'') condensate opacity included.  
Both sets of models use the same gas chemistry and have the same constituent abundances at any given pressure and temperature.  The cloudless models assume that condensates do not contribute to the opacity (but do contribute to the gas chemistry), perhaps because they have ``rained out'' of the atmosphere or occupy only a small fraction of the visible surface area.   The cloudy models explicitly include opacity from cloud grains.
The cloudless models sampled atmospheric parameters
{\teff} = 500--1000~K (50~K steps); {\logg} = 4.0--5.5~cgs (0.5~dex steps); [M/H] = $-$0.3, 0 and +0.3 dex relative to solar; and {\kzz} = 0 and 10$^4$~{\cmms}.
The cloudy models sampled the same parameter space with {\fsed} = 2 and solar metallicity alone.
Hereafter, we refer to sets of parameters by the notation $\bf{k}$ = ({\teff}, {\logg}, [M/H], {\fsed}, $\log{K_{zz}}$), where {\fsed} = NC for the cloudless models.
The model spectra were computed as surface fluxes ($F_{\nu}$ flux units), smoothed 
to the mean resolution of the FIRE prism data ({\ldl} = 300) using a gaussian kernel, and interpolated
onto the same wavelength scale as the data.
We also assigned physical parameters of mass, age and radius to each spectral model, interpolating the appropriate evolutionary tracks from SM08.

The observational data (also in $F_{\nu}$ flux units) were scaled to their apparent UKIDSS $J$-band magnitudes.  Simultaneous fits were made to the 0.85--1.35~$\micron$, 1.42--1.8~$\micron$  and 1.95--2.35~$\micron$ regions to avoid strong telluric absorption.  For each data-model comparison, the goodness-of-fit statistic $G_{\bf k}$ \citep{2008ApJ...678.1372C} was calculated following the same weighting scheme employed in that study, with each pixel weighted by its breadth in wavelength space ($w_j \propto \Delta\lambda_j$).  Model surface fluxes were scaled by the factor $C_{\bf k} \equiv (R/d)^2$ which minimizes $G_{\bf k}$ (Equation~2 in \citealt{2008ApJ...678.1372C}), where $R$ is the radius of the brown dwarf and $d$ its distance from the Earth.  As in \citet{2008ApJ...689L..53B}, we generated
distributions of the fit parameters by weighting each model's contribution according to the F-distribution probability distribution function (PDF):
\begin{equation}
W_i \propto 1-F(G_{\bf k_i}/{\rm min}(G_{\bf k}) \mid \nu_{eff},\nu_{eff}).
\end{equation}
Here, $G_{\bf k_i}/{\rm min}(G_{\bf k})$  is the ratio of goodness-of-fit (effectively a ratio of chi-square residuals) between the best-fit model and the $i^{th}$ model, 
and $F(x{\mid}\nu_{eff},\nu_{eff})$ is the F-distribution PDF for effective degrees
of freedom
\begin{equation}
\nu_{eff} \equiv \left( \frac{1}{{\rm max}(\{w\})}\sum_{j=1}^Nw_j \right) - 1.
\label{equ_nu}
\end{equation}
The sum is over all pixels included in the fit (see \citealt{2010ApJ...710.1142B}).   The F-distribution provides a more robust assessment of the equivalence of different
model fits than the exponential weighting employed in \citet{2008ApJ...689L..53B} and \citet{2010AJ....139.2448B}.
Parameter means and one-sided standard deviations were separately computed as 
\begin{equation}
\langle{\bf k}\rangle \equiv \frac{\sum_iW_i{\bf k}_i}{\sum_iW_i}
\label{equ_meanp}
\end{equation}
and
\begin{equation}
\sigma_{\bf k_{\pm}}^2 = \frac{\sum_{i_{\pm}}W_i({\bf k}_i-\langle{\bf k}\rangle)^2}{\sum_iW_i}.
\label{equ_sigmap}
\end{equation}
The first sum was performed over all models and the second restricted to those models whose parameter values are above (+) or below ($-$) the weighted mean $\langle\bf{k}\rangle$. 

We examined several variations in our fitting method to assess internal systematic effects.  Modest changes to the spectral regions that contributed to computing $G_{\bf k}$ made little difference to the overall best fit.  For example, including the telluric regions in the fits, or expanding the lower and upper ends of the spectral range used, resulted in changes of less than 10~K and 0.1~dex in the mean {\teff} or {\logg} values for all three sources.  Only by excluding whole spectral regions---such as the individual $YJHK$-band flux peaks---did inferred {\teff} and {\logg} values change significantly, by up to 100~K and 0.5~dex (similar to band-to-band variations reported in \citealt{2008ApJ...678.1372C}).  Variations in the pixel weighting scheme also had negligible effect.  Weighting all pixels equally gave identical parameters (to within 10~K and 0.1~dex in {\teff} and {\logg}) as weighting by pixel bandpass.  Finally, using the exponential parameter weighting employed in \citet{2010AJ....139.2448B}, $W_i \propto e^{-0.5G_{\bf k_i}}$, also produced negligible changes in the inferred parameters and their uncertainties.  We therefore assume systematic uncertainties of 20~K in {\teff} and 0.2~dex in {\logg}, although we cannot rule out larger systematic {\em biases} based on the accuracy of the models.

\subsection{Results}

\begin{figure*}
\centering
\epsscale{1.0}
\includegraphics[width=0.9\textwidth]{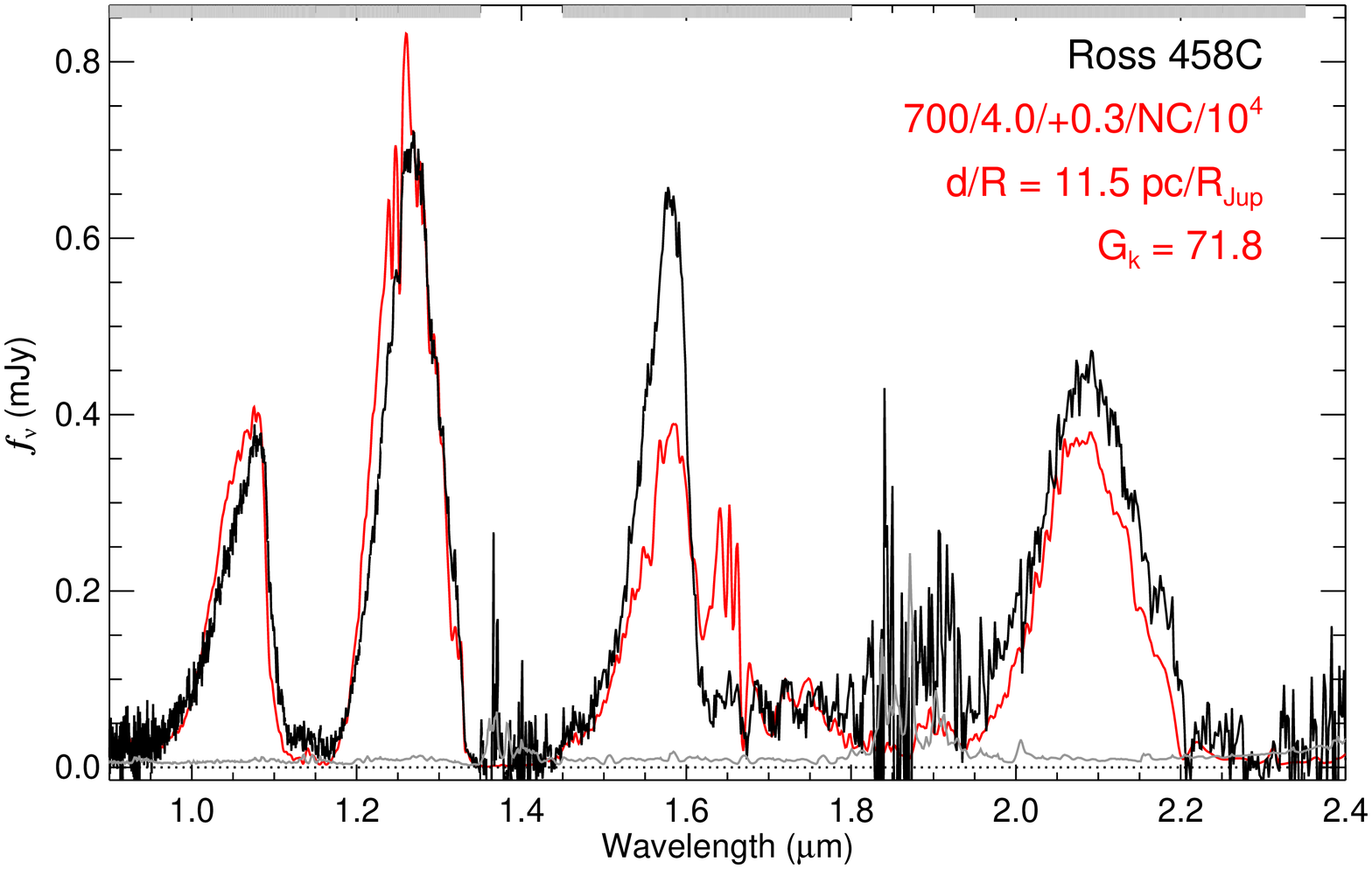}
\includegraphics[width=0.9\textwidth]{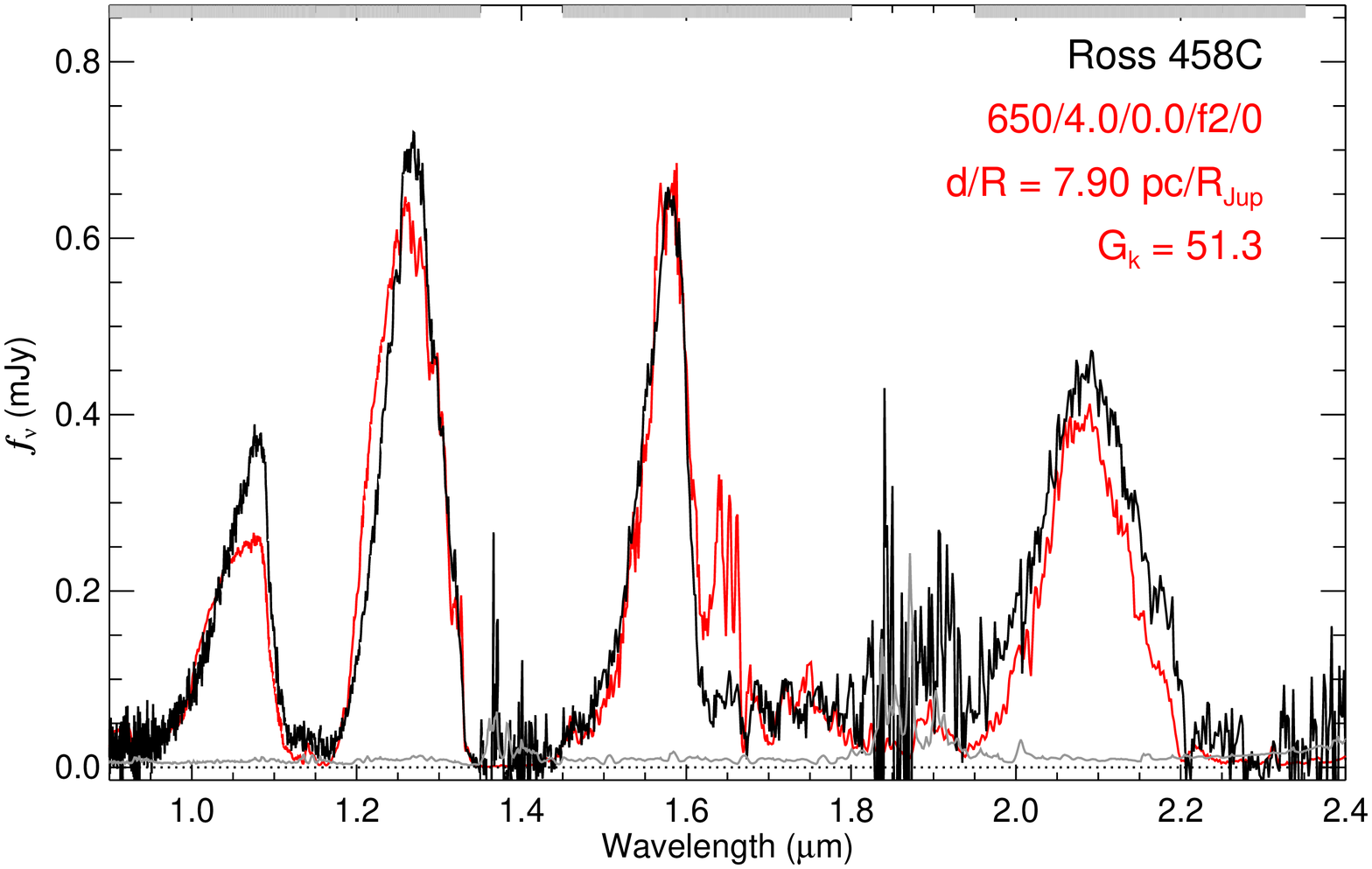}
\caption{Best-fit cloudless (top) and cloudy (bottom) model fits (red lines) to the observed spectrum for {\name} (black lines). Observed data are scaled to the apparent UKIDSS $J$-band magnitude of the source, while the model is scaled to minimize $G_k$ (indicated, along with d/R flux scaling factor). 
The noise spectrum for {\name} is shown by the grey lines, and fit regions indicated
by the shaded boxes at the top of each panel.
 \label{fig_modelfit_ross458}}
\end{figure*}

\begin{figure*}
\centering
\epsscale{1.0}
\includegraphics[width=0.9\textwidth]{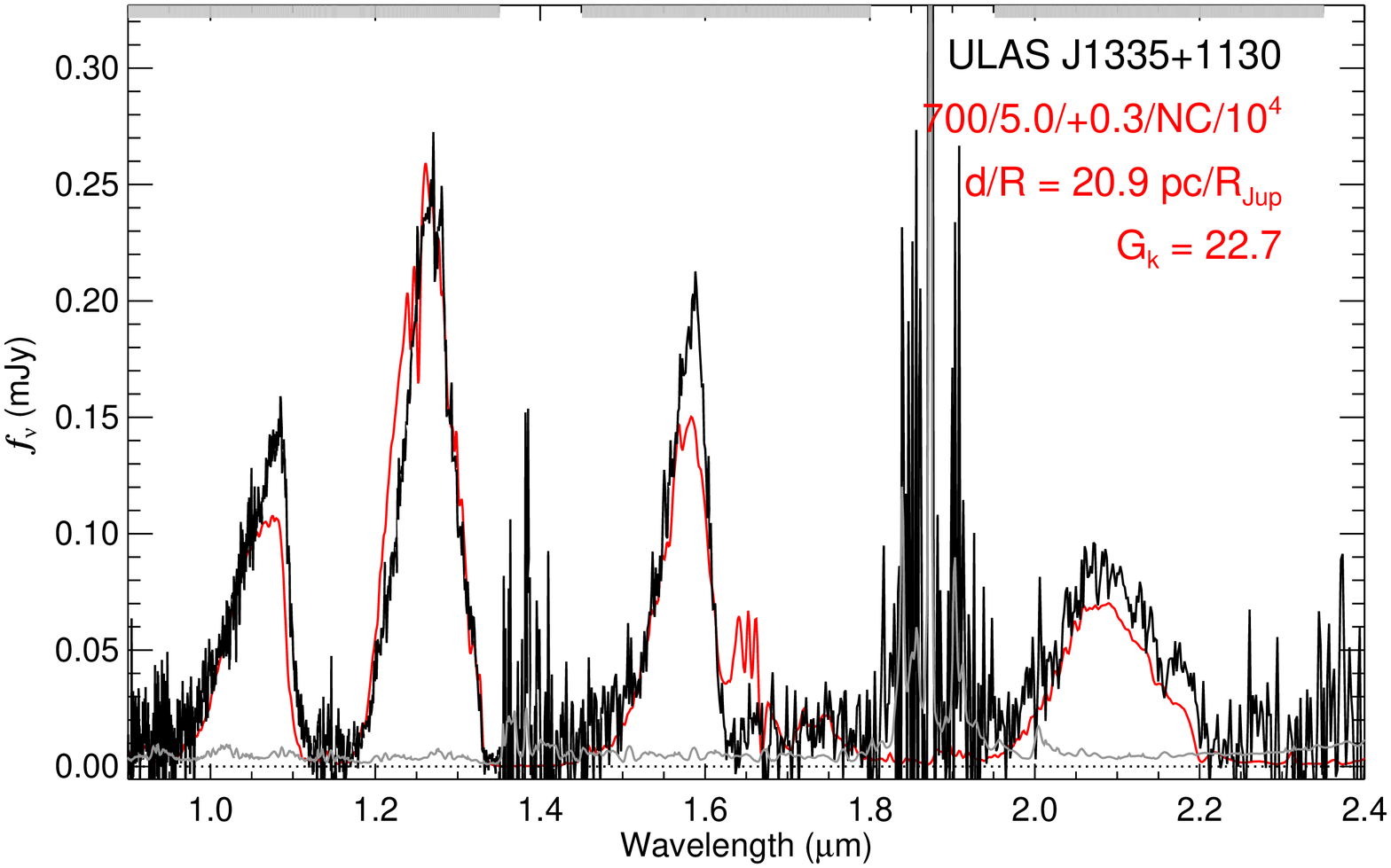}
\includegraphics[width=0.9\textwidth]{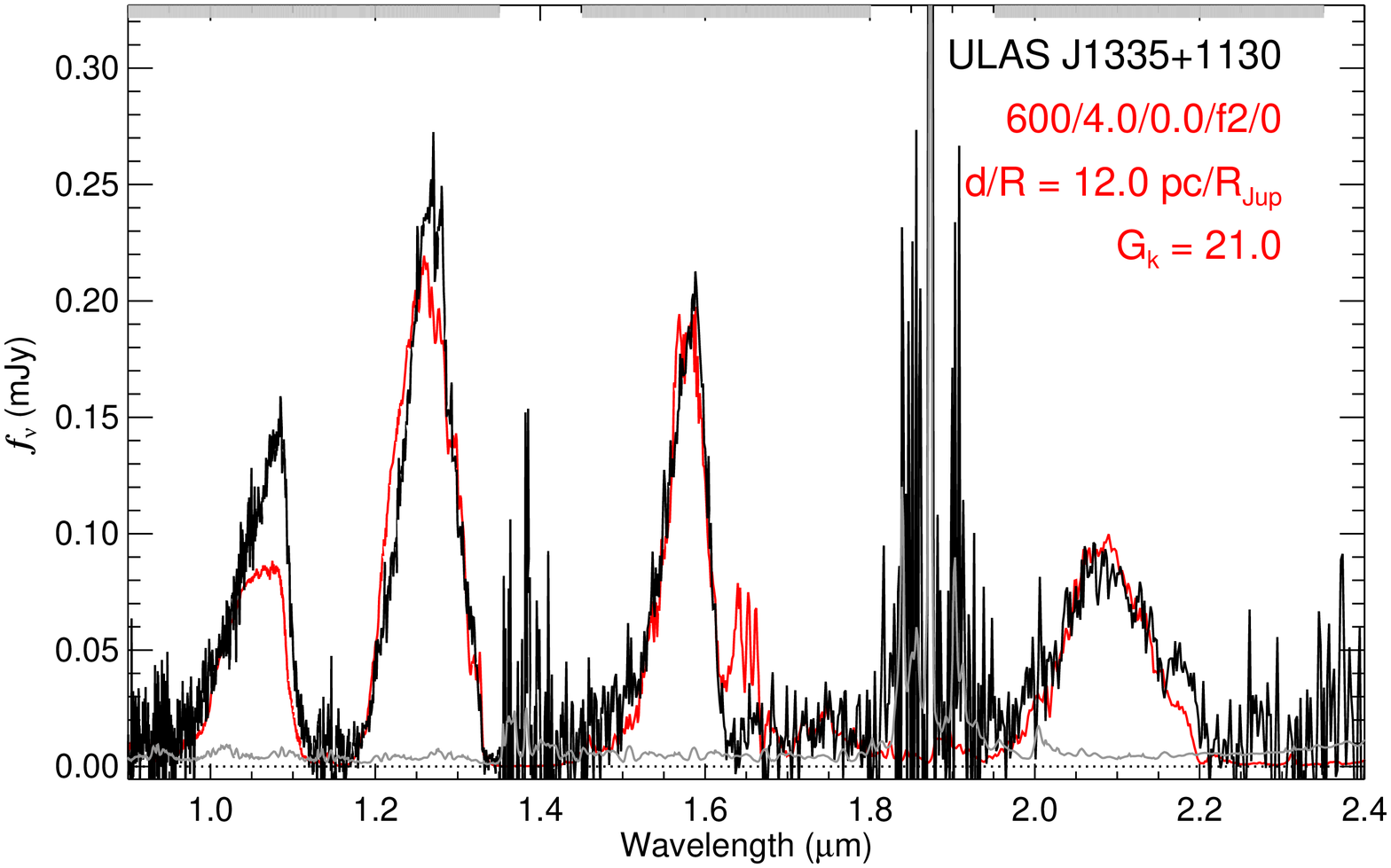}
\caption{Same as Figure~\ref{fig_modelfit_ross458} for ULAS~J1335+1130. 
 \label{fig_modelfit_1335}}
\end{figure*}

\begin{figure*}
\centering
\epsscale{1.0}
\includegraphics[width=0.9\textwidth]{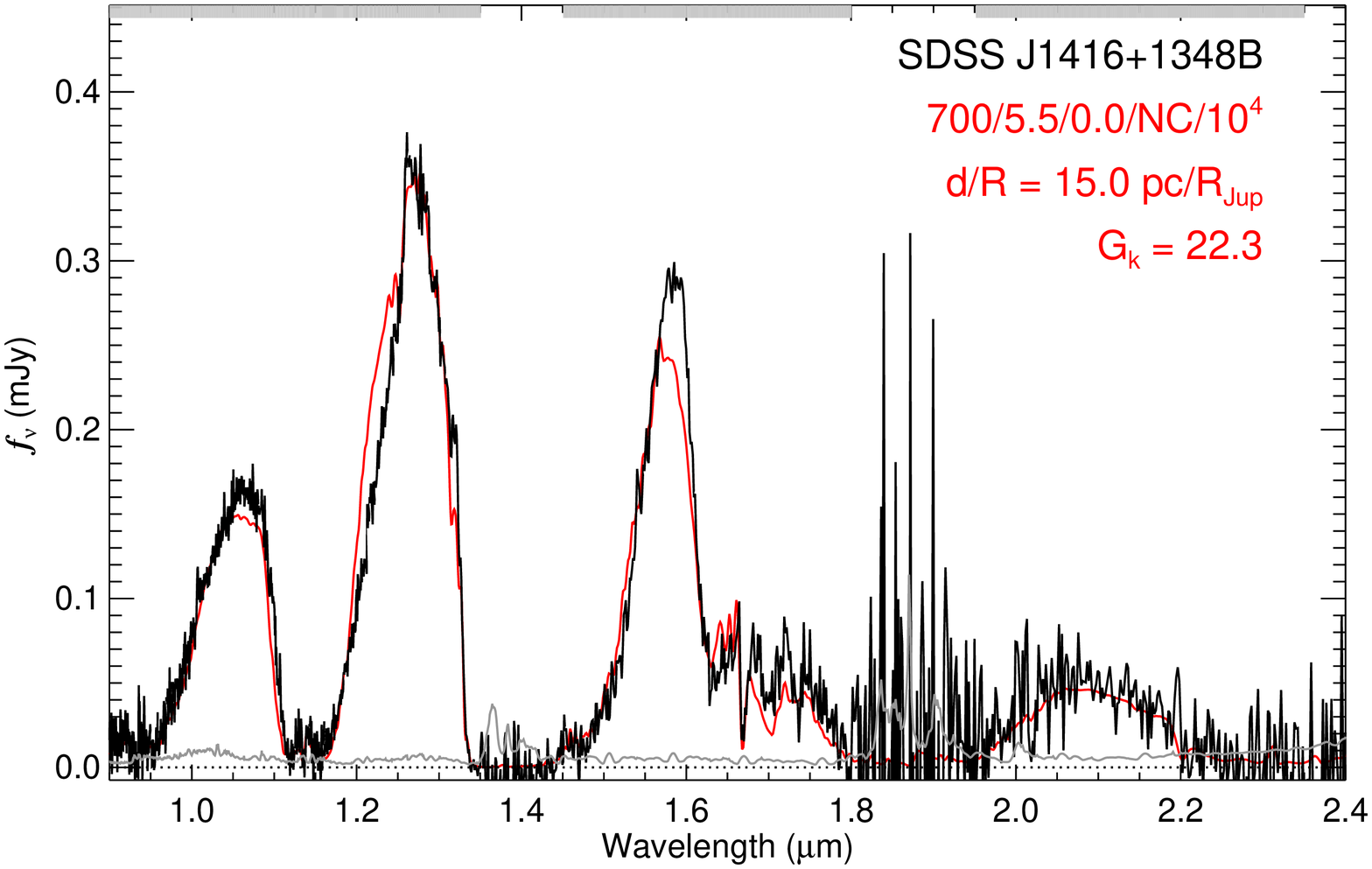}
\includegraphics[width=0.9\textwidth]{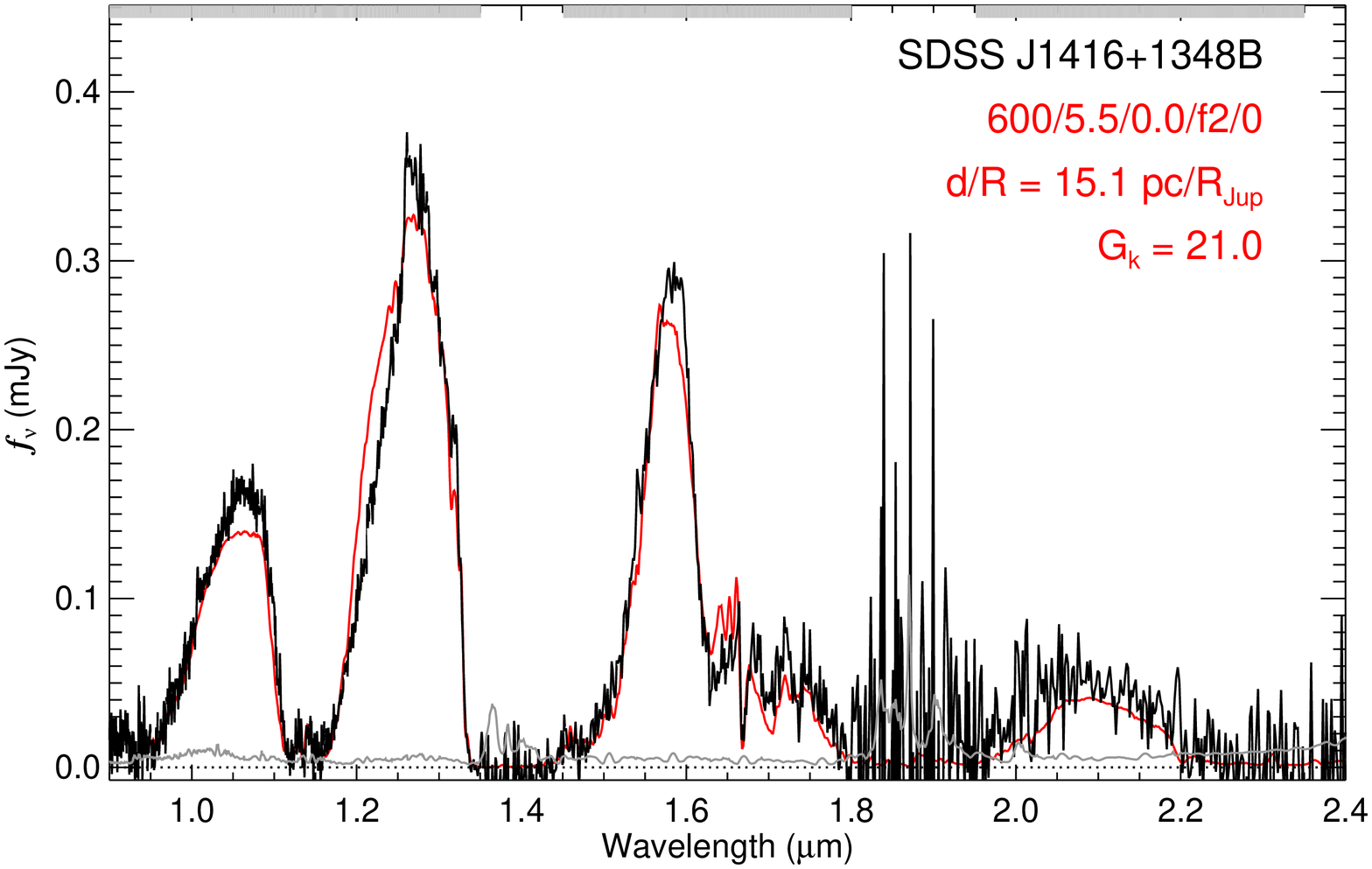}
\caption{Same as Figure~\ref{fig_modelfit_ross458} for SDSS~J1416+1348B. 
 \label{fig_modelfit_1416}}
\end{figure*}

\begin{figure*}
%\centering
\epsscale{1.0}
\includegraphics[width=0.24\textwidth]{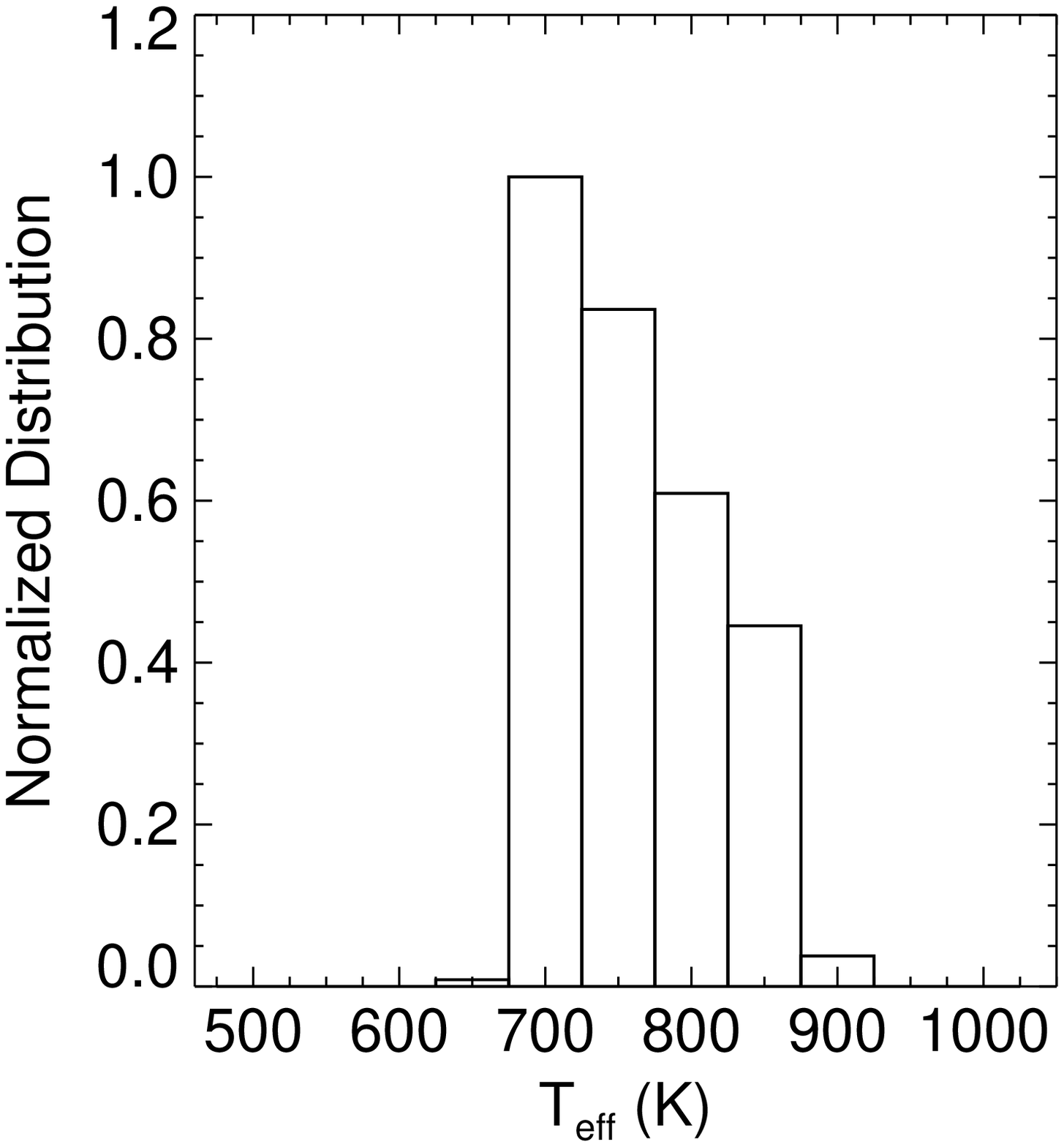}
\includegraphics[width=0.24\textwidth]{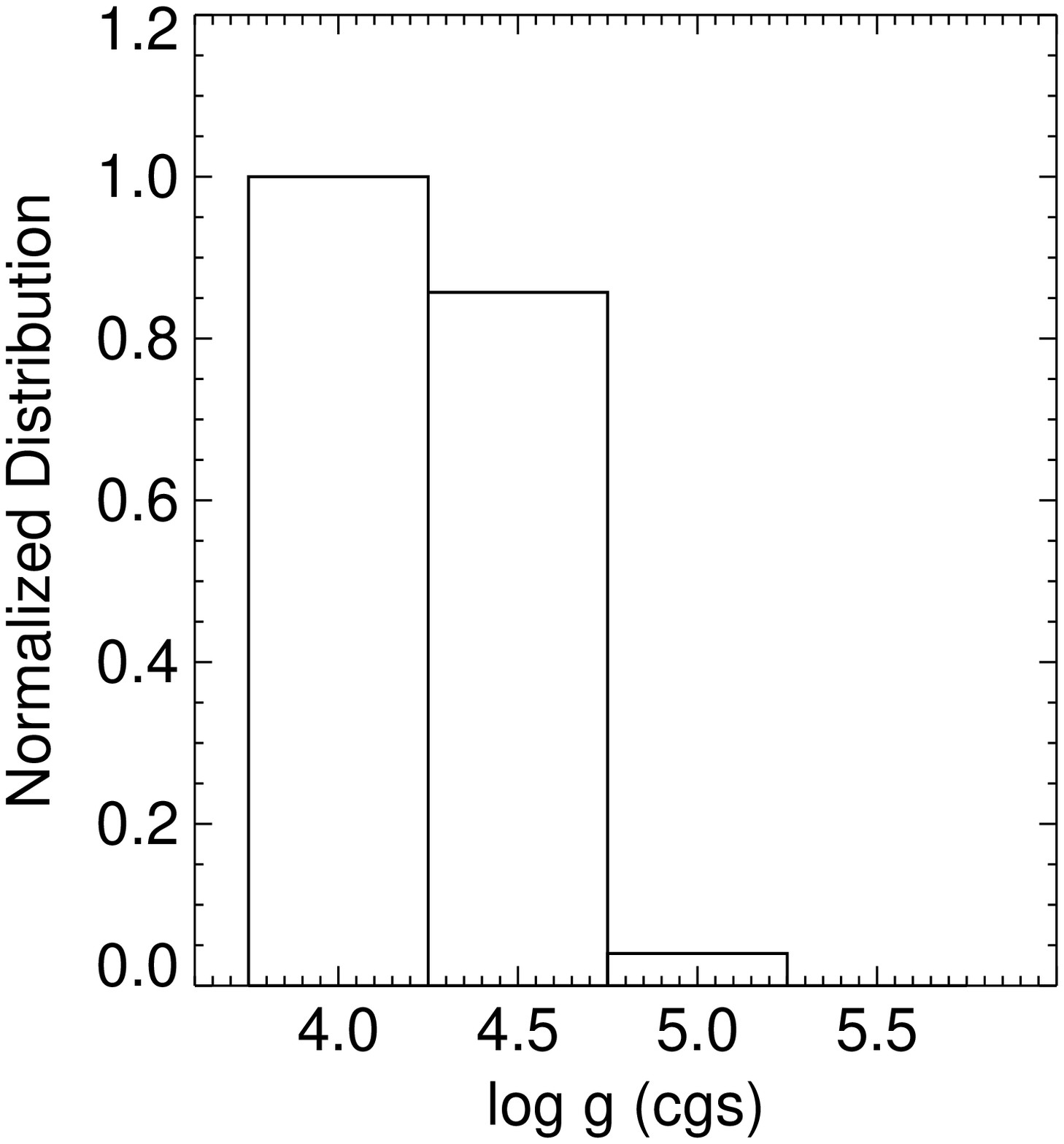}
\includegraphics[width=0.24\textwidth]{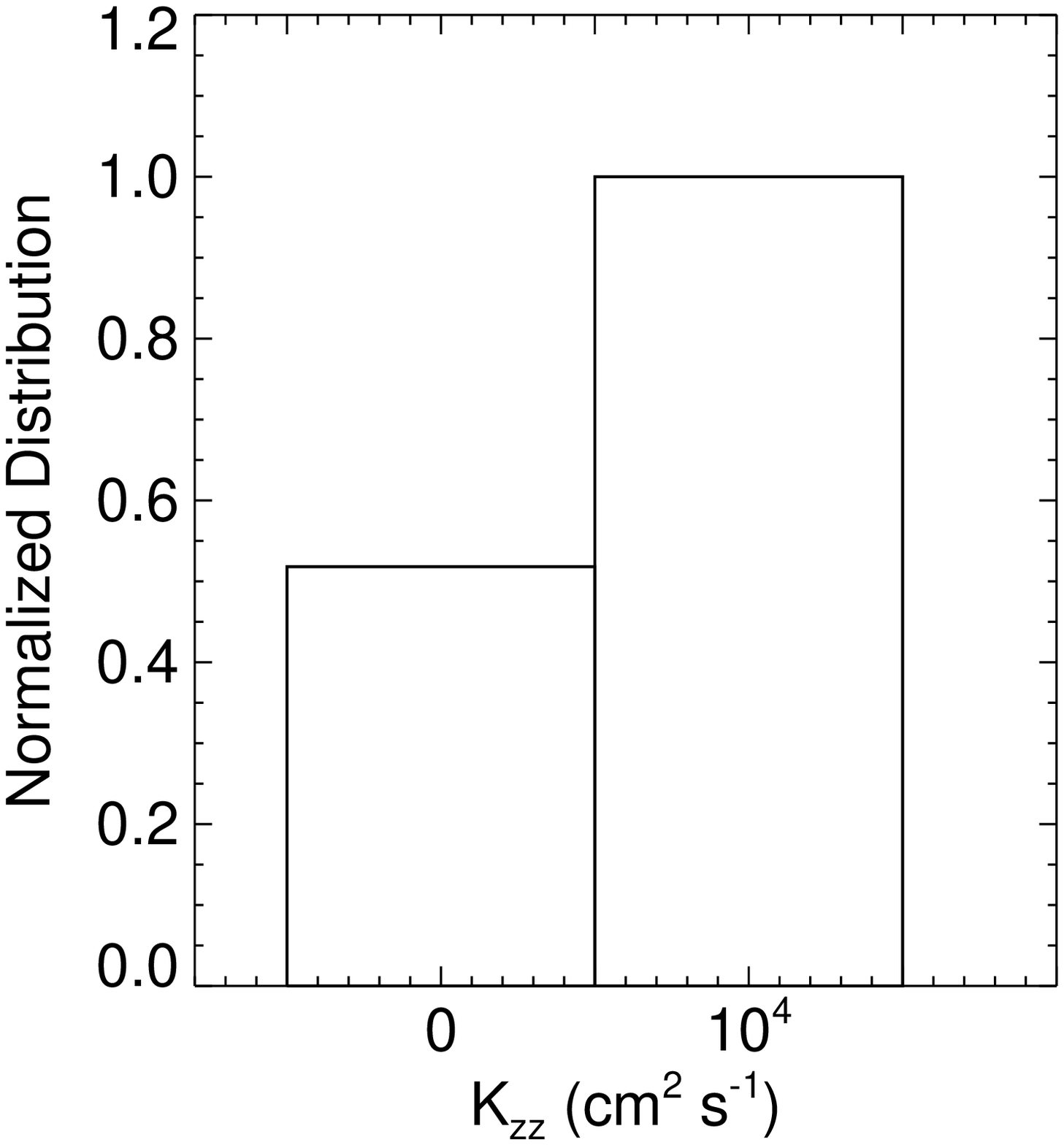}
\includegraphics[width=0.24\textwidth]{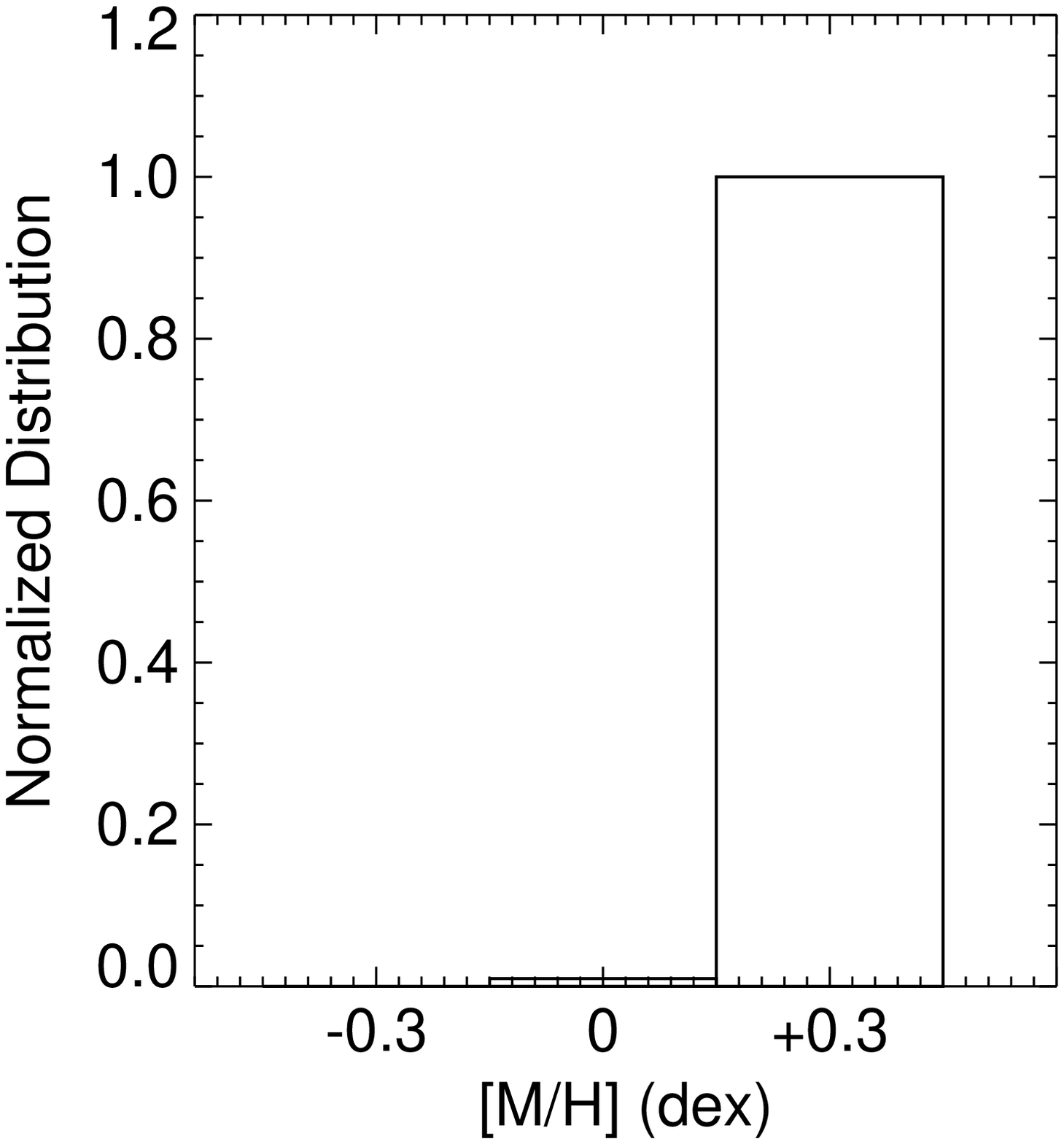} \\
\includegraphics[width=0.24\textwidth]{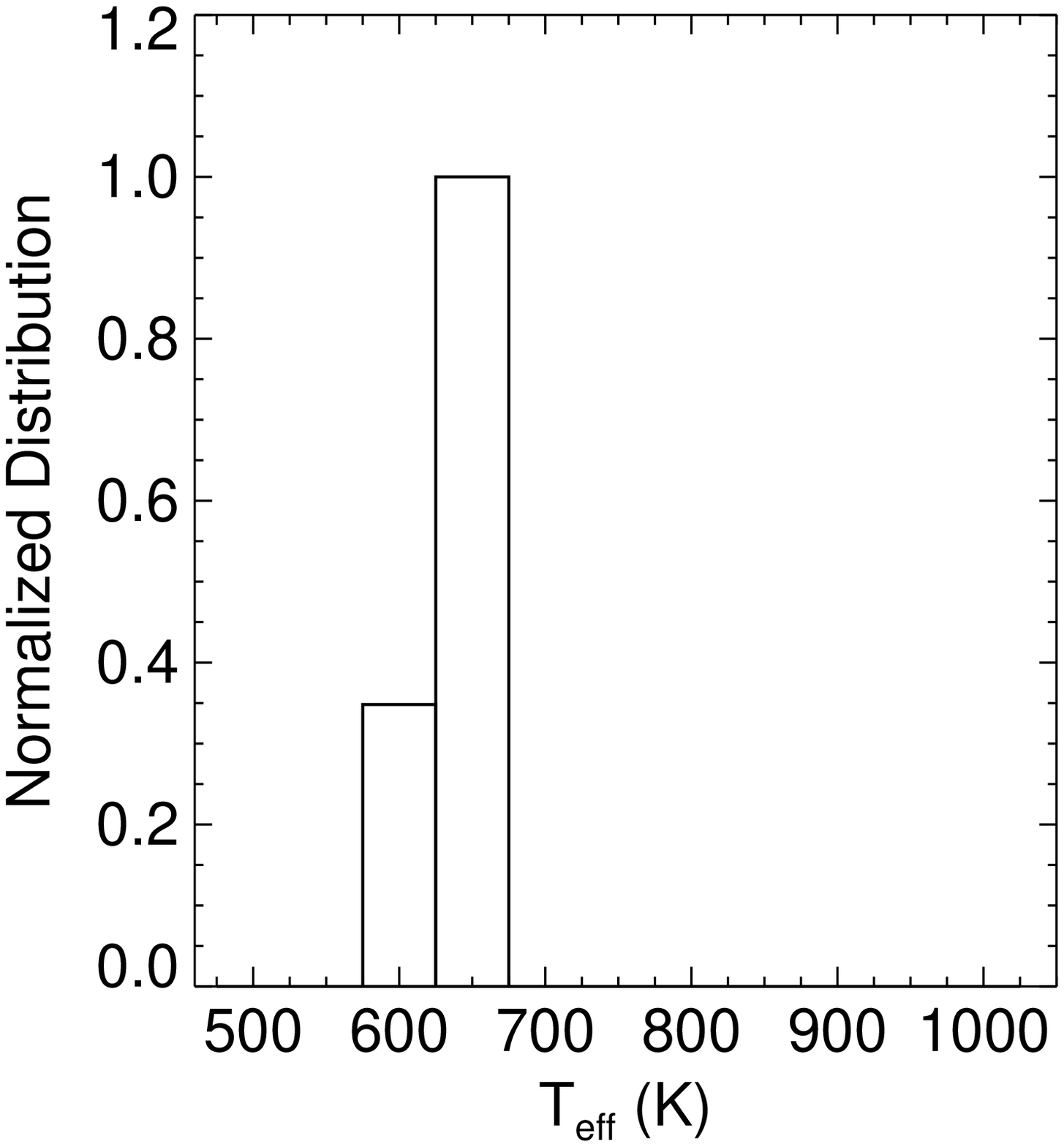}
\includegraphics[width=0.24\textwidth]{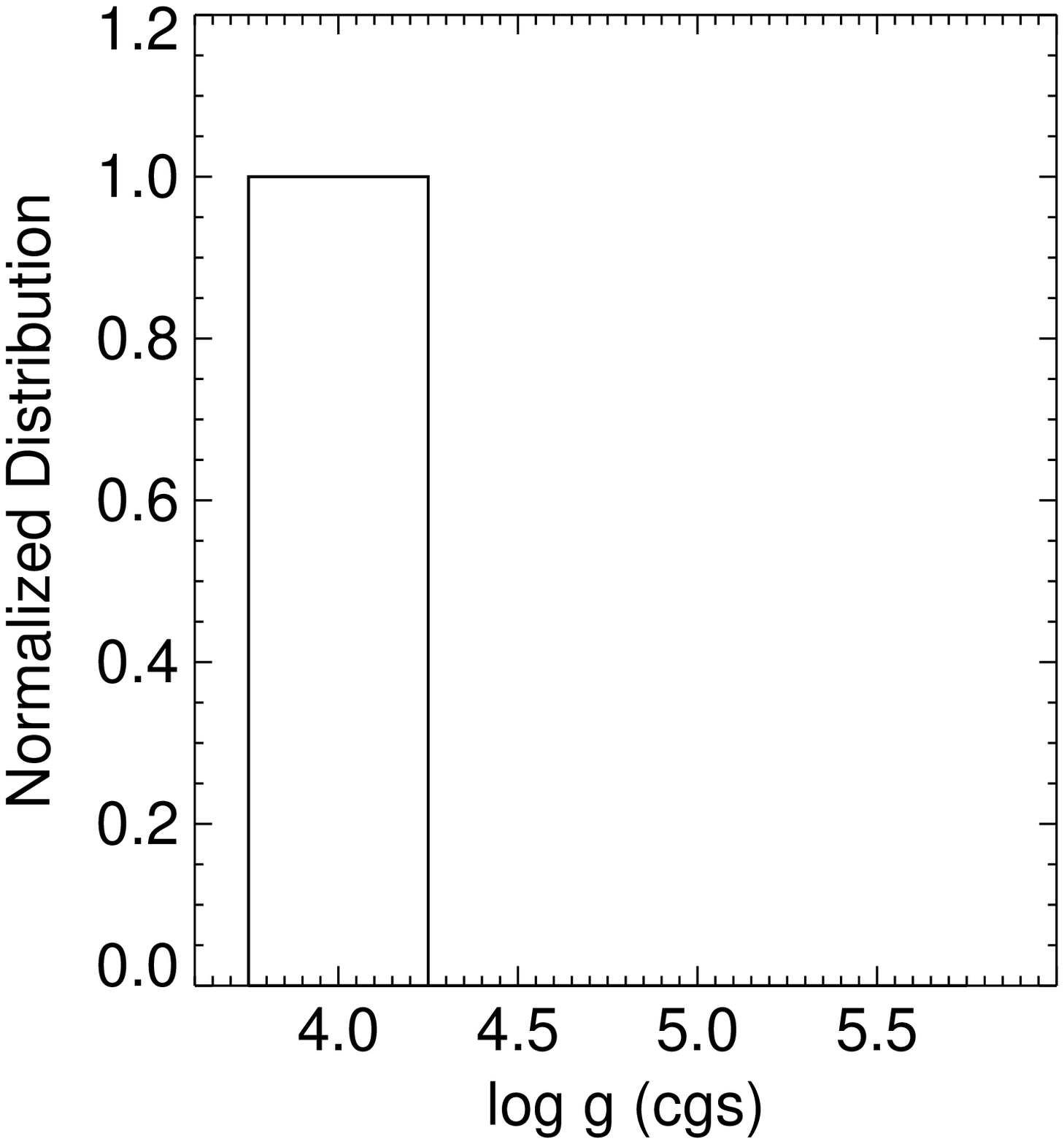}
\includegraphics[width=0.24\textwidth]{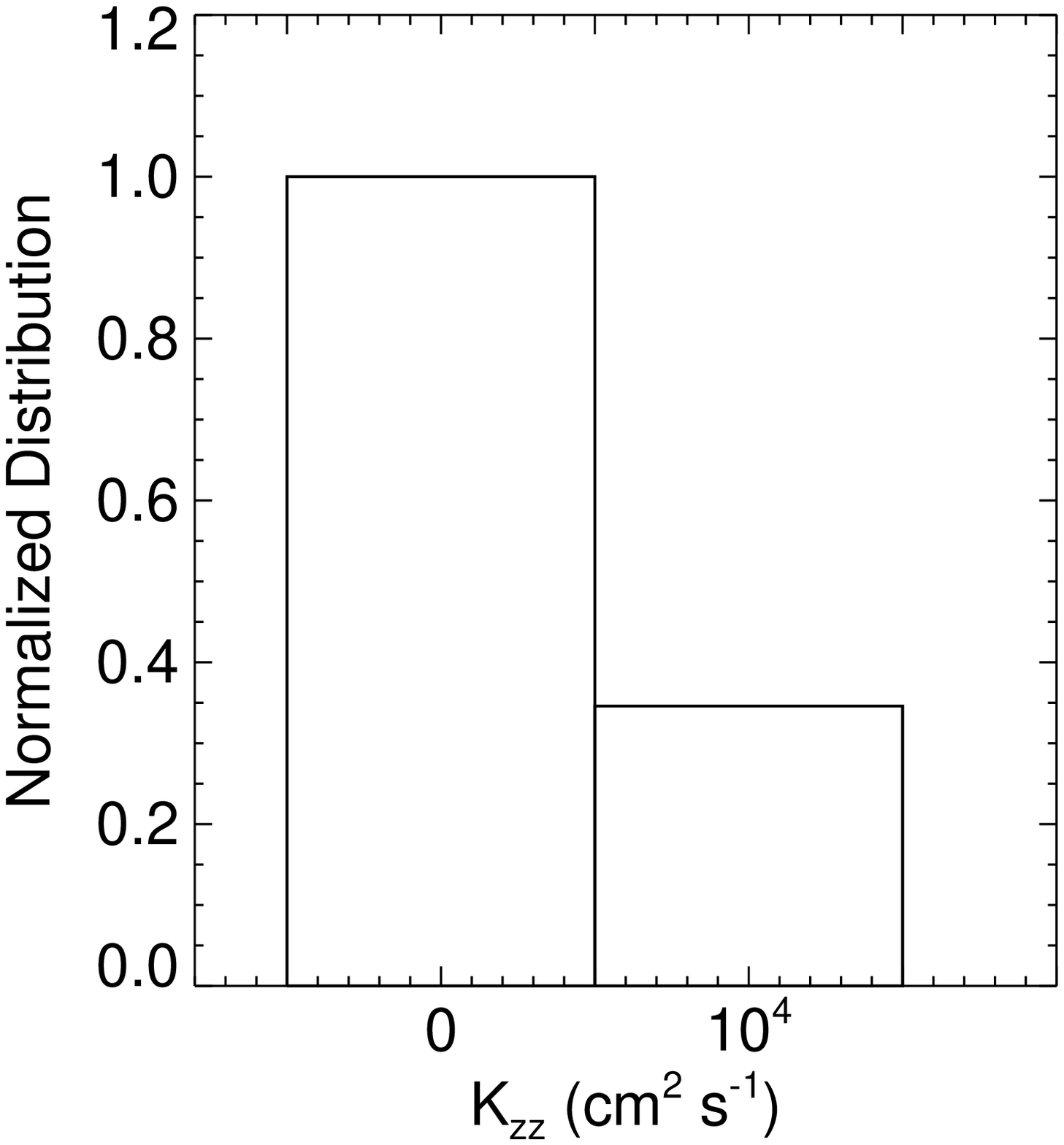}
\caption{Parameter distributions for cloudless (top) and cloudy (bottom) model fits of {\name}, based on the parameter weighting scheme described in the text. From left to right, distributions in {\teff}, {\logg} and {\kzz} are shown.  For the cloudless models, the distribution in [M/H] values is also shown.
 \label{fig_modelfit_ross458_parameters}}
\end{figure*}

\begin{figure*}
%\centering
\epsscale{1.0}
\includegraphics[width=0.24\textwidth]{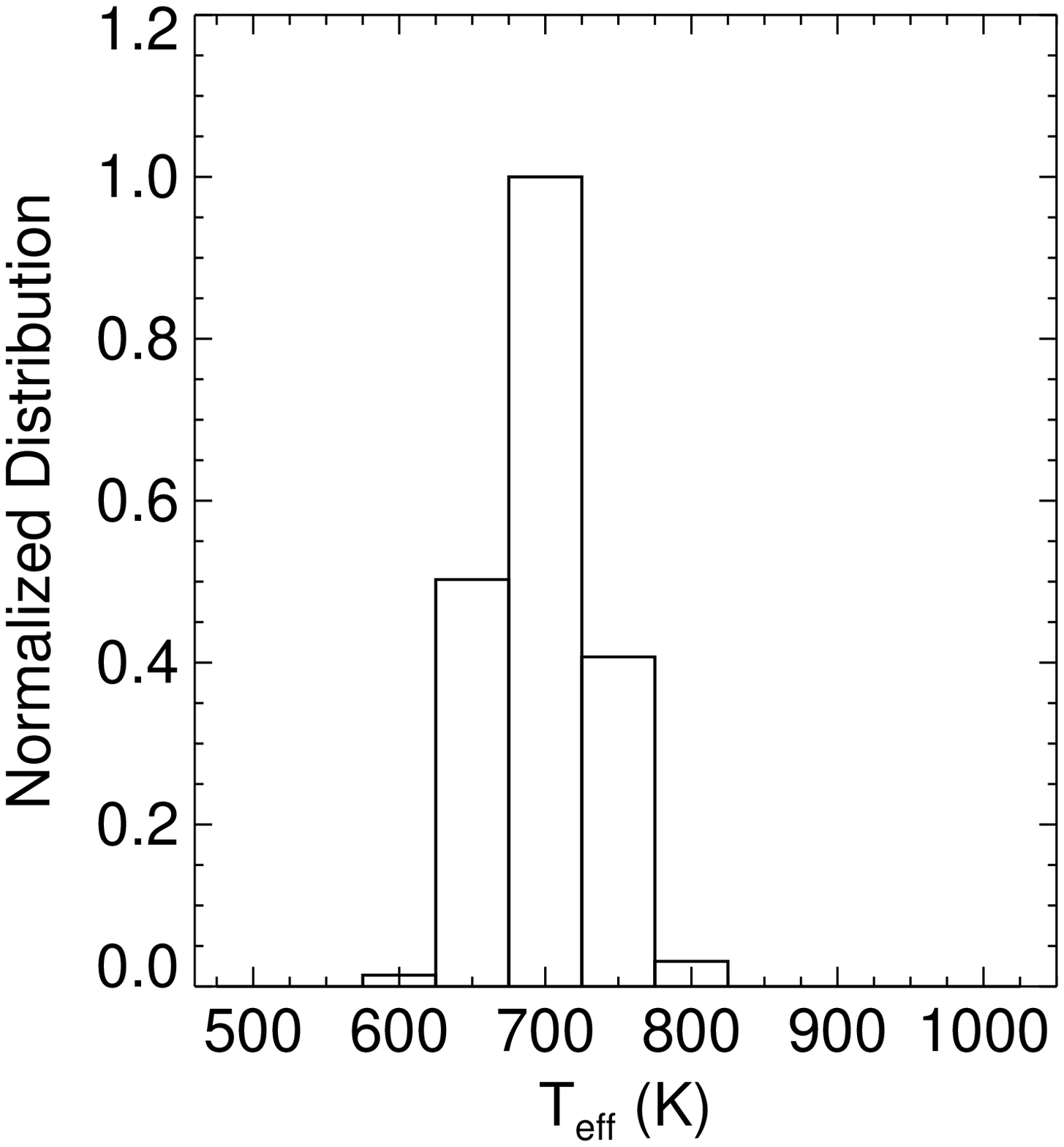}
\includegraphics[width=0.24\textwidth]{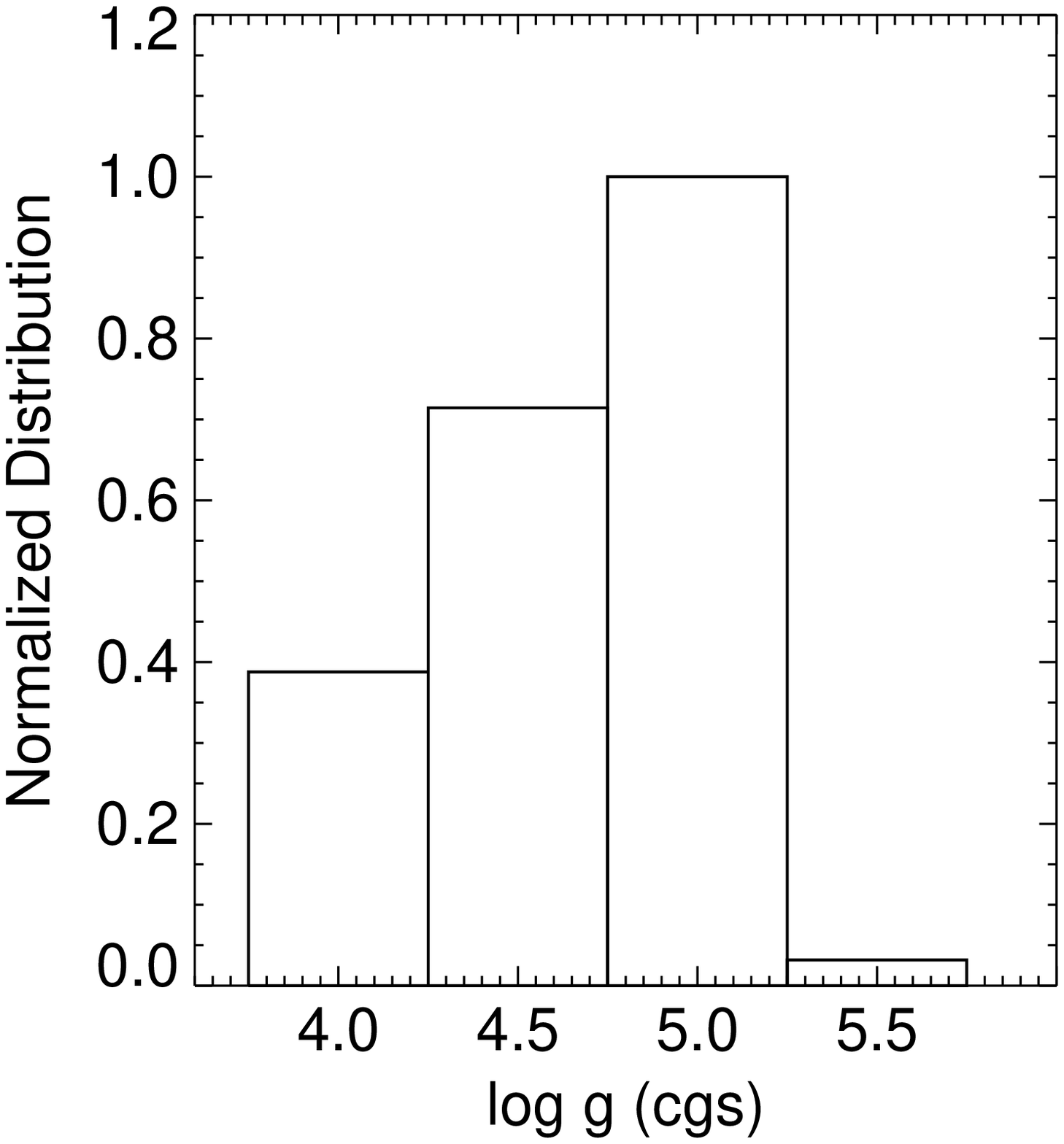}
\includegraphics[width=0.24\textwidth]{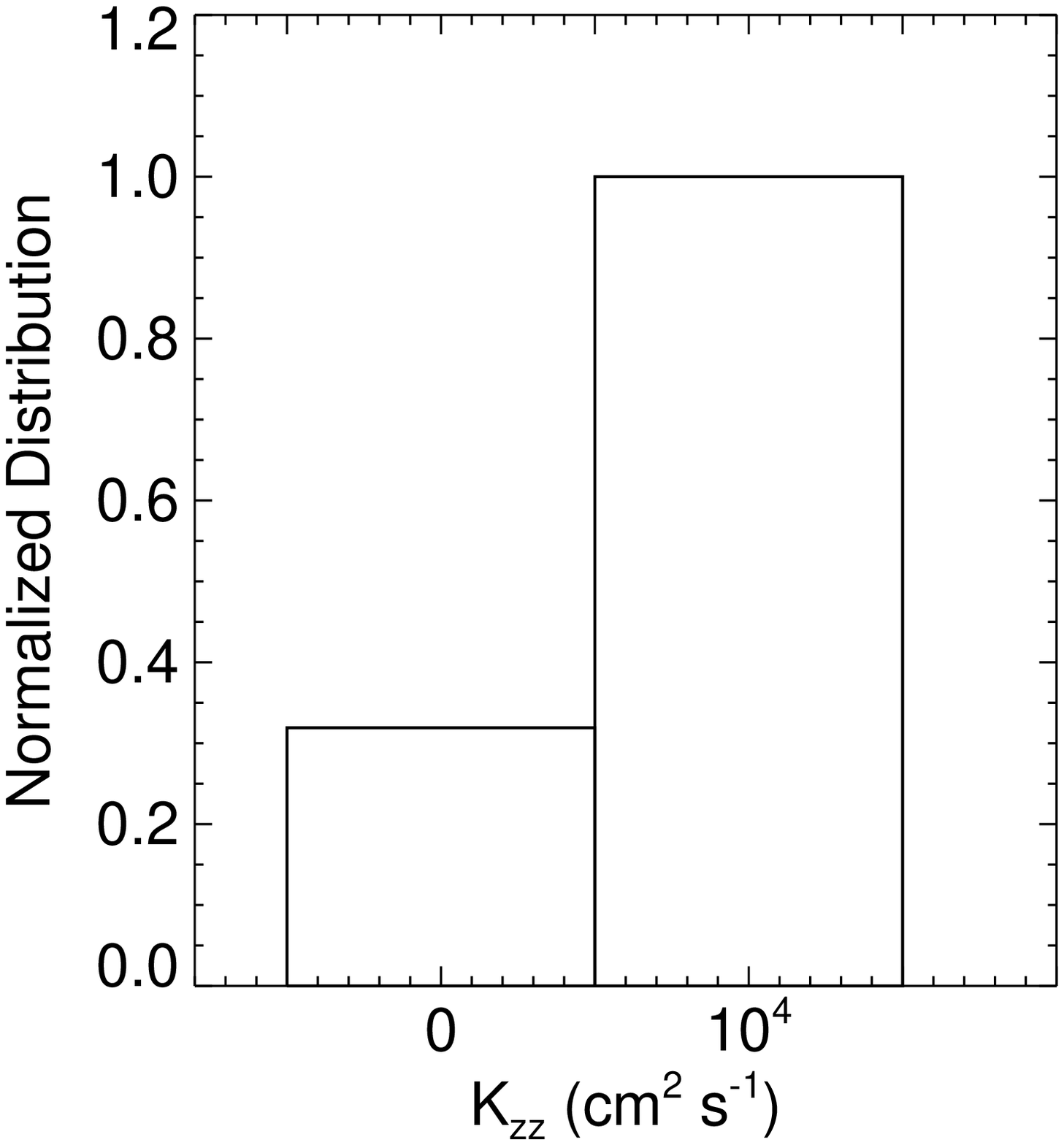}
\includegraphics[width=0.24\textwidth]{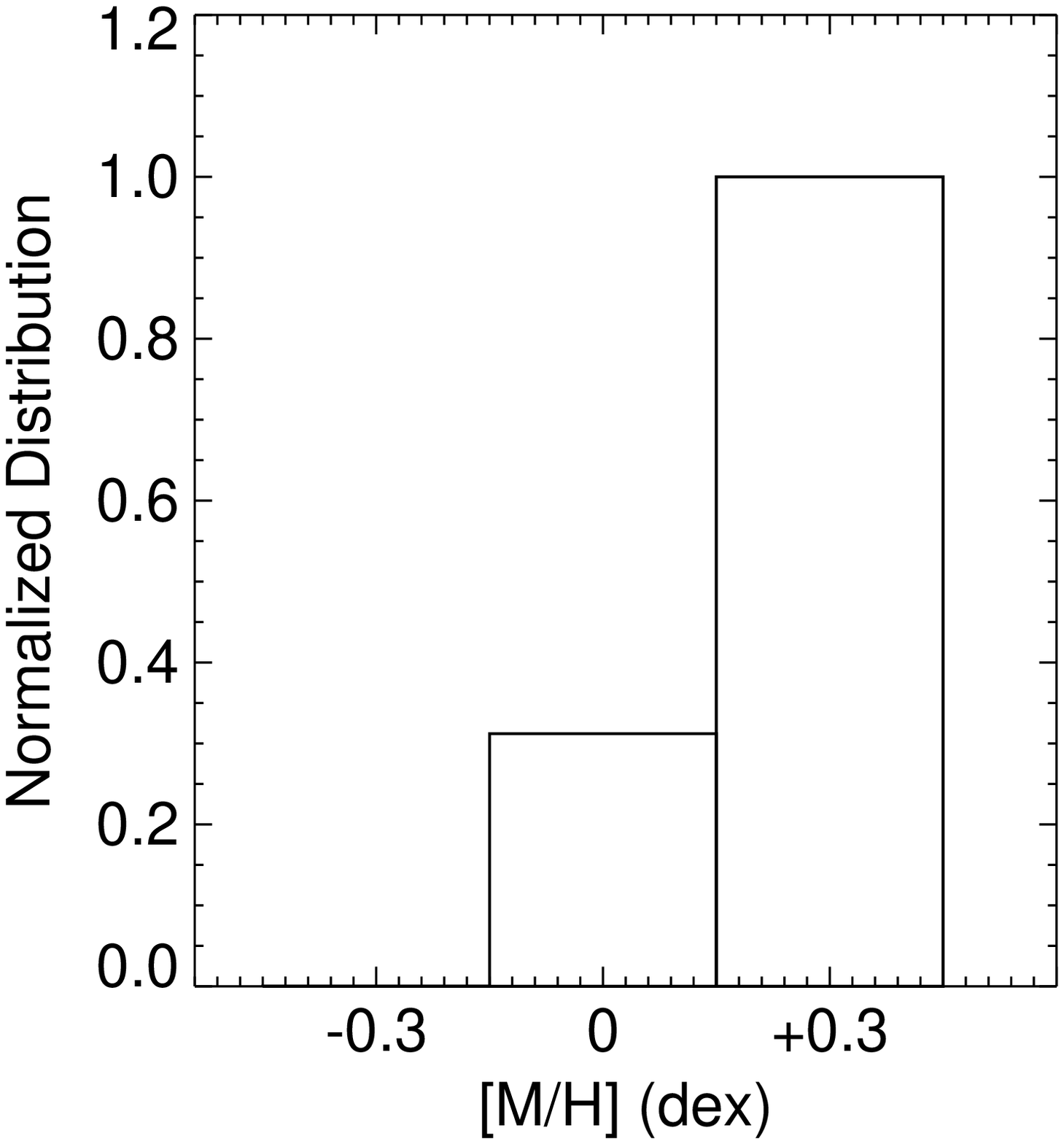} \\
\includegraphics[width=0.24\textwidth]{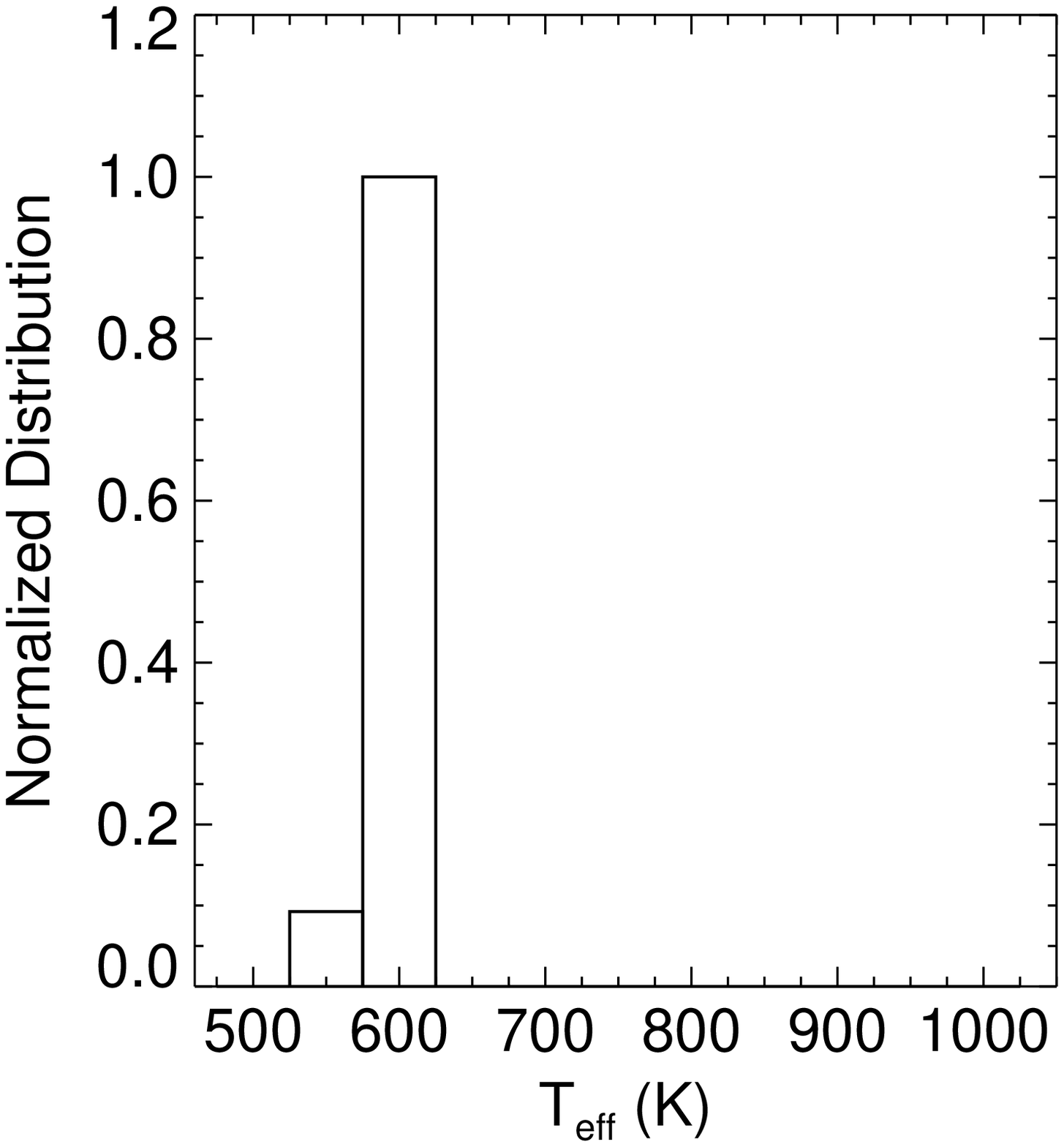}
\includegraphics[width=0.24\textwidth]{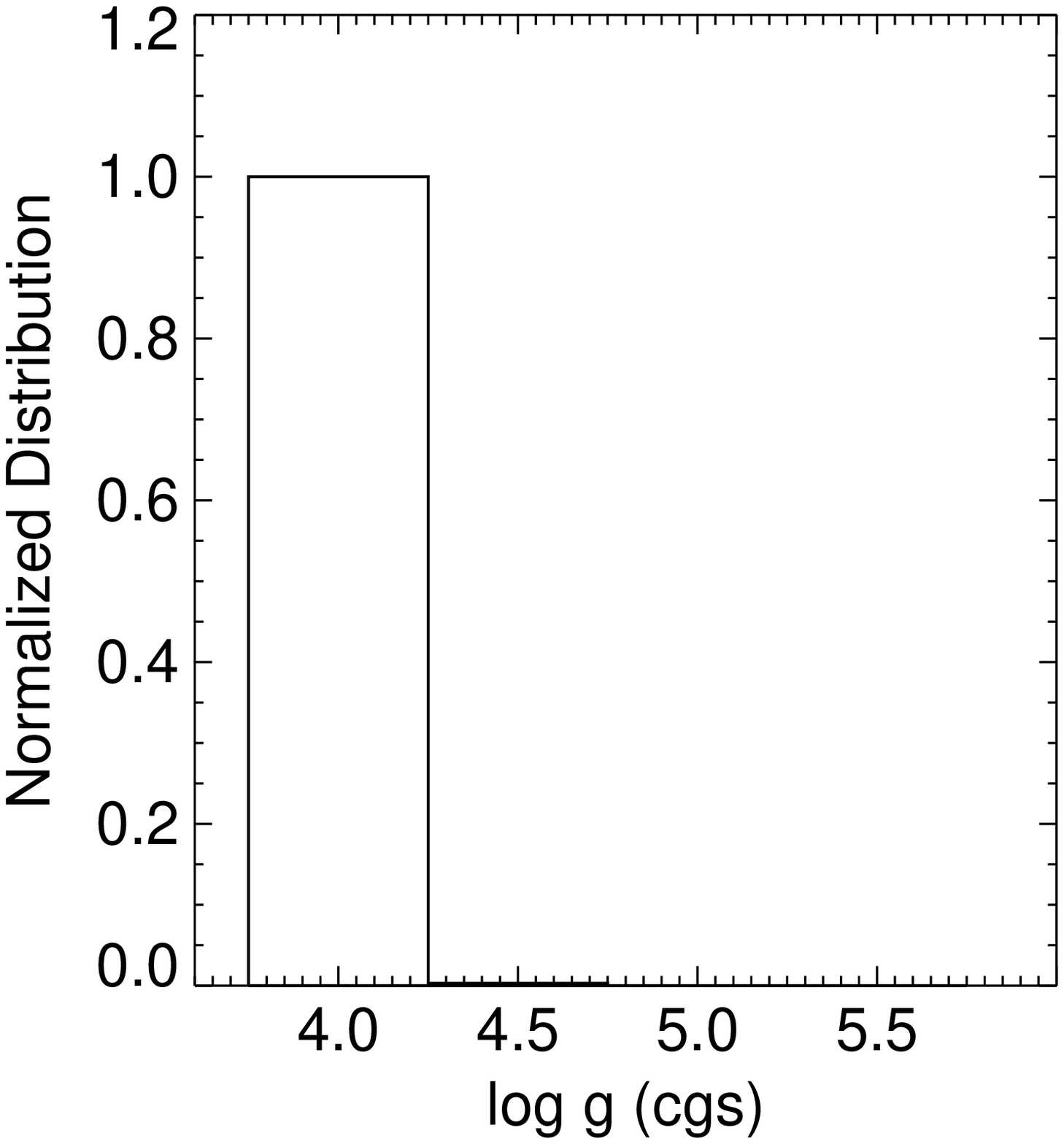}
\includegraphics[width=0.24\textwidth]{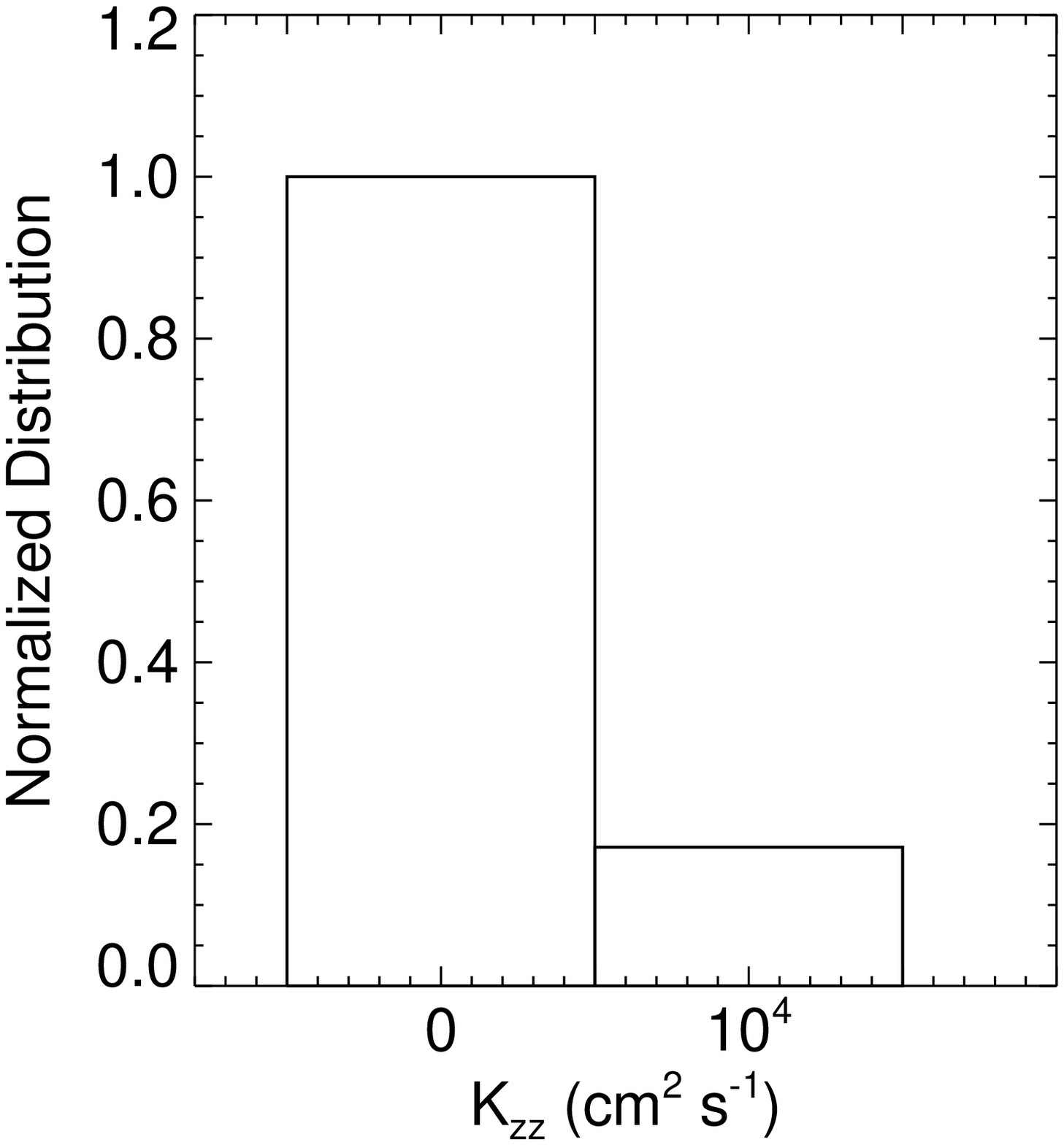}
\caption{Same as Figure~\ref{fig_modelfit_ross458_parameters} for ULAS~J1335+1130.
 \label{fig_modelfit_1335_parameters}}
\end{figure*}

\begin{figure*}
%\centering
\epsscale{1.0}
\includegraphics[width=0.24\textwidth]{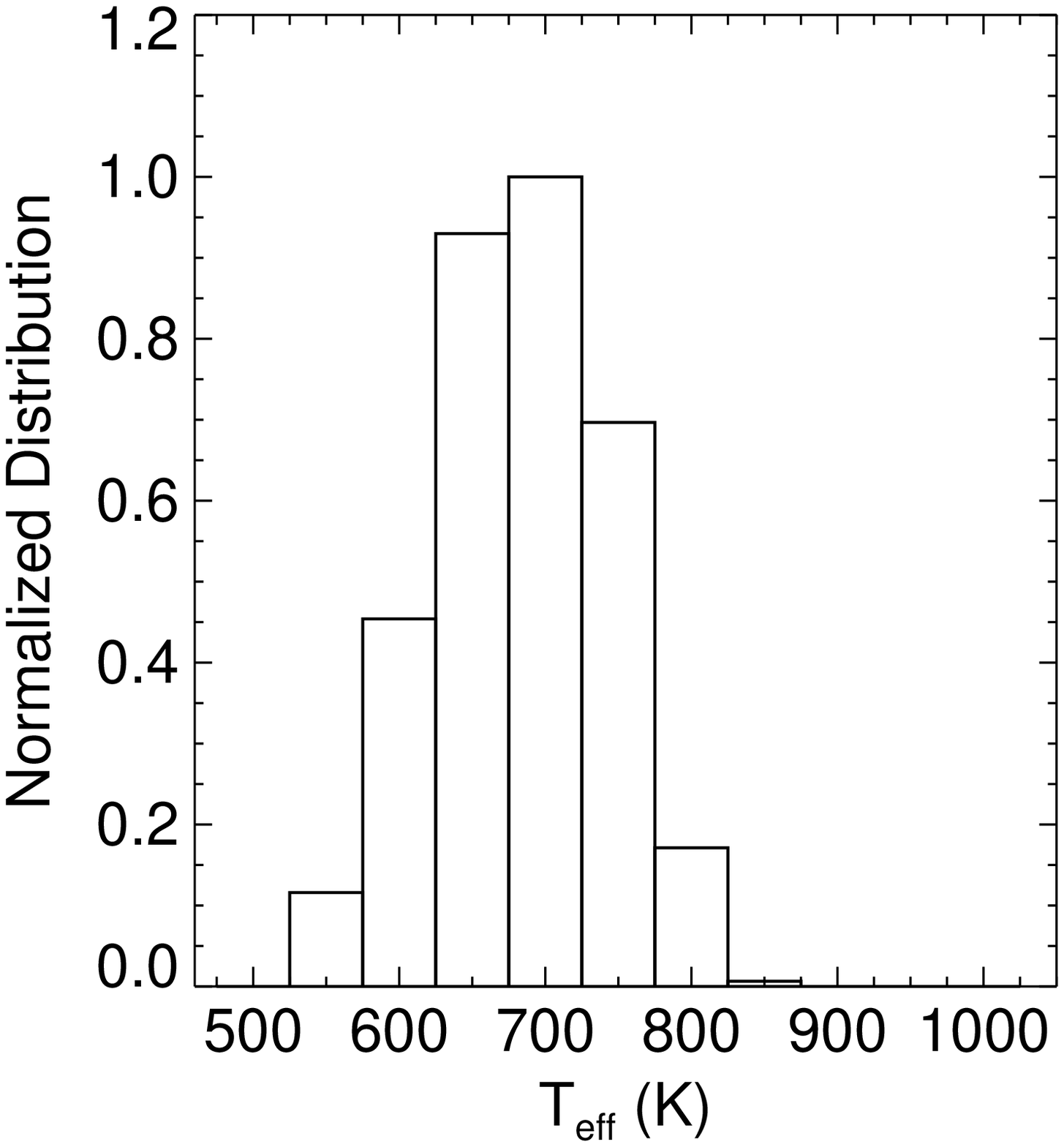}
\includegraphics[width=0.24\textwidth]{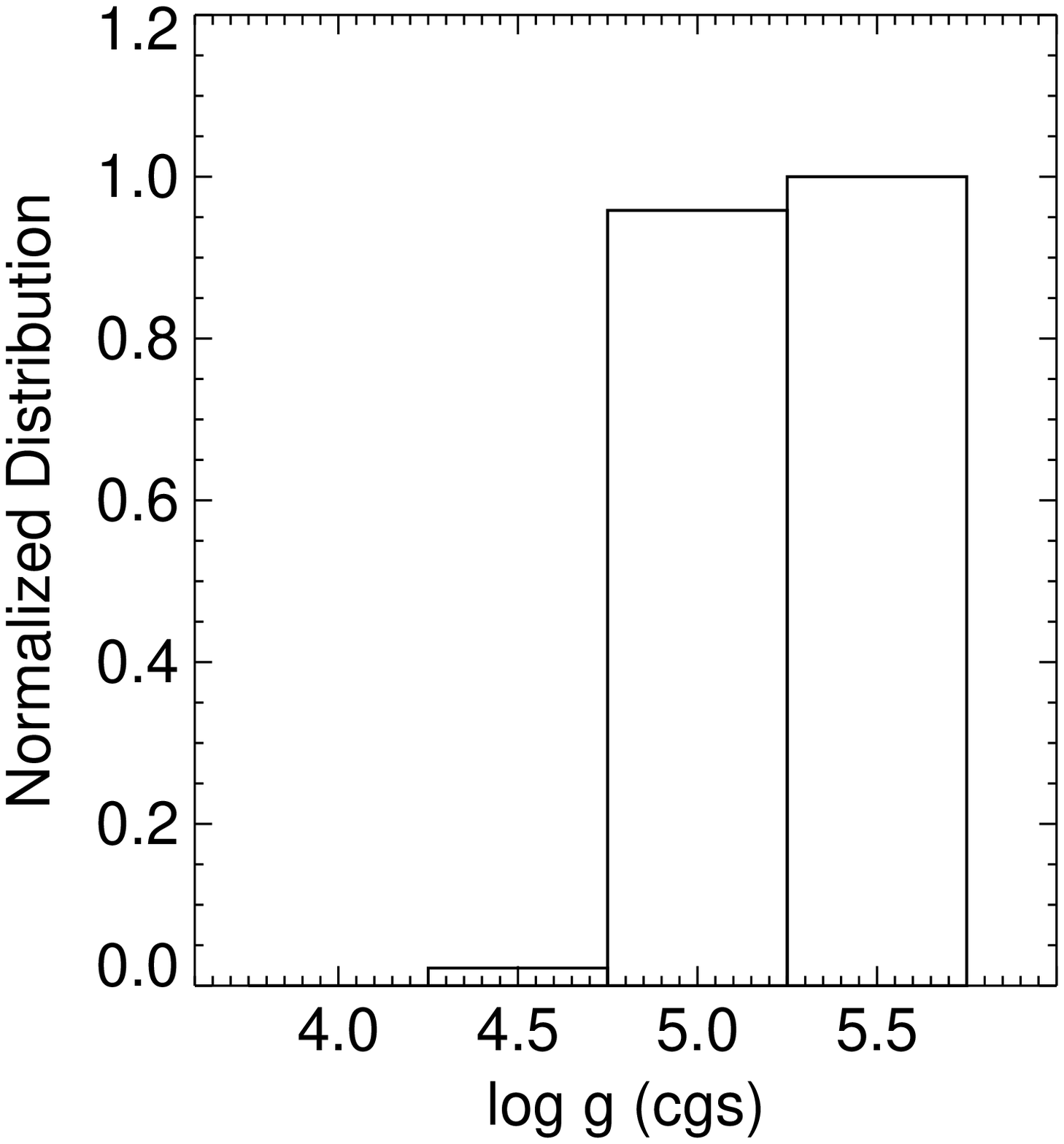}
\includegraphics[width=0.24\textwidth]{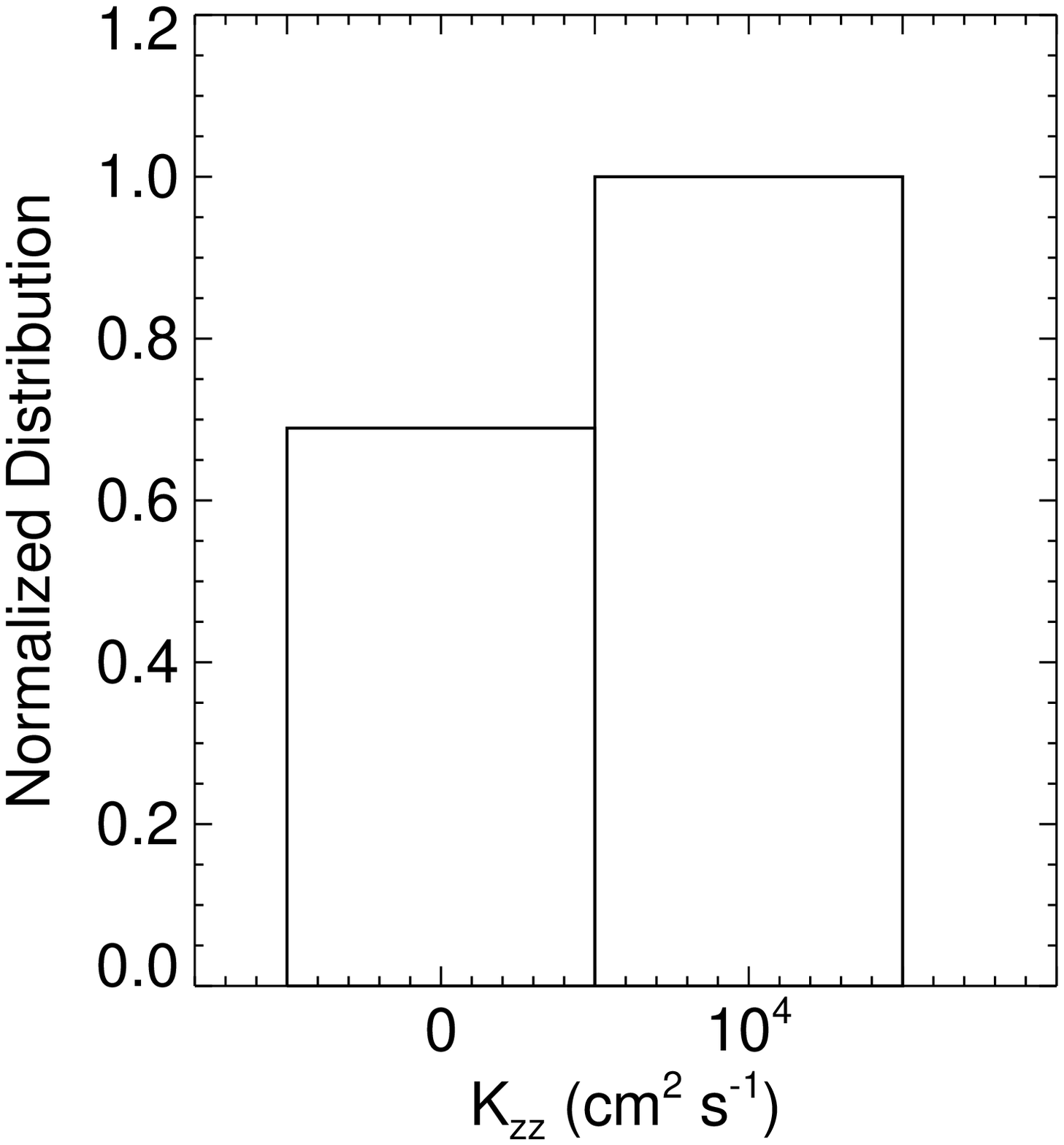}
\includegraphics[width=0.24\textwidth]{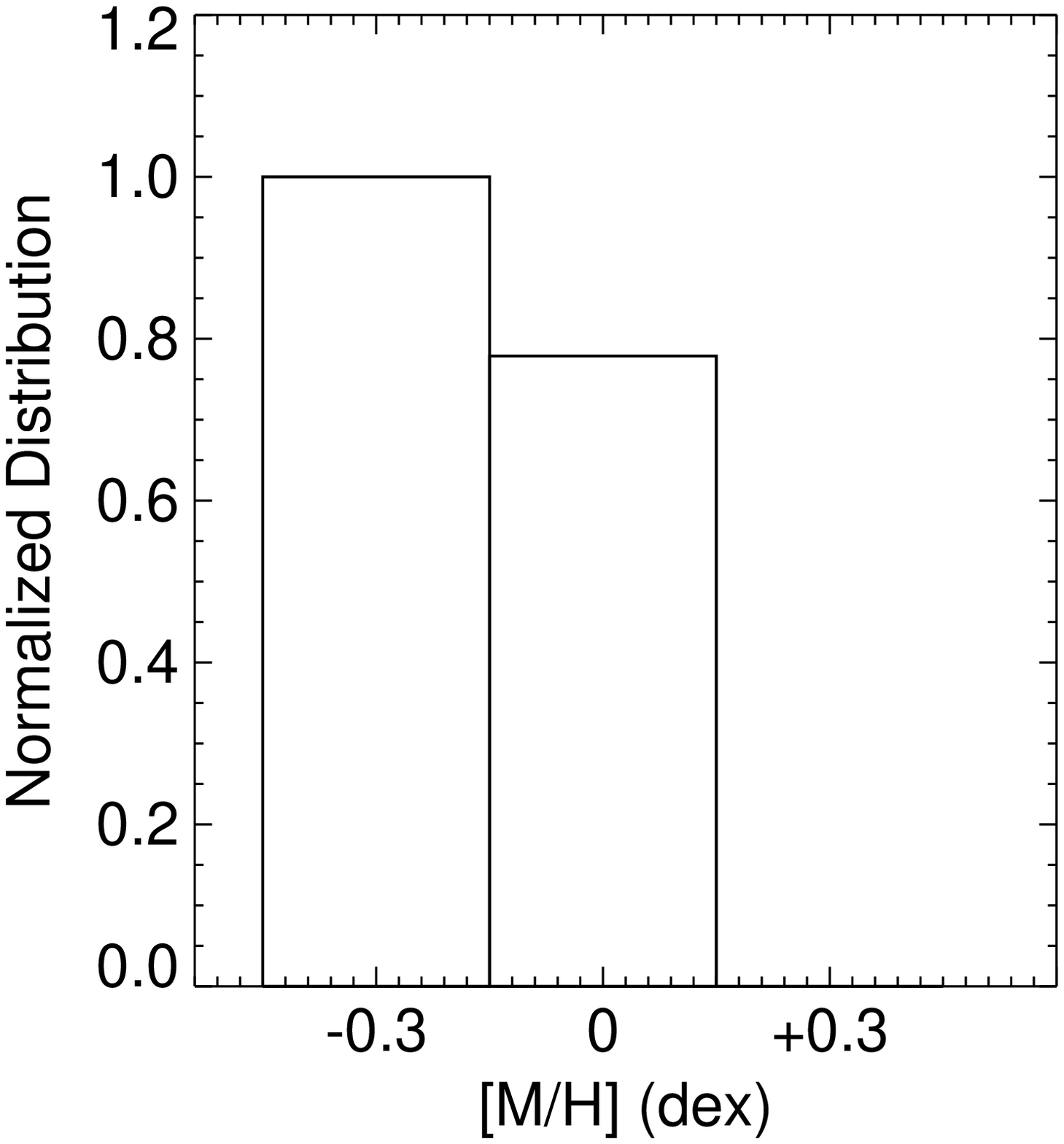} \\
\includegraphics[width=0.24\textwidth]{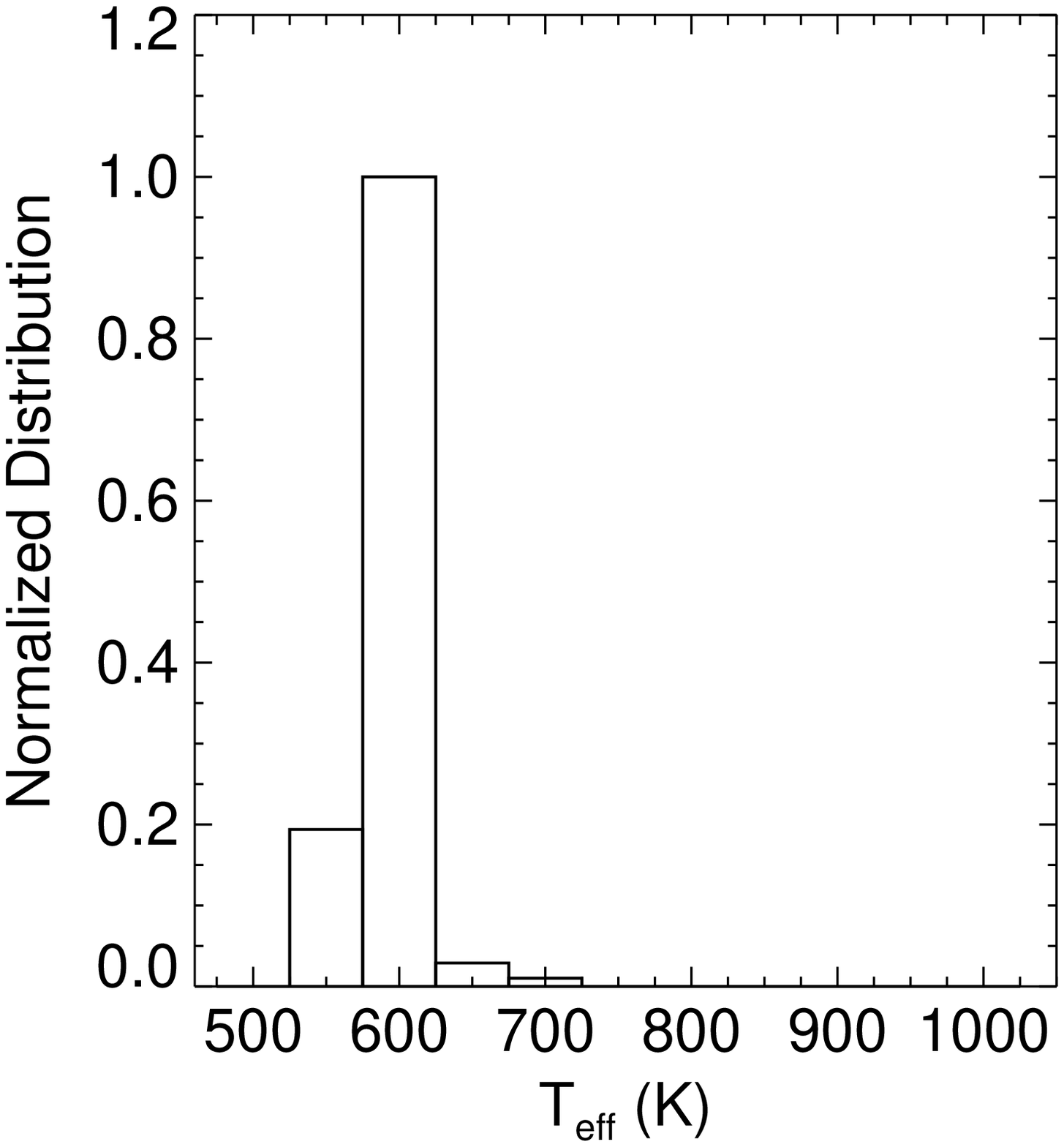}
\includegraphics[width=0.24\textwidth]{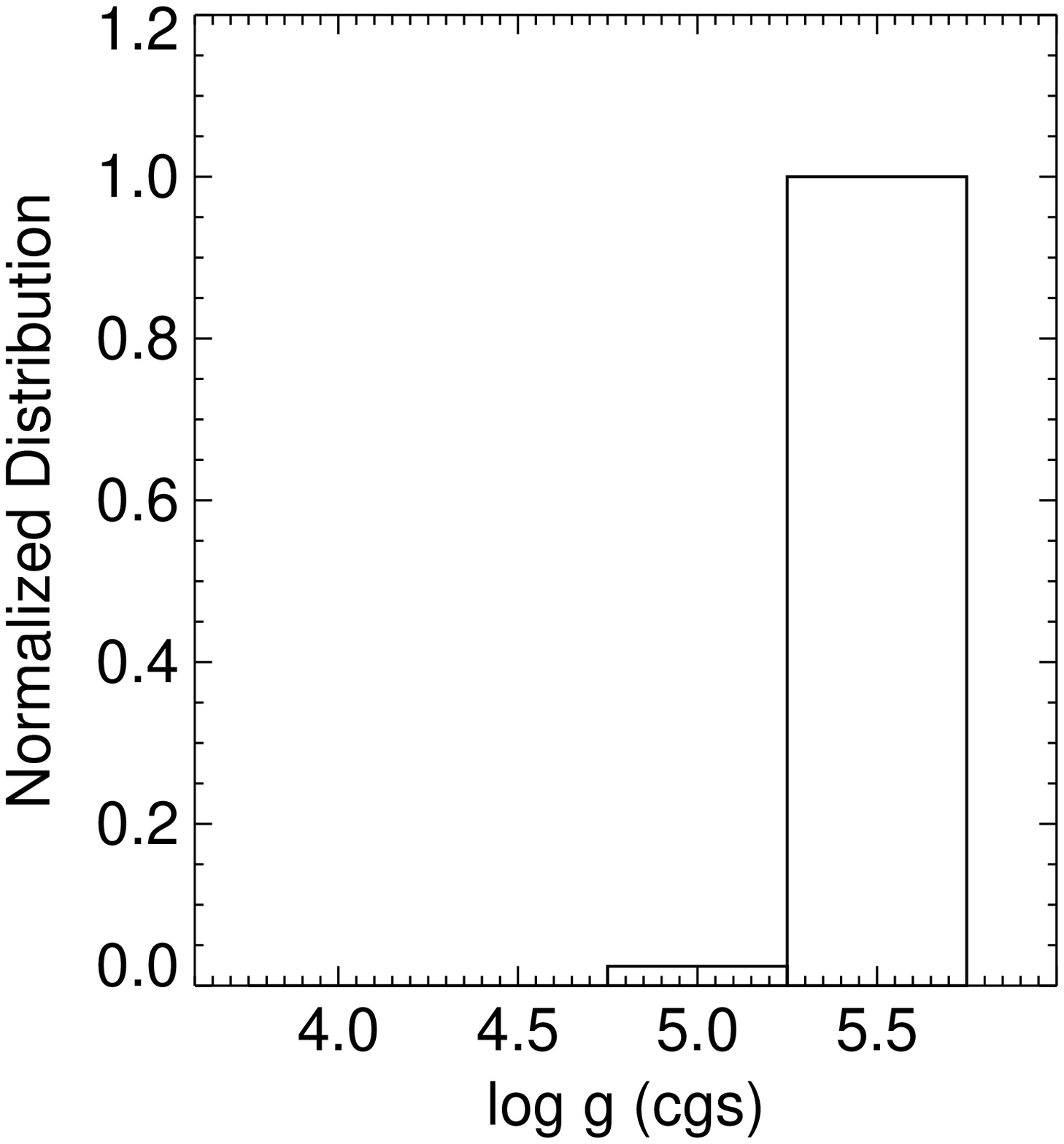}
\includegraphics[width=0.24\textwidth]{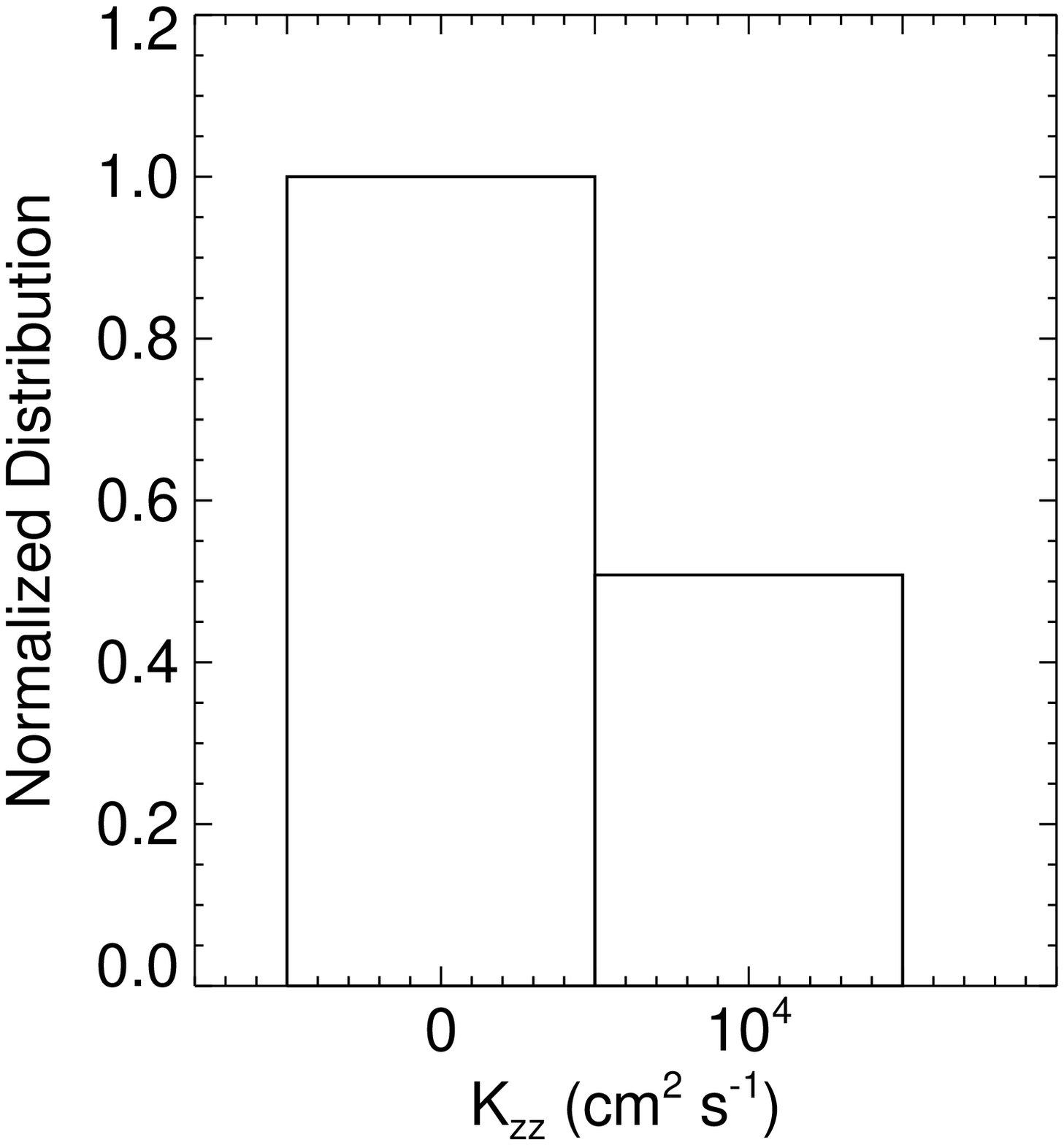}
\caption{Same as Figure~\ref{fig_modelfit_ross458_parameters} for SDSS~J1416+1348B.
 \label{fig_modelfit_1416_parameters}}
\end{figure*}

\begin{deluxetable*}{lcccccl}
\tabletypesize{\small}
\tablecaption{Model-fit Parameters for Observed T Dwarfs. \label{tab_fit}}
\tablewidth{0pt}
\tablehead{
\colhead{Parameter} &
\colhead{Best Fit} &
\colhead{Mean} &
\colhead{Best Fit} &
\colhead{Mean} &
\colhead{Prior Value} &
\colhead{Ref.}  \\
 & 
\colhead{(Cloudless)} &
\colhead{(Cloudless)} &
\colhead{(Cloudy)} &
\colhead{(Cloudy)} &
}
\startdata
\hline
\multicolumn{7}{c}{Ross 458C (T8)} \\
\hline
$G_k$ & 71.8 & \nodata & 51.3 & \nodata & \nodata & \\
{$d/R$ (pc/{\rjup})} & 11.5 & 13.4$\pm$1.9 & 7.9  & 7.5$\pm$0.6 & \nodata \\
{\teff} (K) & 700 & 760$^{+70}_{-45}$   &  650 & 635$^{+25}_{-35}$ & \nodata &  \\
{\logg} (cm~s$^{-2}$) & 4.0 &  4.2$^{+0.3}_{-0.2}$  & 4.0 & $\approx$4.0 & \nodata & \\
{[M/H]} (dex) & +0.3 & $\approx$+0.3  & 0\tablenotemark{a}  & 0\tablenotemark{a} & +0.2...+0.3   & 1  \\
{\kzz} ({\cmms}) & 10$^4$ &  \nodata  & 0 & \nodata & \nodata & \\
{Mass ({\msun})} & 0.006 & 0.010$^{+0.006}_{-0.004}$   & 0.006 & $\lesssim$0.006 & 0.006-0.011\tablenotemark{b} & 1  \\
{Age (Gyr)} &  0.08 & 0.3$^{+0.5}_{-0.2}$ &  0.13 & 0.14$^{+0.03}_{-0.01}$ & 0.4--0.8 & 1 \\
 &  &  & &  &  0.4-2 & 2 \\
{R ({\rjup})} &  1.22 &  1.16$^{+0.07}_{-0.09}$ &  1.25  &  $\approx$1.24 & \nodata \\
{d (pc)} & 14.1 & 15.6$\pm$2.5 & 9.8  & 9.4$\pm$0.7 & 11.7$\pm$0.2 & 3 \\
\hline
\multicolumn{7}{c}{ULAS~J1335+1130 (T9)} \\
\hline
$G_k$ & 22.7 & \nodata & 21.0 & \nodata & \nodata & \\
{$d/R$ (pc/{\rjup})} & 20.9 & 20.2$\pm$2.4  & 12.0  &  11.8$\pm$0.8  & \nodata \\
{\teff} (K) & 700 & 700$^{+30}_{-50}$ &  600  & 595$^{+25}_{-45}$ &  500-600 & 4,5  \\
{\logg} (cm~s$^{-2}$) & 5.0 &  4.7$\pm$0.4 & 4.0 & $\approx$4.0 & 4.0--4.5 & 4,5 \\
{[M/H]} (dex) & +0.3 &  +0.23$^{+0.07}_{-0.23}$  &  0\tablenotemark{a} & 0\tablenotemark{a} & 0...+0.3 & 5 \\
{\kzz} ({\cmms}) & 10$^4$ &  \nodata & 0 & \nodata & \nodata & \\
{Mass ({\msun})} & 0.031  & 0.021$\pm$0.011   & 0.006 & 0.005--0.013 & 0.014--0.030 & 4   \\
 &  &  & &  &  0.005-0.019 & 5 \\
{Age (Gyr)} & 3.3  & 2.0$^{+2.3}_{-1.6}$ & 0.17 &  0.18$^{+0.23}_{-0.01}$ & 0.6--5.3 & 4 \\
 &  &  & & &  0.01-2 & 5 \\
{R ({\rjup})} & 0.90 &  1.01$^{+0.13}_{-0.12}$ &  1.23 & $\approx$1.23 &  1.07--1.17 & 5 \\
{d (pc)} & 18.7 & 20.5$\pm$3.6 & 14.7   &  14.4$\pm$1.1 & 8--12 & 4,5 \\
\hline
\multicolumn{7}{c}{SDSS~J1416+1348B (T7.5p)} \\
\hline
$G_k$ & 22.3 & \nodata & 21.0 & \nodata & \nodata & \\
{$d/R$ (pc/{\rjup})} & 15.0 & 13.2$\pm$3.6 &   15.1 & 13.2$\pm$4.0 &   12.8$\pm$3.0 & 6 \\
{\teff} (K) & 700 & 685$^{+55}_{-65}$ &   600 & 595$^{+25}_{-45}$ &   650$\pm$60 & 6  \\
 &  &  &  &  & 500 & 7 \\
{\logg} (cm~s$^{-2}$) & 5.5 &  5.2$\pm$0.3 &  5.5 & $\approx$5.5 &   5.2$\pm$0.4 & 6,7 \\
{[M/H]} (dex) & 0.0 &  -0.17$^{+0.17}_{-0.13}$ &  0\tablenotemark{a} & 0\tablenotemark{a} & -0.3  & 6,7 \\
{\kzz} ({\cmms}) & 10$^4$ &  \nodata & 0 & \nodata & 10$^4$  & 6,7 \\
{Mass ({\msun})} & 0.052  & 0.041$^{+0.011}_{-0.010}$  & 0.036  & 0.035$^{+0.001}_{-0.003}$  & 0.021--0.045 & 6   \\
 &  &  &  & & 0.029--0.038 & 7 \\
{Age (Gyr)} & 12 & 8$\pm$4 & 12 & 6--12 & 2--10 & 6,7 \\
{R ({\rjup})} & 0.78  &  0.84$\pm$0.06 &  0.86 & $\approx$0.86 &  0.83$^{+0.14}_{-0.10}$ & 6  \\
{d (pc)} & 11.8 & 11.1$\pm$3.2    &  13.0 &  11.4$\pm$3.4    &  7.9$\pm$1.7 & 8 \\
\enddata
\tablenotetext{a}{Solar metallicity is assumed for all cloudy models.}
\tablenotetext{b}{Mass estimate based on the luminosity and estimated age of {\name}, and the cloudy evolutionary models of SM08.}
\tablerefs{(1) This paper; (2) \citet{2010MNRAS.405.1140G}; (3) \citet{2007A&A...474..653V}; (4) \citet{2008MNRAS.391..320B}; (5) \citet{2009ApJ...695.1517L}; (6) \citet{2010AJ....139.2448B}; (7) \citet{2010MNRAS.404.1952B}; (8) \citet{2010A&A...510L...8S}.}
\end{deluxetable*}

Table~\ref{tab_fit} summarizes the inferred parameters for each source, with results for the cloudless and cloudy models reported separately.  Figures~\ref{fig_modelfit_ross458}--\ref{fig_modelfit_1416} show the best-fit models to the individual spectra, while Figures~\ref{fig_modelfit_ross458_parameters}--\ref{fig_modelfit_1416_parameters} show the parameter distributions.
It is first worth noting the quality of the fits.  For both cloudless and cloudy models,
deviations are seen in regions of strong {\meth} absorption, particularly in the 1.6-1.7~$\micron$ region for {\name} and ULAS~J1335+1130 and the wings of the 1.27~{\micron} $J$-band peak for SDSS~J1416+1348B.  As noted previously, these discrepancies are likely the result of incomplete {\meth} opacities \citep{2006ApJ...647..552S}.  
However, there are also differences between the cloudless and cloudy model fits, notably in the relative $YJH$ peak fluxes of {\name} and ULAS~J1335+1130.  In the cloudless model fits of these sources, the $H$-band peak appears to be suppressed relative to $J$, while the $Y$-band peak appears suppressed in the cloudy model fits.  These differences are less pronounced in fits to SDSS~J1416+1348B.
The cloudless models also exhibit excessively strong {\ki} lines at 1.24--1.25~$\micron$, while these lines in the cloudy models are more consistent with the data.  
For both models, the 2.1~$\micron$ $K$-band peak is reasonably well-reproduced for all three sources (see also \citealt{2008ApJ...678.1372C}). 
Nevertheless, it appears that the cloudy models provide better matches to the data than the cloudless ones.

Quantitative differences between the cloudless and cloudy model fits are also seen.  For {\name} (Figures~\ref{fig_modelfit_ross458} and~\ref{fig_modelfit_ross458_parameters}), the cloudy model provides a statistically superior fit to the data, significant at the 99.99\% confidence level based on the F-test PDF.  The cloudy model also predicts a significantly cooler temperature than the cloudless case, {\teff} = 635$^{+25}_{-30}$~K versus {\teff} = 760$^{+70}_{-45}$~K.  The former value is consistent with the luminosity-based {\teff} determined in Section~3.5, while the latter is consistent with a ``normal'' T8 field dwarf like 2MASS~J0415$-$0935 (e.g., \citealt{2004AJ....127.3516G, 2009ApJ...702..154S}).
Inferred surface gravities are comparable for the two models, favoring {\logg} $\approx$ 4.0~cgs, while the cloudless models indicate a metal-rich atmosphere ([M/H] $\approx$ +0.3).  These parameters are consistent with the youth and metallicity of the Ross~458 system, as the mean cloudless and cloudy model ages are 0.3$^{+0.5}_{-0.2}$~Gyr and 0.14$^{+0.03}_{-0.01}$~Gyr, respectively.  Both models also predict a very low mass for {\name}, M = 0.010$^{+0.006}_{-0.004}$~{\msun} and M $\lesssim$0.006~{\msun}, again at or below the deuterium burning mass limit. 
There is a modest difference between whether vertical mixing is slightly favored (cloudless models) or not (cloudy models), but it is not statistically significant.  The cloudless and cloudy models predict distinct distances of 15.6$\pm$2.5~pc and 9.4$\pm$0.7~pc, respectively, based on the optimal model scaling.  These differ by 2.6$\sigma$, and are 1.8$\sigma$ higher and 2.4$\sigma$ lower than the parallactic distance of the Ross~458 system, although the latter is off by only 11\%.

For ULAS~J1335+1130 (Figures~\ref{fig_modelfit_1335} and~\ref{fig_modelfit_1335_parameters}), the best-fit cloudy model also has a smaller $G_k$ value than the best-fit cloudless one, although the statistical significance is reduced (84\% confidence level).  Again, the cloudy models predict a significantly lower effective temperature than the cloudless models ({\teff} = 595$^{+25}_{-45}$~K versus {\teff} = 700$^{+30}_{-50}$~K).  The former is consistent with previously reported values by \citet[550--600~K]{2008MNRAS.391..320B} and \citet[500--550~K]{2009ApJ...695.1517L}.   The cloudy models also indicate a lower surface gravity ({\logg} $\approx$ 4.0~cgs versus {\logg} = 4.7$\pm$0.4~cgs), while differences in vertical mixing are statistically unconstrained.  
Like {\name}, the cloudless models converge on a supersolar metallicity for ULAS~J1335+1130, [M/H] = +0.23$^{+0.07}_{-0.23}$.
The differing parameters between the cloudless and cloudy models translate into marginally overlapping ages (2.0$^{+2.3}_{-1.6}$~Gyr versus 0.18$^{+0.23}_{-0.01}$~Gyr) and masses
(0.010--0.032~{\msun} versus 0.005--0.013~{\msun}), and distinct distances (20.5$\pm$3.6~pc versus 14.4$\pm$1.1~pc, a 1.6$\sigma$ discrepancy).

Finally, for SDSS~J1416+1348B (Figures~\ref{fig_modelfit_1416} and~\ref{fig_modelfit_1416_parameters}), we again find that the best-fit cloudy model is marginally better than the best-fit cloudless one, albeit at even lower statistical significance (confidence level of 77\%) reflecting the similarity in the near-infrared SEDs of these models.  Nevertheless, the inferred temperatures are marginally distinct, with the cloudy models again indicating a cooler temperature ({\teff} = 595$^{+25}_{-45}$~K versus {\teff} = 685$^{+55}_{-65}$~K).  Unlike {\name} and ULAS~J1335+1130, the inferred surface gravity of this source is considerably higher ({\logg} = 5.2$\pm$0.3~cgs cloudless; {\logg} $\approx$ 5.5~cgs cloudy), while the cloudless model indicates a subsolar metallicity ([M/H] = -0.17$^{+0.17}_{-0.13}$). As discussed in \citet{2010MNRAS.404.1952B} and \citet{2010AJ....139.2448B}, high surface gravity and/or subsolar metallicity are required to produce both the suppressed $K$-band peak and broadened $Y$-band peak observed in the spectrum of this source.  
Vertical mixing is again poorly constrained.  The inferred atmospheric parameters of SDSS~J1416+1348B indicate a brown dwarf that is older (4--12~Gyr) and more massive (0.03--0.05~{\msun}) than the other two sources, consistent with prior analyses.
Surprisingly, the cloudless and cloudy models converge on a common distance of 11$\pm$3~pc, larger than but formally consistent with the preliminary parallax measurement of  \citet[7.9$\pm$1.7~pc]{2010A&A...510L...8S}.

\section{Discussion}

\subsection{Clouds in Young, Cold and Metal-Rich T Dwarfs}

The above analysis illustrates how condensate opacity, when included in spectral models, can have a profound effect on the near-infrared spectral energy distributions and inferred {\teff}s of the coldest T dwarfs.  More importantly, the inclusion of condensate opacity provides a statistically significant improvement in fits to the near-infrared spectrum of {\name}, and marginally so for ULAS~J1335+1130 and SDSS~J1416+1348B.  
This is the not the first indication of condensates being present in T dwarf atmospheres.
Early-type T dwarfs exhibit evidence of waning cloud opacity \citep{2002ApJ...568..335M, 2008ApJ...678.1372C, 2008ApJ...682.1256L}, and
SM08 have shown that T dwarf models produce excessively blue near-infrared colors without the inclusion of condensate extinction.
Nevertheless, several theoretical studies have concluded that condensate clouds play little or no role in shaping the near-infrared
spectra of brown dwarfs cooler than $\approx$1000~K, as they reside below the visible photosphere (e.g., \citealt{2001ApJ...556..357A,2006ApJ...640.1063B}).  Instead, we find that clouds are {\em essential}
for reproducing the near-infrared SED of the {\teff} $\approx$ 650~K T dwarf {\name}, and to a lesser degree the {\teff} $\approx$ 600~K T dwarfs ULAS~J1335+1130 and SDSS~J1416+1348B.

The importance of clouds in these two sources is likely related to their youth and associated low surface gravities.  Among the L dwarfs, a correlation between low surface gravity and thick clouds has been noted by \citet{2008ApJ...678.1372C} and \citet{2009ApJ...702..154S} based on broad-band spectral model fits; and by \citet{2006ApJ...651.1166M} and \citet{2009ApJ...699..168D} based on the delayed disruption of clouds at the L dwarf/T dwarf transition.  There is also evidence that
this correlation continues into the T dwarf regime, based on cloudy model fits to broad-band spectral data for the low-gravity T5.5 dwarf SDSSp J111010.01+011613.1 \citep{2009ApJ...702..154S}.
Finally, the 9--11~$\micron$ silicate grain feature has also been shown to be more pronounced in L dwarfs with unusually red $J-K$ colors \citep{2008ApJ...674..451B,2008ApJ...686..528L}, sources whose kinematics reflect youth and low surface gravities  \citep{2009AJ....137....1F, 2010AJ....139.1808S}. 

Atmospheric models show that two complementary effects can drive the persistence of clouds in low gravity, low-temperature brown dwarf atmospheres.  First, the column of
atmosphere above a given pressure level is deeper in a lower
gravity dwarf, and as a consequence the atmosphere is everywhere  
warmer at a fixed {\teff}.  This implies that silicate
clouds lie higher in these atmospheres, providing a larger column depth of extinction
in regions of minimum molecular gas opacity.
Second, cloud models from \citet{2001ApJ...556..872A} and \citet{2008ApJ...675L.105H}
both predict that mean cloud particle sizes are
larger in lower gravity atmospheres.  In
a {\teff} = 650~K SM08 model, for example, the mean particle
radius is roughly ten times larger for $\log g = 4.5$ as compared to 
$\log g = 5.5$.  The sub-micron particles found in higher
gravity models contribute little opacity in the near-infrared because of Mie-scattering effects, rendering the deep
silicate cloud less visible.  Clouds in lower gravity
dwarfs are therefore more visible and scatter more efficiently than those in their older and more massive
counterparts.  This interpretation explains why cloudless models have so far provided adequate fits to the spectra of  {\logg} $\gtrsim$ 5 T dwarfs
(e.g., \citealt{2001ApJ...556..373G, 2007ApJ...656.1136S, 2008ApJ...689L..53B}),
including SDSS~J1416+1348B.

Metallicity is another factor to consider.  
Cloud formation appears to be inhibited in metal-poor L subdwarfs, as exemplified by the retention of Ca, Ti and TiO gas to cooler {\teff}s in these objects \citep{2003ApJ...592.1186B, 2007ApJ...657..494B, 2006AJ....132.2372G}.
Similarly, unusually blue L dwarfs --- including the brighter companion of SDSS~J1416+1348B --- appear to have slightly metal-poor atmospheres with unusually thin clouds \citep{2008ApJ...674..451B, 2010arXiv1009.2802C} or fractured cloud layers \citep{marley10}. 
It is not unreasonable to speculate that supersolar metallicity dwarfs may have thicker
clouds \citep{2008ApJ...686..528L}, yet
theoretical predictions are mixed.
SM08 have shown that warmer cloudy models ({\teff} $\geq$ 1300~K) appear ``cloudier'' (redder) for supersolar metallicities.
However, chemical equilibrium calculations by \citet{2010ApJ...716.1060V} find that MgSiO$_3$ (enstatite) condenses at warmer temperatures in higher metalllicity atmospheres, leading to a deeper cloud that may be buried by gas opacity in cooler brown dwarfs.
Atmospheric model calculations by \citet{2009A&A...506.1367W} find that
grain size and cloud optical depth do not change appreciably between [M/H] = 0 and +0.5
for low-temperature brown dwarfs, although the clouds themselves expand into lower pressure regions of the atmosphere and may have a greater effect at wavelengths corresponding to minimum gas opacity.  Metallicity effects on cloud formation and cloud properties in cool T dwarf atmospheres clearly remains an area of further theoretical exploration.

In summary, the unexpected presence of condensate absorption in the spectrum of {\name} (and possibly ULAS~J1335+1130) can be attributed to its low surface gravity, and possibly its supersolar metallicity.  This hypothesis follows similar trends observed in the L dwarfs.  It also explains why clouds had not previously been needed in model fits of high surface gravity T dwarfs.  The importance of clouds in shaping low gravity T dwarf spectra should nevertheless be a consideration in the characterization of planetary-mass brown dwarf candidates in local young clusters (e.g., \citealt{2002ApJ...578..536Z,2007MNRAS.378.1131C,2009A&A...508..823B, 2010ApJ...709L.158M}) and low-mass exoplanets
recently imaged around nearby young stars (e.g., \citealt{2008Sci...322.1345K, 2008Sci...322.1348M, 2010ApJ...710L..35J, 2010Sci...329...57L}).

\subsection{Clouds and the Convergence of Near-infrared and Mid-infrared T Dwarf Temperature Scales}

{\em Spitzer} has enabled detailed studies of cold T dwarfs in the mid-infrared, where the majority of the SED emerges \citep{2010ApJ...710.1627L}.  Model fits to mid-infrared spectral data have proven particularly robust, with agreement in continuum fluxes and prominent {\wat}, {\meth} and {\ammon} bands (e.g., \citealt{2007ApJ...656.1136S, 2008ApJ...689L..53B, 2009ApJ...695.1517L}).  However, there have been indications that temperatures
inferred from near- and mid-infrared analyses can differ significantly.
 A case in point is SDSS~J1416+1348B.
\citet{2010AJ....139.2448B} inferred {\teff} = 650$\pm$60~K for this source from near-infrared cloudless spectral model fits, similar to the cloudless results presented here. \citet{2010MNRAS.404.1952B} inferred {\teff} $\approx$ 500~K based on the extreme $H-[4.5]$ color of this source \citep{2007MNRAS.381.1400W,2010ApJ...710.1627L}.
We find that inclusion of cloud opacity in the near-infrared modeling of this source brings inferred {\teff}s closer to mid-infrared results, despite having minimal effect on the shape of its spectrum.
Even better convergence is seen for ULAS~J1335+1130, as our near-infrared cloudy model temperature is fully consistent with estimates based on $H-[4.5]$ color \citep{2008MNRAS.391..320B} and near- to mid-infrared spectral model fits \citep{2009ApJ...695.1517L}.
For {\name}, the cloudy model temperature is consistent with that inferred from evolutionary models based on the age and luminosity of the source (Section~3.5).

\begin{figure}
\centering
\includegraphics[width=0.5\textwidth]{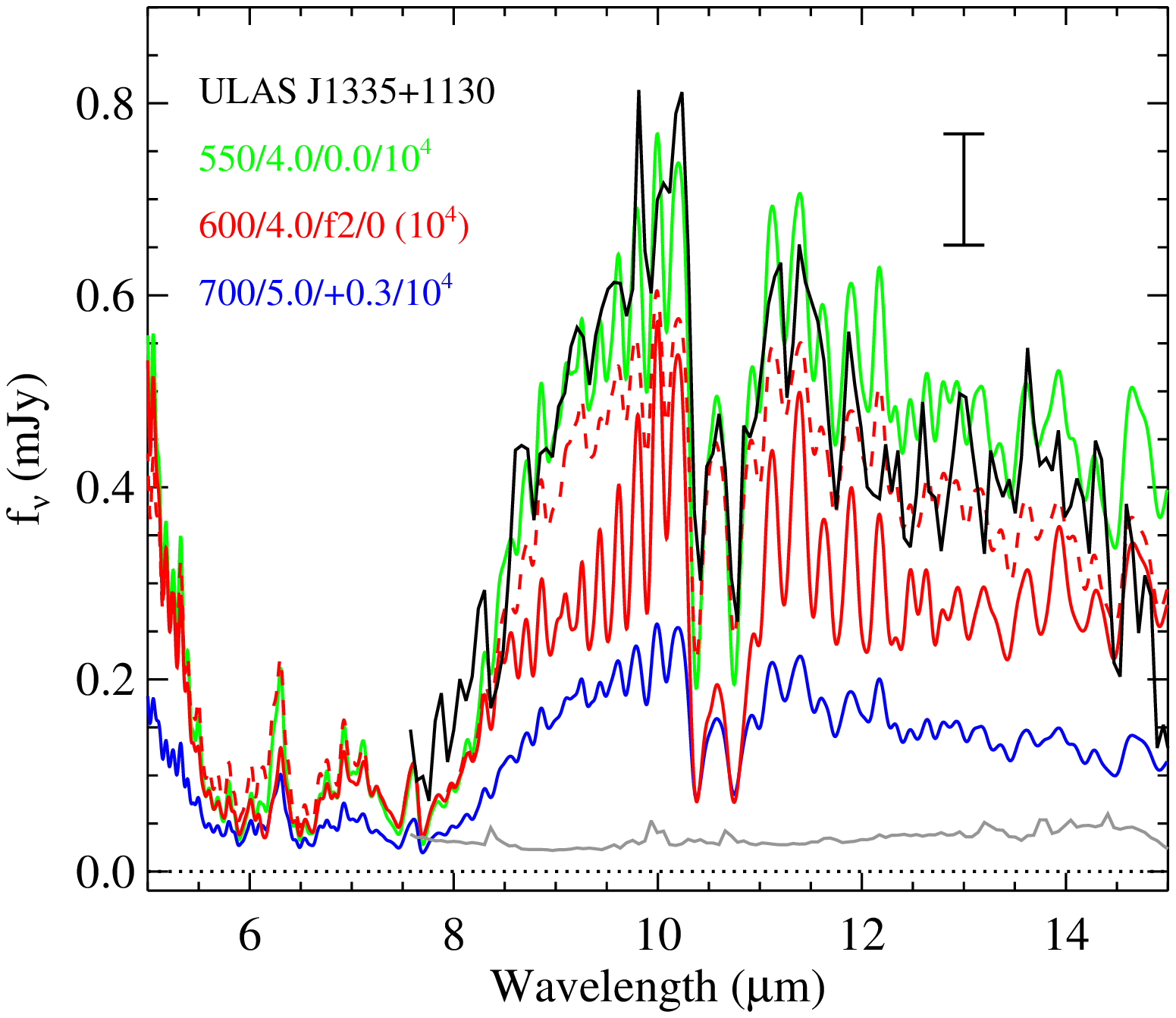}
\caption{Spitzer/IRS spectrum of ULAS~J1335+1130 from \citet[black line, noise spectrum in grey]{2009ApJ...695.1517L}, scaled to its apparent 8.0~$\micron$ magnitude \citep{2008MNRAS.391..320B}, compared to best-fit cloudless (blue line) and cloudy (solid red line) models from our near-infrared analysis, and the best-fit cloudless model from the near- and mid-infrared analysis of \citet[green line]{2009ApJ...695.1517L}. Model surface fluxes have been smoothed to the resolution of the IRS data ({\ldl} $\approx$ 100) using a Gaussian kernel, and scaled according to their respective best-fit distances and radii.  The best-fit cloudy model to the near-infrared data with {\kzz} = 10$^4$~{\cmms} is also shown (dashed red line).  Observational uncertainty in flux at 10~$\micron$ in indicated by the error bar.
\label{fig_mir1135}}
\end{figure}

Prior mid-infrared analyses have been based on cloudless model fits, however, so does the inclusion of clouds affect these results?  To explore this question, Figure~\ref{fig_mir1135} compares the observed mid-infrared spectrum of ULAS~J1335+1130 obtained with the {\em Spitzer} Infrared Spectrograph (IRS; \citealt{2004ApJS..154...18H, 2009ApJ...695.1517L})
to three SM08 models: two sampling the best-fit parameters from our near-infrared cloudless and cloudy model fits, and one sampling the cloudless model fit parameters from \citet{2009ApJ...695.1517L}.\footnote{\citet{2009ApJ...695.1517L} used {\kzz} = 10$^6$~{\cmms} in their best-fit models whereas we use {\kzz} = 10$^4$~{\cmms} in Figure~\ref{fig_mir1135}.  The difference between these models is negligible.}   The first two models are scaled according to the near-infrared normalizations from our analysis, while the Leggett et al.\ model is scaled to the best-fit radius and distance derived in that study.  The 600~K cloudy model comes much closer to reproducing the pseudo-continuum flux of the observed data than the 700~K cloudless model, but is still underluminous and predicts excessively strong {\ammon} absorption in the 9--13~$\micron$ region.  This is not a cloud effect, however, but the absence of vertical mixing in the best-fit cloudy model.  Vertical mixing reduces {\ammon} and CO photospheric abundances through non-equilibrium chemistry, but has little impact on the near-infrared spectrum for {\teff} $\lesssim$ 1000~K \citep{1999ApJ...519L..85G,2006ApJ...647..552S}.  This is confirmed in our spectral modeling, as the presence or absence of vertical mixing is statistically ambiguous (Figures~\ref{fig_modelfit_ross458_parameters}--\ref{fig_modelfit_1416_parameters}).  
It has a profound effect on mid-infrared spectra, however, as the {\teff} = 600~K, {\logg} = 4.0, {\kzz} = 10$^4$~{\cmms} cloudy model is a good match to data for ULAS~J1335+1130 (i.e., consistent within the observational uncertainties).
As such, the inclusion of cloud opacity does not appear to modify mid-infrared analyses for the coldest T dwarfs.

Convergence between near-infrared and mid-infrared spectral model fit parameters and temperature scales for cold T dwarfs is important for both near- and long-term study of these objects. With {\em Spitzer} now depleted of cryogen, our ability to obtain mid-infrared spectroscopy for faint brown dwarfs is restricted pending deployment of the James Webb Space Telescope ({\em JWST}) Mid-Infrared Instrument (MIRI; \citealt{2008SPIE.7010E..27W}) later this decade.  In addition, characterization of several dozens of cold brown dwarf discoveries expected from UKIDSS, the Canada France Hawaii Telescope Legacy Survey (CFHTLS; \citealt{2008A&A...484..469D}), the Wide Field Infrared Explorer (WISE; \citealt{2008SPIE.7017E..16L}) and the Visible and Infrared Survey Telescope for Astronomy (VISTA; \citealt{2006SPIE.6267E...7M}) cannot rely on space-based spectroscopy.  Rather, accurate reproduction of ground-based near-infrared spectroscopy, facilitated by sensitive instruments such as FIRE, will be necessary to study the physical characteristics and demographics of these intrinsically faint sources.

\subsection{Is {\name} a Planet?}

Finally, we consider whether the properties of {\name} qualify it as a ``planet'' \citep{2006AREPS..34..193B, 2006AJ....132.2513S, 2007ApJ...666..475M}.  In terms of characteristics, the planet moniker appears appropriate.
{\name} is gravitationally bound to stellar (hydrogen-fusing) companions, 
and itself appears to have a mass 
below which no fusion is expected.  Its relative mass is roughly 1:50 compared to its companions. Its surface temperature is well below those of
several well-studied, so-called planets; e.g., 
AB Pic b (2000$^{+100}_{-300}$~K; \citealt{2010A&A...512A..52B}),
1RXS J160929.1-210524 b (1800$\pm$200~K; \citealt{2010ApJ...719..497L}), 
2MASS 1207-3932b (1600$\pm$100~K; \citealt{2007ApJ...657.1064M}), 
and HR 8799 b (1300--1700~K, \citealt{2010arXiv1008.4582B}).
Evidence of clouds also
lends {\name} a decidedly planetary-like character. 

However, cosmogony seems to indicate otherwise.  {\name}'s 1100~AU projected separation makes it highly
unlikely to have formed according to the ``standard'' model; i.e., the accumulation of gas and dust in a circumstellar (perhaps in this case circumbinary) disk \citep{1996Icar..124...62P}.
Indeed, if {\name} is a planet, it has the widest and longest ($\sim$50,000~yr) planetary orbit known, 2-4 times wider than 
AB Pic b (M $\approx$ 0.011~{\msun}, $a$ $\approx$ 260~AU; \citealt{2005A&A...438L..29C, 2010A&A...512A..52B}),
1RXS J160929.1-210524 (M $\approx$ 0.008~{\msun}, $a$ $\approx$ 330~AU; \citealt{2008ApJ...689L.153L, 2010ApJ...719..497L}),
CT Cha b (M = 0.016$\pm$0.006~AU, $a$ $\approx$ 440~AU; \citealt{2008A&A...491..311S})
and UScoCTIO 106b (M = 0.013$^{+0.002}_{-0.008}$, $a$ $\approx$ 670~AU; \citealt{2008ApJ...673L.185B}).  Can planets form at these separations?
Some disk fragmentation simulations have been able to produce very wide ($>$1000~AU) substellar companions, albeit with extreme initial conditions (e.g., M$_{disk}$/M$_*$ $\approx$ 1; \citealt{2009MNRAS.392..413S, 2010ApJ...714L.133V}; see also \citealt{2006ApJ...637L.137B}).
Planet-planet interactions have been shown to scatter sources out to orbits 100--1000~AU in size, although they are exceedingly rare ($<$1\%) and short-lived in simulations (e.g., \citealt{2009ApJ...707...79D,2009ApJ...696.1600V}).  The triple nature of Ross~458 may provide an angular momentum sink for such a scenario, allowing a planet to be scattered out without being ejected \citep{1999AJ....117..621H, 2003MNRAS.345..233N, 2008A&A...483..633P}.  All of these scenarios require dramatic---and highly fortuitous---dynamic interactions to send a planet formed in the region of a circumstellar/circumbinary disk to an orbital radius 100 times wider than Jupiter's.  Of course, a simpler alternative may be that Ross~458 formed as a ``normal'' hierarchical triple star system, albeit one with very low mass components.  

{\name} therefore stands as a benchmark not just for studies of low-temperature, planetary-like atmospheres, but for the definition of the word planet in general.  Better characterization of the age of this system and the orbital and physical properties of its components will ultimately 
help us address what parameter space defines a planet, and whether ``well-defined boundaries, empirical verifiability, or logical consistency ...[can] overcome gut feelings'' \citep{2006AREPS..34..193B}.

\section{Summary}

We have obtained FIRE near-infrared spectra of the faint
companion {\name} and the late-type field T dwarfs ULAS~J1335+1130 and SDSS~J1416+1348B.  Data for the first source confirm its substellar nature
and identify it as a T8 dwarf, with an inferred {\lbol} = $-$5.62$\pm$0.03 and {\teff} = 650$\pm$25~K
based on empirical bolometric corrections and evolutionary models.  These data also show a pronounced $K$-band peak and relatively red $J-K$ spectrophotometric colors, indicative of supersolar metallicity and youth that coincide with the properties of the Ross~458 system (age = 150--800~Myr,  [Fe/H] = +0.2 to +0.3).  Fits of the data to synthetic spectra
from SM08 confirm these secondary parameters, albeit with discrepant effective temperatures and distances depending on the presence or absence of condensate opacity.  In fact, cloudy models provide significantly better fits to the near-infrared
spectrum of {\name} and closer agreement with its parallax distance, while also providing marginally better fits to data for ULAS~J1335+1130 and SDSS~J1416+1348B.  The fact that cloud opacity effects are more pronounced in the spectra of {\name} and ULAS~J1335+1130
indicates that clouds are particularly relevant for shaping the spectra of 
young, low-gravity brown dwarfs, and/or those with supersolar metallicities, following trends seen in the warmer L dwarfs.  

Despite the improvement, model fits to brown dwarf spectra remain a work in progress.  The cloudy models ``overcorrect'' $Y$- and $J$-band fluxes and underestimate absolute fluxes in the case of {\name}. Both shortcomings suggest that thinner clouds (e.g., {\fsed} = 3 or 4) may be needed to accurately account for condensate extinction, in line with values inferred for L/T transition objects \citep{2008ApJ...678.1372C, 2009ApJ...702..154S}.  Furthermore, the limited sample examined here may not be representative of the greater population of low-temperature, low surface gravity T dwarfs.  A comprehensive modeling effort can more fully explore the properties of clouds in these objects, while accurate parallax measurements would provide the necessary empirical constraints on surface fluxes.  Nevertheless, our results indicate that by including condensate opacity in T dwarf spectral models, more accurate determinations of the physical parameters of cold brown dwarfs and exoplanets are possible with near-infrared spectroscopy alone.

\acknowledgements

The authors acknowledges the efforts of the Las Campanas Observatory
staff in helping us bring FIRE to the first light and science operations.  
Specifically, we thank Frank Per\'{e}z, Alan Uomoto, Mark Phillips, Dave Osip, and Povilas Palunas
for their assistance during commissioning.
We also thank our telescope operators Jorge Araya, Mauricio Martinez, and Hern\'{a}n Nu\~{n}ez for their assistance during the observations.
We acknowledge helpful comments from G.\ Bjorn, B.\ Bowler, J.\ Faherty, J.\ Gizis, D.\ Looper and R.-D.\ Scholz; 
and thank our referee Channon Visscher for his prompt and helpful review of our original manuscript.
The authors also thank B.\ Burningham and S.\ Leggett for providing their spectral data of
ULAS~J1335+1130, and J.\ D.\ Kirkpatrick for providing his spectrum of Ross~458AB used in the analysis.
Support for the modeling work of D.S.\ was provided by NASA
through the Spitzer Science Center.
This research has benefitted from the M, L, and T dwarf compendium housed at DwarfArchives.org and maintained by Chris Gelino, Davy Kirkpatrick, and Adam Burgasser.

Facilities: \facility{Magellan~(FIRE)}

%\bibliography{../../biblibrary}

\begin{thebibliography}{163}
\expandafter\ifx\csname natexlab\endcsname\relax\def\natexlab#1{#1}\fi

\bibitem[{{Ackerman} \& {Marley}(2001)}]{2001ApJ...556..872A}
{Ackerman}, A.~S., \& {Marley}, M.~S. 2001, \apj, 556, 872

\bibitem[{{Allard} {et~al.}(2001){Allard}, {Hauschildt}, {Alexander},
  {Tamanai}, \& {Schweitzer}}]{2001ApJ...556..357A}
{Allard}, F., {Hauschildt}, P.~H., {Alexander}, D.~R., {Tamanai}, A., \&
  {Schweitzer}, A. 2001, \apj, 556, 357

\bibitem[{{Allard} {et~al.}(2003){Allard}, {Allard}, {Hauschildt}, {Kielkopf},
  \& {Machin}}]{2003A&A...411L.473A}
{Allard}, N.~F., {Allard}, F., {Hauschildt}, P.~H., {Kielkopf}, J.~F., \&
  {Machin}, L. 2003, \aap, 411, L473

\bibitem[{{Antoja} {et~al.}(2008){Antoja}, {Figueras}, {Fern{\'a}ndez}, \&
  {Torra}}]{2008A&A...490..135A}
{Antoja}, T., {Figueras}, F., {Fern{\'a}ndez}, D., \& {Torra}, J. 2008, \aap,
  490, 135

\bibitem[{{Barnes}(2003)}]{2003ApJ...586..464B}
{Barnes}, S.~A. 2003, \apj, 586, 464

\bibitem[{{Basri} \& {Brown}(2006)}]{2006AREPS..34..193B}
{Basri}, G., \& {Brown}, M.~E. 2006, Annual Review of Earth and Planetary
  Sciences, 34, 193

\bibitem[{{B{\'e}jar} {et~al.}(2008){B{\'e}jar}, {Zapatero Osorio},
  {P{\'e}rez-Garrido}, {{\'A}lvarez}, {Mart{\'{\i}}n}, {Rebolo},
  {Vill{\'o}-P{\'e}rez}, \& {D{\'{\i}}az-S{\'a}nchez}}]{2008ApJ...673L.185B}
{B{\'e}jar}, V.~J.~S., {Zapatero Osorio}, M.~R., {P{\'e}rez-Garrido}, A.,
  {{\'A}lvarez}, C., {Mart{\'{\i}}n}, E.~L., {Rebolo}, R.,
  {Vill{\'o}-P{\'e}rez}, I., \& {D{\'{\i}}az-S{\'a}nchez}, A. 2008, \apjl, 673,
  L185

\bibitem[{{Beuzit} {et~al.}(2004){Beuzit}, {S{\'e}gransan}, {Forveille},
  {Udry}, {Delfosse}, {Mayor}, {Perrier}, {Hainaut}, {Roddier}, {Roddier}, \&
  {Mart{\'{\i}}n}}]{2004A&A...425..997B}
{Beuzit}, J.-L., {S{\'e}gransan}, D., {Forveille}, T., {Udry}, S., {Delfosse},
  X., {Mayor}, M., {Perrier}, C., {Hainaut}, M.-C., {Roddier}, C., {Roddier},
  F., \& {Mart{\'{\i}}n}, E.~L. 2004, \aap, 425, 997

\bibitem[{{Bildsten} {et~al.}(1997){Bildsten}, {Brown}, {Matzner}, \&
  {Ushomirsky}}]{1997ApJ...482..442B}
{Bildsten}, L., {Brown}, E.~F., {Matzner}, C.~D., \& {Ushomirsky}, G. 1997,
  \apj, 482, 442

\bibitem[{{Bochanski} {et~al.}(2007){Bochanski}, {West}, {Hawley}, \&
  {Covey}}]{2007AJ....133..531B}
{Bochanski}, J.~J., {West}, A.~A., {Hawley}, S.~L., \& {Covey}, K.~R. 2007,
  \aj, 133, 531

\bibitem[{{Bonfils} {et~al.}(2005){Bonfils}, {Delfosse}, {Udry}, {Santos},
  {Forveille}, \& {S{\'e}gransan}}]{2005A&A...442..635B}
{Bonfils}, X., {Delfosse}, X., {Udry}, S., {Santos}, N.~C., {Forveille}, T., \&
  {S{\'e}gransan}, D. 2005, \aap, 442, 635

\bibitem[{{Bonnefoy} {et~al.}(2010){Bonnefoy}, {Chauvin}, {Rojo}, {Allard},
  {Lagrange}, {Homeier}, {Dumas}, \& {Beuzit}}]{2010A&A...512A..52B}
{Bonnefoy}, M., {Chauvin}, G., {Rojo}, P., {Allard}, F., {Lagrange}, A.,
  {Homeier}, D., {Dumas}, C., \& {Beuzit}, J. 2010, \aap, 512, A52+

\bibitem[{{Borysow}(2002)}]{2002A&A...390..779B}
{Borysow}, A. 2002, \aap, 390, 779

\bibitem[{{Boss}(2006)}]{2006ApJ...637L.137B}
{Boss}, A.~P. 2006, \apjl, 637, L137

\bibitem[{{Bowler} {et~al.}(2009){Bowler}, {Liu}, \&
  {Cushing}}]{2009ApJ...706.1114B}
{Bowler}, B.~P., {Liu}, M.~C., \& {Cushing}, M.~C. 2009, \apj, 706, 1114

\bibitem[{{Bowler} {et~al.}(2010){Bowler}, {Liu}, {Dupuy}, \&
  {Cushing}}]{2010arXiv1008.4582B}
{Bowler}, B.~P., {Liu}, M.~C., {Dupuy}, T.~J., \& {Cushing}, M.~C. 2010, ArXiv
  e-prints

\bibitem[{{Browning} {et~al.}(2010){Browning}, {Basri}, {Marcy}, {West}, \&
  {Zhang}}]{2010AJ....139..504B}
{Browning}, M.~K., {Basri}, G., {Marcy}, G.~W., {West}, A.~A., \& {Zhang}, J.
  2010, \aj, 139, 504

\bibitem[{{Burgasser}(2007)}]{2007ApJ...659..655B}
{Burgasser}, A.~J. 2007, \apj, 659, 655

\bibitem[{{Burgasser} {et~al.}(2006{\natexlab{a}}){Burgasser}, {Burrows}, \&
  {Kirkpatrick}}]{2006ApJ...639.1095B}
{Burgasser}, A.~J., {Burrows}, A., \& {Kirkpatrick}, J.~D. 2006{\natexlab{a}},
  \apj, 639, 1095

\bibitem[{{Burgasser} {et~al.}(2010{\natexlab{a}}){Burgasser}, {Cruz},
  {Cushing}, {Gelino}, {Looper}, {Faherty}, {Kirkpatrick}, \&
  {Reid}}]{2010ApJ...710.1142B}
{Burgasser}, A.~J., {Cruz}, K.~L., {Cushing}, M., {Gelino}, C.~R., {Looper},
  D.~L., {Faherty}, J.~K., {Kirkpatrick}, J.~D., \& {Reid}, I.~N.
  2010{\natexlab{a}}, \apj, 710, 1142

\bibitem[{{Burgasser} {et~al.}(2007){Burgasser}, {Cruz}, \&
  {Kirkpatrick}}]{2007ApJ...657..494B}
{Burgasser}, A.~J., {Cruz}, K.~L., \& {Kirkpatrick}, J.~D. 2007, \apj, 657, 494

\bibitem[{{Burgasser} {et~al.}(2006{\natexlab{b}}){Burgasser}, {Geballe},
  {Leggett}, {Kirkpatrick}, \& {Golimowski}}]{2006ApJ...637.1067B}
{Burgasser}, A.~J., {Geballe}, T.~R., {Leggett}, S.~K., {Kirkpatrick}, J.~D.,
  \& {Golimowski}, D.~A. 2006{\natexlab{b}}, \apj, 637, 1067

\bibitem[{{Burgasser} {et~al.}(2002{\natexlab{a}}){Burgasser}, {Kirkpatrick},
  {Brown}, {Reid}, {Burrows}, {Liebert}, {Matthews}, {Gizis}, {Dahn}, {Monet},
  {Cutri}, \& {Skrutskie}}]{2002ApJ...564..421B}
{Burgasser}, A.~J., {Kirkpatrick}, J.~D., {Brown}, M.~E., {Reid}, I.~N.,
  {Burrows}, A., {Liebert}, J., {Matthews}, K., {Gizis}, J.~E., {Dahn}, C.~C.,
  {Monet}, D.~G., {Cutri}, R.~M., \& {Skrutskie}, M.~F. 2002{\natexlab{a}},
  \apj, 564, 421

\bibitem[{{Burgasser} {et~al.}(2003){Burgasser}, {Kirkpatrick}, {Burrows},
  {Liebert}, {Reid}, {Gizis}, {McGovern}, {Prato}, \&
  {McLean}}]{2003ApJ...592.1186B}
{Burgasser}, A.~J., {Kirkpatrick}, J.~D., {Burrows}, A., {Liebert}, J., {Reid},
  I.~N., {Gizis}, J.~E., {McGovern}, M.~R., {Prato}, L., \& {McLean}, I.~S.
  2003, \apj, 592, 1186

\bibitem[{{Burgasser} {et~al.}(2010{\natexlab{b}}){Burgasser}, {Looper}, \&
  {Rayner}}]{2010AJ....139.2448B}
{Burgasser}, A.~J., {Looper}, D., \& {Rayner}, J.~T. 2010{\natexlab{b}}, \aj,
  139, 2448

\bibitem[{{Burgasser} {et~al.}(2008{\natexlab{a}}){Burgasser}, {Looper},
  {Kirkpatrick}, {Cruz}, \& {Swift}}]{2008ApJ...674..451B}
{Burgasser}, A.~J., {Looper}, D.~L., {Kirkpatrick}, J.~D., {Cruz}, K.~L., \&
  {Swift}, B.~J. 2008{\natexlab{a}}, \apj, 674, 451

\bibitem[{{Burgasser} {et~al.}(2002{\natexlab{b}}){Burgasser}, {Marley},
  {Ackerman}, {Saumon}, {Lodders}, {Dahn}, {Harris}, \&
  {Kirkpatrick}}]{2002ApJ...571L.151B}
{Burgasser}, A.~J., {Marley}, M.~S., {Ackerman}, A.~S., {Saumon}, D.,
  {Lodders}, K., {Dahn}, C.~C., {Harris}, H.~C., \& {Kirkpatrick}, J.~D.
  2002{\natexlab{b}}, \apjl, 571, L151

\bibitem[{{Burgasser} {et~al.}(2008{\natexlab{b}}){Burgasser}, {Tinney},
  {Cushing}, {Saumon}, {Marley}, {Bennett}, \&
  {Kirkpatrick}}]{2008ApJ...689L..53B}
{Burgasser}, A.~J., {Tinney}, C.~G., {Cushing}, M.~C., {Saumon}, D., {Marley},
  M.~S., {Bennett}, C.~S., \& {Kirkpatrick}, J.~D. 2008{\natexlab{b}}, \apjl,
  689, L53

\bibitem[{{Burgasser} {et~al.}(2000)}]{2000ApJ...531L..57B}
{Burgasser}, A.~J., {et~al.} 2000, \apjl, 531, L57

\bibitem[{{Burgess} {et~al.}(2009){Burgess}, {Moraux}, {Bouvier}, {Marmo},
  {Albert}, \& {Bouy}}]{2009A&A...508..823B}
{Burgess}, A.~S.~M., {Moraux}, E., {Bouvier}, J., {Marmo}, C., {Albert}, L., \&
  {Bouy}, H. 2009, \aap, 508, 823

\bibitem[{{Burningham} {et~al.}(2010){Burningham}, {Leggett}, {Lucas},
  {Pinfield}, {Smart}, {Day-Jones}, {Jones}, {Murray}, {Nickson}, {Tamura},
  {Zhang}, {Lodieu}, {Tinney}, \& {Zapatero Osorio}}]{2010MNRAS.404.1952B}
{Burningham}, B., {Leggett}, S.~K., {Lucas}, P.~W., {Pinfield}, D.~J., {Smart},
  R.~L., {Day-Jones}, A.~C., {Jones}, H.~R.~A., {Murray}, D., {Nickson}, E.,
  {Tamura}, M., {Zhang}, Z., {Lodieu}, N., {Tinney}, C.~G., \& {Zapatero
  Osorio}, M.~R. 2010, \mnras, 404, 1952

\bibitem[{{Burningham} {et~al.}(2008)}]{2008MNRAS.391..320B}
{Burningham}, B., {et~al.} 2008, \mnras, 391, 320

\bibitem[{{Burningham} {et~al.}(2009)}]{2009MNRAS.395.1237B}
---. 2009, \mnras, 395, 1237

\bibitem[{{Burrows} {et~al.}(2002){Burrows}, {Burgasser}, {Kirkpatrick},
  {Liebert}, {Milsom}, {Sudarsky}, \& {Hubeny}}]{2002ApJ...573..394B}
{Burrows}, A., {Burgasser}, A.~J., {Kirkpatrick}, J.~D., {Liebert}, J.,
  {Milsom}, J.~A., {Sudarsky}, D., \& {Hubeny}, I. 2002, \apj, 573, 394

\bibitem[{{Burrows} {et~al.}(2006){Burrows}, {Sudarsky}, \&
  {Hubeny}}]{2006ApJ...640.1063B}
{Burrows}, A., {Sudarsky}, D., \& {Hubeny}, I. 2006, \apj, 640, 1063

\bibitem[{{Burrows} \& {Volobuyev}(2003)}]{2003ApJ...583..985B}
{Burrows}, A., \& {Volobuyev}, M. 2003, \apj, 583, 985

\bibitem[{{Casewell} {et~al.}(2007){Casewell}, {Dobbie}, {Hodgkin}, {Moraux},
  {Jameson}, {Hambly}, {Irwin}, \& {Lodieu}}]{2007MNRAS.378.1131C}
{Casewell}, S.~L., {Dobbie}, P.~D., {Hodgkin}, S.~T., {Moraux}, E., {Jameson},
  R.~F., {Hambly}, N.~C., {Irwin}, J., \& {Lodieu}, N. 2007, \mnras, 378, 1131

\bibitem[{{Chabrier} \& {Baraffe}(2000)}]{2000ARA&A..38..337C}
{Chabrier}, G., \& {Baraffe}, I. 2000, \araa, 38, 337

\bibitem[{{Chabrier} {et~al.}(2000){Chabrier}, {Baraffe}, {Allard}, \&
  {Hauschildt}}]{2000ApJ...542..464C}
{Chabrier}, G., {Baraffe}, I., {Allard}, F., \& {Hauschildt}, P. 2000, \apj,
  542, 464

\bibitem[{{Chabrier} {et~al.}(1996){Chabrier}, {Baraffe}, \&
  {Plez}}]{1996ApJ...459L..91C}
{Chabrier}, G., {Baraffe}, I., \& {Plez}, B. 1996, \apjl, 459, L91+

\bibitem[{{Chauvin} {et~al.}(2005){Chauvin}, {Lagrange}, {Zuckerman}, {Dumas},
  {Mouillet}, {Song}, {Beuzit}, {Lowrance}, \& {Bessell}}]{2005A&A...438L..29C}
{Chauvin}, G., {Lagrange}, A.-M., {Zuckerman}, B., {Dumas}, C., {Mouillet}, D.,
  {Song}, I., {Beuzit}, J.-L., {Lowrance}, P., \& {Bessell}, M.~S. 2005, \aap,
  438, L29

\bibitem[{{Cooper} {et~al.}(2003){Cooper}, {Sudarsky}, {Milsom}, {Lunine}, \&
  {Burrows}}]{2003ApJ...586.1320C}
{Cooper}, C.~S., {Sudarsky}, D., {Milsom}, J.~A., {Lunine}, J.~I., \&
  {Burrows}, A. 2003, \apj, 586, 1320

\bibitem[{{Cushing} {et~al.}(2008){Cushing}, {Marley}, {Saumon}, {Kelly},
  {Vacca}, {Rayner}, {Freedman}, {Lodders}, \& {Roellig}}]{2008ApJ...678.1372C}
{Cushing}, M.~C., {Marley}, M.~S., {Saumon}, D., {Kelly}, B.~C., {Vacca},
  W.~D., {Rayner}, J.~T., {Freedman}, R.~S., {Lodders}, K., \& {Roellig}, T.~L.
  2008, \apj, 678, 1372

\bibitem[{{Cushing} {et~al.}(2005){Cushing}, {Rayner}, \&
  {Vacca}}]{2005ApJ...623.1115C}
{Cushing}, M.~C., {Rayner}, J.~T., \& {Vacca}, W.~D. 2005, \apj, 623, 1115

\bibitem[{{Cushing} {et~al.}(2006){Cushing}, {Roellig}, {Marley}, {Saumon},
  {Leggett}, {Kirkpatrick}, {Wilson}, {Sloan}, {Mainzer}, {Van Cleve}, \&
  {Houck}}]{2006ApJ...648..614C}
{Cushing}, M.~C., {Roellig}, T.~L., {Marley}, M.~S., {Saumon}, D., {Leggett},
  S.~K., {Kirkpatrick}, J.~D., {Wilson}, J.~C., {Sloan}, G.~C., {Mainzer},
  A.~K., {Van Cleve}, J.~E., \& {Houck}, J.~R. 2006, \apj, 648, 614

\bibitem[{{Cushing} {et~al.}(2010){Cushing}, {Saumon}, \&
  {Marley}}]{2010arXiv1009.2802C}
{Cushing}, M.~C., {Saumon}, D., \& {Marley}, M.~S. 2010, ArXiv e-prints

\bibitem[{{Cushing} {et~al.}(2004){Cushing}, {Vacca}, \&
  {Rayner}}]{2004PASP..116..362C}
{Cushing}, M.~C., {Vacca}, W.~D., \& {Rayner}, J.~T. 2004, \pasp, 116, 362

\bibitem[{{de Bruijne} {et~al.}(2001){de Bruijne}, {Hoogerwerf}, \& {de
  Zeeuw}}]{2001A&A...367..111D}
{de Bruijne}, J.~H.~J., {Hoogerwerf}, R., \& {de Zeeuw}, P.~T. 2001, \aap, 367,
  111

\bibitem[{{Dehnen} \& {Binney}(1998)}]{1998MNRAS.298..387D}
{Dehnen}, W., \& {Binney}, J.~J. 1998, \mnras, 298, 387

\bibitem[{{Delorme} {et~al.}(2008){Delorme}, {Willott}, {Forveille},
  {Delfosse}, {Reyl{\'e}}, {Bertin}, {Albert}, {Artigau}, {Robin}, {Allard},
  {Doyon}, \& {Hill}}]{2008A&A...484..469D}
{Delorme}, P., {Willott}, C.~J., {Forveille}, T., {Delfosse}, X., {Reyl{\'e}},
  C., {Bertin}, E., {Albert}, L., {Artigau}, E., {Robin}, A.~C., {Allard}, F.,
  {Doyon}, R., \& {Hill}, G.~J. 2008, \aap, 484, 469

\bibitem[{{Dodson-Robinson} {et~al.}(2009){Dodson-Robinson}, {Veras}, {Ford},
  \& {Beichman}}]{2009ApJ...707...79D}
{Dodson-Robinson}, S.~E., {Veras}, D., {Ford}, E.~B., \& {Beichman}, C.~A.
  2009, \apj, 707, 79

\bibitem[{{Dupuy} {et~al.}(2009{\natexlab{a}}){Dupuy}, {Liu}, \&
  {Ireland}}]{2009ApJ...692..729D}
{Dupuy}, T.~J., {Liu}, M.~C., \& {Ireland}, M.~J. 2009{\natexlab{a}}, \apj,
  692, 729

\bibitem[{{Dupuy} {et~al.}(2009{\natexlab{b}}){Dupuy}, {Liu}, \&
  {Ireland}}]{2009ApJ...699..168D}
---. 2009{\natexlab{b}}, \apj, 699, 168

\bibitem[{{Eggen}(1960)}]{1960MNRAS.120..540E}
{Eggen}, O.~J. 1960, \mnras, 120, 540

\bibitem[{{Faherty} {et~al.}(2009){Faherty}, {Burgasser}, {Cruz}, {Shara},
  {Walter}, \& {Gelino}}]{2009AJ....137....1F}
{Faherty}, J.~K., {Burgasser}, A.~J., {Cruz}, K.~L., {Shara}, M.~M., {Walter},
  F.~M., \& {Gelino}, C.~R. 2009, \aj, 137, 1

\bibitem[{{Famaey} {et~al.}(2005){Famaey}, {Jorissen}, {Luri}, {Mayor}, {Udry},
  {Dejonghe}, \& {Turon}}]{2005A&A...430..165F}
{Famaey}, B., {Jorissen}, A., {Luri}, X., {Mayor}, M., {Udry}, S., {Dejonghe},
  H., \& {Turon}, C. 2005, \aap, 430, 165

\bibitem[{{Feast} \& {Whitelock}(1997)}]{1997MNRAS.291..683F}
{Feast}, M., \& {Whitelock}, P. 1997, \mnras, 291, 683

\bibitem[{{Fegley} \& {Lodders}(1996)}]{1996ApJ...472L..37F}
{Fegley}, B.~J., \& {Lodders}, K. 1996, \apjl, 472, L37+

\bibitem[{{Fleming} {et~al.}(1995){Fleming}, {Schmitt}, \&
  {Giampapa}}]{1995ApJ...450..401F}
{Fleming}, T.~A., {Schmitt}, J.~H.~M.~M., \& {Giampapa}, M.~S. 1995, \apj, 450,
  401

\bibitem[{{Freedman} {et~al.}(2008){Freedman}, {Marley}, \&
  {Lodders}}]{2008ApJS..174..504F}
{Freedman}, R.~S., {Marley}, M.~S., \& {Lodders}, K. 2008, \apjs, 174, 504

\bibitem[{{Geballe} {et~al.}(2002){Geballe}, {Knapp}, {Leggett}, {Fan},
  {Golimowski}, {Anderson}, {Brinkmann}, {Csabai}, {Gunn}, {Hawley},
  {Hennessy}, {Henry}, {Hill}, {Hindsley}, {Ivezi{\'c}}, {Lupton}, {McDaniel},
  {Munn}, {Narayanan}, {Peng}, {Pier}, {Rockosi}, {Schneider}, {Smith},
  {Strauss}, {Tsvetanov}, {Uomoto}, {York}, \& {Zheng}}]{2002ApJ...564..466G}
{Geballe}, T.~R., {Knapp}, G.~R., {Leggett}, S.~K., {Fan}, X., {Golimowski},
  D.~A., {Anderson}, S., {Brinkmann}, J., {Csabai}, I., {Gunn}, J.~E.,
  {Hawley}, S.~L., {Hennessy}, G., {Henry}, T.~J., {Hill}, G.~J., {Hindsley},
  R.~B., {Ivezi{\'c}}, {\v Z}., {Lupton}, R.~H., {McDaniel}, A., {Munn}, J.~A.,
  {Narayanan}, V.~K., {Peng}, E., {Pier}, J.~R., {Rockosi}, C.~M., {Schneider},
  D.~P., {Smith}, J.~A., {Strauss}, M.~A., {Tsvetanov}, Z.~I., {Uomoto}, A.,
  {York}, D.~G., \& {Zheng}, W. 2002, \apj, 564, 466

\bibitem[{{Geballe} {et~al.}(2001){Geballe}, {Saumon}, {Leggett}, {Knapp},
  {Marley}, \& {Lodders}}]{2001ApJ...556..373G}
{Geballe}, T.~R., {Saumon}, D., {Leggett}, S.~K., {Knapp}, G.~R., {Marley},
  M.~S., \& {Lodders}, K. 2001, \apj, 556, 373

\bibitem[{{Gizis} \& {Harvin}(2006)}]{2006AJ....132.2372G}
{Gizis}, J.~E., \& {Harvin}, J. 2006, \aj, 132, 2372

\bibitem[{{Goldman} {et~al.}(2010){Goldman}, {Marsat}, {Henning}, {Clemens}, \&
  {Greiner}}]{2010MNRAS.405.1140G}
{Goldman}, B., {Marsat}, S., {Henning}, T., {Clemens}, C., \& {Greiner}, J.
  2010, \mnras, 405, 1140

\bibitem[{{Golimowski} {et~al.}(2004)}]{2004AJ....127.3516G}
{Golimowski}, D.~A., {et~al.} 2004, \aj, 127, 3516

\bibitem[{{Greiner} {et~al.}(2008){Greiner}, {Bornemann}, {Clemens}, {Deuter},
  {Hasinger}, {Honsberg}, {Huber}, {Huber}, {Krauss}, {Kr{\"u}hler},
  {K{\"u}pc{\"u} Yolda{\c s}}, {Mayer-Hasselwander}, {Mican}, {Primak},
  {Schrey}, {Steiner}, {Szokoly}, {Th{\"o}ne}, {Yolda{\c s}}, {Klose}, {Laux},
  \& {Winkler}}]{2008PASP..120..405G}
{Greiner}, J., {Bornemann}, W., {Clemens}, C., {Deuter}, M., {Hasinger}, G.,
  {Honsberg}, M., {Huber}, H., {Huber}, S., {Krauss}, M., {Kr{\"u}hler}, T.,
  {K{\"u}pc{\"u} Yolda{\c s}}, A., {Mayer-Hasselwander}, H., {Mican}, B.,
  {Primak}, N., {Schrey}, F., {Steiner}, I., {Szokoly}, G., {Th{\"o}ne}, C.~C.,
  {Yolda{\c s}}, A., {Klose}, S., {Laux}, U., \& {Winkler}, J. 2008, \pasp,
  120, 405

\bibitem[{{Griffith} \& {Yelle}(1999)}]{1999ApJ...519L..85G}
{Griffith}, C.~A., \& {Yelle}, R.~V. 1999, \apjl, 519, L85

\bibitem[{{Helling} {et~al.}(2008){Helling}, {Dehn}, {Woitke}, \&
  {Hauschildt}}]{2008ApJ...675L.105H}
{Helling}, C., {Dehn}, M., {Woitke}, P., \& {Hauschildt}, P.~H. 2008, \apjl,
  675, L105

\bibitem[{{Helling} {et~al.}(2006){Helling}, {Thi}, {Woitke}, \&
  {Fridlund}}]{2006A&A...451L...9H}
{Helling}, C., {Thi}, W.-F., {Woitke}, P., \& {Fridlund}, M. 2006, \aap, 451,
  L9

\bibitem[{{Hodapp} {et~al.}(2003){Hodapp}, {Jensen}, {Irwin}, {Yamada},
  {Chung}, {Fletcher}, {Robertson}, {Hora}, {Simons}, {Mays}, {Nolan}, {Bec},
  {Merrill}, \& {Fowler}}]{2003PASP..115.1388H}
{Hodapp}, K.~W., {Jensen}, J.~B., {Irwin}, E.~M., {Yamada}, H., {Chung}, R.,
  {Fletcher}, K., {Robertson}, L., {Hora}, J.~L., {Simons}, D.~A., {Mays}, W.,
  {Nolan}, R., {Bec}, M., {Merrill}, M., \& {Fowler}, A.~M. 2003, \pasp, 115,
  1388

\bibitem[{{H{\o}g} {et~al.}(2000){H{\o}g}, {Fabricius}, {Makarov}, {Urban},
  {Corbin}, {Wycoff}, {Bastian}, {Schwekendiek}, \&
  {Wicenec}}]{2000A&A...355L..27H}
{H{\o}g}, E., {Fabricius}, C., {Makarov}, V.~V., {Urban}, S., {Corbin}, T.,
  {Wycoff}, G., {Bastian}, U., {Schwekendiek}, P., \& {Wicenec}, A. 2000, \aap,
  355, L27

\bibitem[{{Holman} \& {Wiegert}(1999)}]{1999AJ....117..621H}
{Holman}, M.~J., \& {Wiegert}, P.~A. 1999, \aj, 117, 621

\bibitem[{{Houck} {et~al.}(2004){Houck}, {Roellig}, {van Cleve}, {Forrest},
  {Herter}, {Lawrence}, {Matthews}, {Reitsema}, {Soifer}, {Watson}, {Weedman},
  {Huisjen}, {Troeltzsch}, {Barry}, {Bernard-Salas}, {Blacken}, {Brandl},
  {Charmandaris}, {Devost}, {Gull}, {Hall}, {Henderson}, {Higdon}, {Pirger},
  {Schoenwald}, {Sloan}, {Uchida}, {Appleton}, {Armus}, {Burgdorf},
  {Fajardo-Acosta}, {Grillmair}, {Ingalls}, {Morris}, \&
  {Teplitz}}]{2004ApJS..154...18H}
{Houck}, J.~R., {Roellig}, T.~L., {van Cleve}, J., {Forrest}, W.~J., {Herter},
  T., {Lawrence}, C.~R., {Matthews}, K., {Reitsema}, H.~J., {Soifer}, B.~T.,
  {Watson}, D.~M., {Weedman}, D., {Huisjen}, M., {Troeltzsch}, J., {Barry},
  D.~J., {Bernard-Salas}, J., {Blacken}, C.~E., {Brandl}, B.~R.,
  {Charmandaris}, V., {Devost}, D., {Gull}, G.~E., {Hall}, P., {Henderson},
  C.~P., {Higdon}, S.~J.~U., {Pirger}, B.~E., {Schoenwald}, J., {Sloan}, G.~C.,
  {Uchida}, K.~I., {Appleton}, P.~N., {Armus}, L., {Burgdorf}, M.~J.,
  {Fajardo-Acosta}, S.~B., {Grillmair}, C.~J., {Ingalls}, J.~G., {Morris},
  P.~W., \& {Teplitz}, H.~I. 2004, \apjs, 154, 18

\bibitem[{{Hubeny} \& {Burrows}(2007)}]{2007ApJ...669.1248H}
{Hubeny}, I., \& {Burrows}, A. 2007, \apj, 669, 1248

\bibitem[{{Janson} {et~al.}(2010){Janson}, {Bergfors}, {Goto}, {Brandner}, \&
  {Lafreni{\`e}re}}]{2010ApJ...710L..35J}
{Janson}, M., {Bergfors}, C., {Goto}, M., {Brandner}, W., \& {Lafreni{\`e}re},
  D. 2010, \apjl, 710, L35

\bibitem[{{Johnson} \& {Apps}(2009)}]{2009ApJ...699..933J}
{Johnson}, J.~A., \& {Apps}, K. 2009, \apj, 699, 933

\bibitem[{{Jones} \& {Tsuji}(1997)}]{1997ApJ...480L..39J}
{Jones}, H.~R.~A., \& {Tsuji}, T. 1997, \apjl, 480, L39+

\bibitem[{{Kalas} {et~al.}(2008){Kalas}, {Graham}, {Chiang}, {Fitzgerald},
  {Clampin}, {Kite}, {Stapelfeldt}, {Marois}, \& {Krist}}]{2008Sci...322.1345K}
{Kalas}, P., {Graham}, J.~R., {Chiang}, E., {Fitzgerald}, M.~P., {Clampin}, M.,
  {Kite}, E.~S., {Stapelfeldt}, K., {Marois}, C., \& {Krist}, J. 2008, Science,
  322, 1345

\bibitem[{{Kiraga} \& {Stepien}(2007)}]{2007AcA....57..149K}
{Kiraga}, M., \& {Stepien}, K. 2007, Acta Astronomica, 57, 149

\bibitem[{{Kirkpatrick}(2005)}]{2005ARA&A..43..195K}
{Kirkpatrick}, J.~D. 2005, \araa, 43, 195

\bibitem[{{Kirkpatrick} {et~al.}(1999){Kirkpatrick}, {Reid}, {Liebert},
  {Cutri}, {Nelson}, {Beichman}, {Dahn}, {Monet}, {Gizis}, \&
  {Skrutskie}}]{1999ApJ...519..802K}
{Kirkpatrick}, J.~D., {Reid}, I.~N., {Liebert}, J., {Cutri}, R.~M., {Nelson},
  B., {Beichman}, C.~A., {Dahn}, C.~C., {Monet}, D.~G., {Gizis}, J.~E., \&
  {Skrutskie}, M.~F. 1999, \apj, 519, 802

\bibitem[{{Kirkpatrick} {et~al.}(2008)}]{2008ApJ...689.1295K}
{Kirkpatrick}, J.~D., {et~al.} 2008, \apj, 689, 1295

\bibitem[{{Knapp} {et~al.}(2004){Knapp}, {Leggett}, {Fan}, {Marley}, {Geballe},
  {Golimowski}, {Finkbeiner}, {Gunn}, {Hennawi}, {Ivezi{\'c}}, {Lupton},
  {Schlegel}, {Strauss}, {Tsvetanov}, {Chiu}, {Hoversten}, {Glazebrook},
  {Zheng}, {Hendrickson}, {Williams}, {Uomoto}, {Vrba}, {Henden}, {Luginbuhl},
  {Guetter}, {Munn}, {Canzian}, {Schneider}, \&
  {Brinkmann}}]{2004AJ....127.3553K}
{Knapp}, G.~R., {Leggett}, S.~K., {Fan}, X., {Marley}, M.~S., {Geballe}, T.~R.,
  {Golimowski}, D.~A., {Finkbeiner}, D., {Gunn}, J.~E., {Hennawi}, J.,
  {Ivezi{\'c}}, Z., {Lupton}, R.~H., {Schlegel}, D.~J., {Strauss}, M.~A.,
  {Tsvetanov}, Z.~I., {Chiu}, K., {Hoversten}, E.~A., {Glazebrook}, K.,
  {Zheng}, W., {Hendrickson}, M., {Williams}, C.~C., {Uomoto}, A., {Vrba},
  F.~J., {Henden}, A.~A., {Luginbuhl}, C.~B., {Guetter}, H.~H., {Munn}, J.~A.,
  {Canzian}, B., {Schneider}, D.~P., \& {Brinkmann}, J. 2004, \aj, 127, 3553

\bibitem[{{Lafreni{\`e}re} {et~al.}(2008){Lafreni{\`e}re}, {Jayawardhana}, \&
  {van Kerkwijk}}]{2008ApJ...689L.153L}
{Lafreni{\`e}re}, D., {Jayawardhana}, R., \& {van Kerkwijk}, M.~H. 2008, \apjl,
  689, L153

\bibitem[{{Lafreni{\`e}re} {et~al.}(2010){Lafreni{\`e}re}, {Jayawardhana}, \&
  {van Kerkwijk}}]{2010ApJ...719..497L}
---. 2010, \apj, 719, 497

\bibitem[{{Lagrange} {et~al.}(2010){Lagrange}, {Bonnefoy}, {Chauvin}, {Apai},
  {Ehrenreich}, {Boccaletti}, {Gratadour}, {Rouan}, {Mouillet}, {Lacour}, \&
  {Kasper}}]{2010Sci...329...57L}
{Lagrange}, A., {Bonnefoy}, M., {Chauvin}, G., {Apai}, D., {Ehrenreich}, D.,
  {Boccaletti}, A., {Gratadour}, D., {Rouan}, D., {Mouillet}, D., {Lacour}, S.,
  \& {Kasper}, M. 2010, Science, 329, 57

\bibitem[{{Lawrence} {et~al.}(2007)}]{2007MNRAS.379.1599L}
{Lawrence}, A., {et~al.} 2007, \mnras, 379, 1599

\bibitem[{{Leggett} {et~al.}(2010){Leggett}, {Burningham}, {Saumon}, {Marley},
  {Warren}, {Smart}, {Jones}, {Lucas}, {Pinfield}, \&
  {Tamura}}]{2010ApJ...710.1627L}
{Leggett}, S.~K., {Burningham}, B., {Saumon}, D., {Marley}, M.~S., {Warren},
  S.~J., {Smart}, R.~L., {Jones}, H.~R.~A., {Lucas}, P.~W., {Pinfield}, D.~J.,
  \& {Tamura}, M. 2010, \apj, 710, 1627

\bibitem[{{Leggett} {et~al.}(2009){Leggett}, {Cushing}, {Saumon}, {Marley},
  {Roellig}, {Warren}, {Burningham}, {Jones}, {Kirkpatrick}, {Lodieu}, {Lucas},
  {Mainzer}, {Mart{\'{\i}}n}, {McCaughrean}, {Pinfield}, {Sloan}, {Smart},
  {Tamura}, \& {Van Cleve}}]{2009ApJ...695.1517L}
{Leggett}, S.~K., {Cushing}, M.~C., {Saumon}, D., {Marley}, M.~S., {Roellig},
  T.~L., {Warren}, S.~J., {Burningham}, B., {Jones}, H.~R.~A., {Kirkpatrick},
  J.~D., {Lodieu}, N., {Lucas}, P.~W., {Mainzer}, A.~K., {Mart{\'{\i}}n},
  E.~L., {McCaughrean}, M.~J., {Pinfield}, D.~J., {Sloan}, G.~C., {Smart},
  R.~L., {Tamura}, M., \& {Van Cleve}, J. 2009, \apj, 695, 1517

\bibitem[{{Leggett} {et~al.}(2007){Leggett}, {Marley}, {Freedman}, {Saumon},
  {Liu}, {Geballe}, {Golimowski}, \& {Stephens}}]{2007ApJ...667..537L}
{Leggett}, S.~K., {Marley}, M.~S., {Freedman}, R., {Saumon}, D., {Liu}, M.~C.,
  {Geballe}, T.~R., {Golimowski}, D.~A., \& {Stephens}, D.~C. 2007, \apj, 667,
  537

\bibitem[{{Leggett} {et~al.}(2008){Leggett}, {Saumon}, {Albert}, {Cushing},
  {Liu}, {Luhman}, {Marley}, {Kirkpatrick}, {Roellig}, \&
  {Allers}}]{2008ApJ...682.1256L}
{Leggett}, S.~K., {Saumon}, D., {Albert}, L., {Cushing}, M.~C., {Liu}, M.~C.,
  {Luhman}, K.~L., {Marley}, M.~S., {Kirkpatrick}, J.~D., {Roellig}, T.~L., \&
  {Allers}, K.~N. 2008, \apj, 682, 1256

\bibitem[{{Linsky}(1969)}]{1969ApJ...156..989L}
{Linsky}, J.~L. 1969, \apj, 156, 989

\bibitem[{{Liu} {et~al.}(2008){Liu}, {Cutri}, {Greanias}, {Duval},
  {Eisenhardt}, {Elwell}, {Heinrichsen}, {Howard}, {Irace}, {Mainzer},
  {Razzaghi}, {Royer}, \& {Wright}}]{2008SPIE.7017E..16L}
{Liu}, F., {Cutri}, R., {Greanias}, G., {Duval}, V., {Eisenhardt}, P.,
  {Elwell}, J., {Heinrichsen}, I., {Howard}, J., {Irace}, W., {Mainzer}, A.,
  {Razzaghi}, A., {Royer}, D., \& {Wright}, E.~L. 2008, in Society of
  Photo-Optical Instrumentation Engineers (SPIE) Conference Series, Vol. 7017,
  Society of Photo-Optical Instrumentation Engineers (SPIE) Conference Series

\bibitem[{{Liu} {et~al.}(2010){Liu}, {Dupuy}, \&
  {Leggett}}]{2010arXiv1008.2200L}
{Liu}, M.~C., {Dupuy}, T.~J., \& {Leggett}, S.~K. 2010, ArXiv e-prints

\bibitem[{{Lodders}(1999)}]{1999ApJ...519..793L}
{Lodders}, K. 1999, \apj, 519, 793

\bibitem[{{Lodders}(2002)}]{2002ApJ...577..974L}
---. 2002, \apj, 577, 974

\bibitem[{{Lodders}(2003)}]{2003ApJ...591.1220L}
---. 2003, \apj, 591, 1220

\bibitem[{{Lodders} \& {Fegley}(2002)}]{2002Icar..155..393L}
{Lodders}, K., \& {Fegley}, B. 2002, Icarus, 155, 393

\bibitem[{{Lodders} \& {Fegley}(2006)}]{2006asup.book....1L}
{Lodders}, K., \& {Fegley}, Jr., B. 2006, {Chemistry of Low Mass Substellar
  Objects} (Astrophysics Update 2), 1--+

\bibitem[{{Looper} {et~al.}(2008){Looper}, {Kirkpatrick}, {Cutri}, {Barman},
  {Burgasser}, {Cushing}, {Roellig}, {McGovern}, {McLean}, {Rice}, {Swift}, \&
  {Schurr}}]{2008ApJ...686..528L}
{Looper}, D.~L., {Kirkpatrick}, J.~D., {Cutri}, R.~M., {Barman}, T.,
  {Burgasser}, A.~J., {Cushing}, M.~C., {Roellig}, T., {McGovern}, M.~R.,
  {McLean}, I.~S., {Rice}, E., {Swift}, B.~J., \& {Schurr}, S.~D. 2008, \apj,
  686, 528

\bibitem[{{L{\'o}pez-Santiago} {et~al.}(2009){L{\'o}pez-Santiago}, {Micela}, \&
  {Montes}}]{2009A&A...499..129L}
{L{\'o}pez-Santiago}, J., {Micela}, G., \& {Montes}, D. 2009, \aap, 499, 129

\bibitem[{{Luhman} {et~al.}(2007)}]{2007ApJ...654..570L}
{Luhman}, K.~L., {et~al.} 2007, \apj, 654, 570

\bibitem[{{Madsen}(2003)}]{2003A&A...401..565M}
{Madsen}, S. 2003, \aap, 401, 565

\bibitem[{{Magazzu} {et~al.}(1993){Magazzu}, {Martin}, \&
  {Rebolo}}]{1993ApJ...404L..17M}
{Magazzu}, A., {Martin}, E.~L., \& {Rebolo}, R. 1993, \apjl, 404, L17

\bibitem[{{Marchi}(2007)}]{2007ApJ...666..475M}
{Marchi}, S. 2007, \apj, 666, 475

\bibitem[{{Marley}(2000)}]{2000ASPC..212..152M}
{Marley}, M. 2000, in Astronomical Society of the Pacific Conference Series,
  Vol. 212, From Giant Planets to Cool Stars, ed. {C.~A.~Griffith \&
  M.~S.~Marley}, 152--+

\bibitem[{{Marley} {et~al.}(submitted){Marley}, {Saumon}, \&
  {Goldblatt}}]{marley10}
{Marley}, M.~S., {Saumon}, D., \& {Goldblatt}, C. submitted, ApJ

\bibitem[{{Marley} {et~al.}(2002){Marley}, {Seager}, {Saumon}, {Lodders},
  {Ackerman}, {Freedman}, \& {Fan}}]{2002ApJ...568..335M}
{Marley}, M.~S., {Seager}, S., {Saumon}, D., {Lodders}, K., {Ackerman}, A.~S.,
  {Freedman}, R.~S., \& {Fan}, X. 2002, \apj, 568, 335

\bibitem[{{Marois} {et~al.}(2008){Marois}, {Macintosh}, {Barman}, {Zuckerman},
  {Song}, {Patience}, {Lafreni{\`e}re}, \& {Doyon}}]{2008Sci...322.1348M}
{Marois}, C., {Macintosh}, B., {Barman}, T., {Zuckerman}, B., {Song}, I.,
  {Patience}, J., {Lafreni{\`e}re}, D., \& {Doyon}, R. 2008, Science, 322, 1348

\bibitem[{{Marsh} {et~al.}(2010){Marsh}, {Kirkpatrick}, \&
  {Plavchan}}]{2010ApJ...709L.158M}
{Marsh}, K.~A., {Kirkpatrick}, J.~D., \& {Plavchan}, P. 2010, \apjl, 709, L158

\bibitem[{{Martin} {et~al.}(1996){Martin}, {Rebolo}, \&
  {Zapatero-Osorio}}]{1996ApJ...469..706M}
{Martin}, E.~L., {Rebolo}, R., \& {Zapatero-Osorio}, M.~R. 1996, \apj, 469, 706

\bibitem[{{McLean} {et~al.}(2003){McLean}, {McGovern}, {Burgasser},
  {Kirkpatrick}, {Prato}, \& {Kim}}]{2003ApJ...596..561M}
{McLean}, I.~S., {McGovern}, M.~R., {Burgasser}, A.~J., {Kirkpatrick}, J.~D.,
  {Prato}, L., \& {Kim}, S.~S. 2003, \apj, 596, 561

\bibitem[{{McLean} {et~al.}(2007){McLean}, {Prato}, {McGovern}, {Burgasser},
  {Kirkpatrick}, {Rice}, \& {Kim}}]{2007ApJ...658.1217M}
{McLean}, I.~S., {Prato}, L., {McGovern}, M.~R., {Burgasser}, A.~J.,
  {Kirkpatrick}, J.~D., {Rice}, E.~L., \& {Kim}, S.~S. 2007, \apj, 658, 1217

\bibitem[{{McPherson} {et~al.}(2006){McPherson}, {Born}, {Sutherland},
  {Emerson}, {Little}, {Jeffers}, {Stewart}, {Murray}, \&
  {Ward}}]{2006SPIE.6267E...7M}
{McPherson}, A.~M., {Born}, A., {Sutherland}, W., {Emerson}, J., {Little}, B.,
  {Jeffers}, P., {Stewart}, M., {Murray}, J., \& {Ward}, K. 2006, in Society of
  Photo-Optical Instrumentation Engineers (SPIE) Conference Series, Vol. 6267,
  Society of Photo-Optical Instrumentation Engineers (SPIE) Conference Series

\bibitem[{{Metchev} \& {Hillenbrand}(2006)}]{2006ApJ...651.1166M}
{Metchev}, S.~A., \& {Hillenbrand}, L.~A. 2006, \apj, 651, 1166

\bibitem[{{Mohanty} {et~al.}(2007){Mohanty}, {Jayawardhana}, {Hu{\'e}lamo}, \&
  {Mamajek}}]{2007ApJ...657.1064M}
{Mohanty}, S., {Jayawardhana}, R., {Hu{\'e}lamo}, N., \& {Mamajek}, E. 2007,
  \apj, 657, 1064

\bibitem[{{Montes} {et~al.}(2001){Montes}, {L{\'o}pez-Santiago},
  {Fern{\'a}ndez-Figueroa}, \& {G{\'a}lvez}}]{2001A&A...379..976M}
{Montes}, D., {L{\'o}pez-Santiago}, J., {Fern{\'a}ndez-Figueroa}, M.~J., \&
  {G{\'a}lvez}, M.~C. 2001, \aap, 379, 976

\bibitem[{{Moorwood} {et~al.}(1998){Moorwood}, {Cuby}, {Biereichel}, {Brynnel},
  {Delabre}, {Devillard}, {van Dijsseldonk}, {Finger}, {Gemperlein},
  {Gilmozzi}, {Herlin}, {Huster}, {Knudstrup}, {Lidman}, {Lizon}, {Mehrgan},
  {Meyer}, {Nicolini}, {Petr}, {Spyromilio}, \&
  {Stegmeier}}]{1998Msngr..94....7M}
{Moorwood}, A., {Cuby}, J., {Biereichel}, P., {Brynnel}, J., {Delabre}, B.,
  {Devillard}, N., {van Dijsseldonk}, A., {Finger}, G., {Gemperlein}, H.,
  {Gilmozzi}, R., {Herlin}, T., {Huster}, G., {Knudstrup}, J., {Lidman}, C.,
  {Lizon}, J., {Mehrgan}, H., {Meyer}, M., {Nicolini}, G., {Petr}, M.,
  {Spyromilio}, J., \& {Stegmeier}, J. 1998, The Messenger, 94, 7

\bibitem[{{Nakajima} {et~al.}(1995){Nakajima}, {Oppenheimer}, {Kulkarni},
  {Golimowski}, {Matthews}, \& {Durrance}}]{1995Natur.378..463N}
{Nakajima}, T., {Oppenheimer}, B.~R., {Kulkarni}, S.~R., {Golimowski}, D.~A.,
  {Matthews}, K., \& {Durrance}, S.~T. 1995, \nat, 378, 463

\bibitem[{{Nelson}(2003)}]{2003MNRAS.345..233N}
{Nelson}, R.~P. 2003, \mnras, 345, 233

\bibitem[{{Nidever} {et~al.}(2002){Nidever}, {Marcy}, {Butler}, {Fischer}, \&
  {Vogt}}]{2002ApJS..141..503N}
{Nidever}, D.~L., {Marcy}, G.~W., {Butler}, R.~P., {Fischer}, D.~A., \& {Vogt},
  S.~S. 2002, \apjs, 141, 503

\bibitem[{{Perryman} {et~al.}(1998){Perryman}, {Brown}, {Lebreton}, {Gomez},
  {Turon}, {Cayrel de Strobel}, {Mermilliod}, {Robichon}, {Kovalevsky}, \&
  {Crifo}}]{1998A&A...331...81P}
{Perryman}, M.~A.~C., {Brown}, A.~G.~A., {Lebreton}, Y., {Gomez}, A., {Turon},
  C., {Cayrel de Strobel}, G., {Mermilliod}, J.~C., {Robichon}, N.,
  {Kovalevsky}, J., \& {Crifo}, F. 1998, \aap, 331, 81

\bibitem[{{Pierens} \& {Nelson}(2008)}]{2008A&A...483..633P}
{Pierens}, A., \& {Nelson}, R.~P. 2008, \aap, 483, 633

\bibitem[{{Pinfield} {et~al.}(2006){Pinfield}, {Jones}, {Lucas}, {Kendall},
  {Folkes}, {Day-Jones}, {Chappelle}, \& {Steele}}]{2006MNRAS.368.1281P}
{Pinfield}, D.~J., {Jones}, H.~R.~A., {Lucas}, P.~W., {Kendall}, T.~R.,
  {Folkes}, S.~L., {Day-Jones}, A.~C., {Chappelle}, R.~J., \& {Steele}, I.~A.
  2006, \mnras, 368, 1281

\bibitem[{{Pizzolato} {et~al.}(2003){Pizzolato}, {Maggio}, {Micela},
  {Sciortino}, \& {Ventura}}]{2003A&A...397..147P}
{Pizzolato}, N., {Maggio}, A., {Micela}, G., {Sciortino}, S., \& {Ventura}, P.
  2003, \aap, 397, 147

\bibitem[{{Pollack} {et~al.}(1996){Pollack}, {Hubickyj}, {Bodenheimer},
  {Lissauer}, {Podolak}, \& {Greenzweig}}]{1996Icar..124...62P}
{Pollack}, J.~B., {Hubickyj}, O., {Bodenheimer}, P., {Lissauer}, J.~J.,
  {Podolak}, M., \& {Greenzweig}, Y. 1996, Icarus, 124, 62

\bibitem[{{Reid} {et~al.}(1995){Reid}, {Hawley}, \&
  {Gizis}}]{1995AJ....110.1838R}
{Reid}, I.~N., {Hawley}, S.~L., \& {Gizis}, J.~E. 1995, \aj, 110, 1838

\bibitem[{{Saumon} {et~al.}(1994){Saumon}, {Bergeron}, {Lunine}, {Hubbard}, \&
  {Burrows}}]{1994ApJ...424..333S}
{Saumon}, D., {Bergeron}, P., {Lunine}, J.~I., {Hubbard}, W.~B., \& {Burrows},
  A. 1994, \apj, 424, 333

\bibitem[{{Saumon} \& {Marley}(2008)}]{2008ApJ...689.1327S}
{Saumon}, D., \& {Marley}, M.~S. 2008, \apj, 689, 1327

\bibitem[{{Saumon} {et~al.}(2006){Saumon}, {Marley}, {Cushing}, {Leggett},
  {Roellig}, {Lodders}, \& {Freedman}}]{2006ApJ...647..552S}
{Saumon}, D., {Marley}, M.~S., {Cushing}, M.~C., {Leggett}, S.~K., {Roellig},
  T.~L., {Lodders}, K., \& {Freedman}, R.~S. 2006, \apj, 647, 552

\bibitem[{{Saumon} {et~al.}(2007){Saumon}, {Marley}, {Leggett}, {Geballe},
  {Stephens}, {Golimowski}, {Cushing}, {Fan}, {Rayner}, {Lodders}, \&
  {Freedman}}]{2007ApJ...656.1136S}
{Saumon}, D., {Marley}, M.~S., {Leggett}, S.~K., {Geballe}, T.~R., {Stephens},
  D., {Golimowski}, D.~A., {Cushing}, M.~C., {Fan}, X., {Rayner}, J.~T.,
  {Lodders}, K., \& {Freedman}, R.~S. 2007, \apj, 656, 1136

\bibitem[{{Schlaufman} \& {Laughlin}(2010)}]{2010arXiv1006.2850S}
{Schlaufman}, K.~C., \& {Laughlin}, G. 2010, ArXiv e-prints

\bibitem[{{Schmidt} {et~al.}(2010){Schmidt}, {West}, {Hawley}, \&
  {Pineda}}]{2010AJ....139.1808S}
{Schmidt}, S.~J., {West}, A.~A., {Hawley}, S.~L., \& {Pineda}, J.~S. 2010, \aj,
  139, 1808

\bibitem[{{Schmidt} {et~al.}(2008){Schmidt}, {Neuh{\"a}user}, {Seifahrt},
  {Vogt}, {Bedalov}, {Helling}, {Witte}, \& {Hauschildt}}]{2008A&A...491..311S}
{Schmidt}, T.~O.~B., {Neuh{\"a}user}, R., {Seifahrt}, A., {Vogt}, N.,
  {Bedalov}, A., {Helling}, C., {Witte}, S., \& {Hauschildt}, P.~H. 2008, \aap,
  491, 311

\bibitem[{{Scholz}(2010{\natexlab{a}})}]{2010A&A...515A..92S}
{Scholz}, R. 2010{\natexlab{a}}, \aap, 515, A92+

\bibitem[{{Scholz}(2010{\natexlab{b}})}]{2010A&A...510L...8S}
---. 2010{\natexlab{b}}, \aap, 510, L8+

\bibitem[{{Simcoe} {et~al.}(2008){Simcoe}, {Burgasser}, {Bernstein}, {Bigelow},
  {Fishner}, {Forrest}, {McMurtry}, {Pipher}, {Schechter}, \&
  {Smith}}]{2008SPIE.7014E..27S}
{Simcoe}, R.~A., {Burgasser}, A.~J., {Bernstein}, R.~A., {Bigelow}, B.~C.,
  {Fishner}, J., {Forrest}, W.~J., {McMurtry}, C., {Pipher}, J.~L.,
  {Schechter}, P.~L., \& {Smith}, M. 2008, in Society of Photo-Optical
  Instrumentation Engineers (SPIE) Conference Series, Vol. 7014, Society of
  Photo-Optical Instrumentation Engineers (SPIE) Conference Series

\bibitem[{{Simcoe} {et~al.}(2010){Simcoe}, {Burgasser}, {Bochanski},
  {Schechter}, {Smith}, {Fishner}, {Bernstein}, {Bigelow}, {Pipher}, {Forrest},
  {McMurtry}, {Blank}, \& {Kumler}}]{simcoe_fire}
{Simcoe}, R.~A., {Burgasser}, A.~J., {Bochanski}, J.~J., {Schechter}, P.~L.,
  {Smith}, M.~J., {Fishner}, J., {Bernstein}, R.~A., {Bigelow}, B.~C.,
  {Pipher}, J.~L., {Forrest}, W.~J., {McMurtry}, C., {Blank}, R., \& {Kumler},
  J. 2010, PASP

\bibitem[{{Simons} \& {Tokunaga}(2002)}]{2002PASP..114..169S}
{Simons}, D.~A., \& {Tokunaga}, A. 2002, \pasp, 114, 169

\bibitem[{{Soter}(2006)}]{2006AJ....132.2513S}
{Soter}, S. 2006, \aj, 132, 2513

\bibitem[{{Spiegel} {et~al.}(2010){Spiegel}, {Burrows}, \&
  {Milsom}}]{2010arXiv1008.5150S}
{Spiegel}, D.~S., {Burrows}, A., \& {Milsom}, J.~A. 2010, ArXiv e-prints

\bibitem[{{Stamatellos} \& {Whitworth}(2009)}]{2009MNRAS.392..413S}
{Stamatellos}, D., \& {Whitworth}, A.~P. 2009, \mnras, 392, 413

\bibitem[{{Stephens} {et~al.}(2009){Stephens}, {Leggett}, {Cushing}, {Marley},
  {Saumon}, {Geballe}, {Golimowski}, {Fan}, \& {Noll}}]{2009ApJ...702..154S}
{Stephens}, D.~C., {Leggett}, S.~K., {Cushing}, M.~C., {Marley}, M.~S.,
  {Saumon}, D., {Geballe}, T.~R., {Golimowski}, D.~A., {Fan}, X., \& {Noll},
  K.~S. 2009, \apj, 702, 154

\bibitem[{{Takeda} {et~al.}(2007){Takeda}, {Ford}, {Sills}, {Rasio}, {Fischer},
  \& {Valenti}}]{2007ApJS..168..297T}
{Takeda}, G., {Ford}, E.~B., {Sills}, A., {Rasio}, F.~A., {Fischer}, D.~A., \&
  {Valenti}, J.~A. 2007, \apjs, 168, 297

\bibitem[{{Tinney} {et~al.}(2005){Tinney}, {Burgasser}, {Kirkpatrick}, \&
  {McElwain}}]{2005AJ....130.2326T}
{Tinney}, C.~G., {Burgasser}, A.~J., {Kirkpatrick}, J.~D., \& {McElwain}, M.~W.
  2005, \aj, 130, 2326

\bibitem[{{Tody}(1986)}]{1986SPIE..627..733T}
{Tody}, D. 1986, in Society of Photo-Optical Instrumentation Engineers (SPIE)
  Conference Series, Vol. 627, Society of Photo-Optical Instrumentation
  Engineers (SPIE) Conference Series, ed. D.~L. {Crawford}, 733--+

\bibitem[{{Tokunaga} {et~al.}(2002){Tokunaga}, {Simons}, \&
  {Vacca}}]{2002PASP..114..180T}
{Tokunaga}, A.~T., {Simons}, D.~A., \& {Vacca}, W.~D. 2002, \pasp, 114, 180

\bibitem[{{Tsuji}(2005)}]{2005ApJ...621.1033T}
{Tsuji}, T. 2005, \apj, 621, 1033

\bibitem[{{Tsuji} {et~al.}(1999){Tsuji}, {Ohnaka}, \&
  {Aoki}}]{1999ApJ...520L.119T}
{Tsuji}, T., {Ohnaka}, K., \& {Aoki}, W. 1999, \apjl, 520, L119

\bibitem[{{Vacca} {et~al.}(2003){Vacca}, {Cushing}, \&
  {Rayner}}]{2003PASP..115..389V}
{Vacca}, W.~D., {Cushing}, M.~C., \& {Rayner}, J.~T. 2003, \pasp, 115, 389

\bibitem[{{van Leeuwen}(2007)}]{2007A&A...474..653V}
{van Leeuwen}, F. 2007, \aap, 474, 653

\bibitem[{{Veras} {et~al.}(2009){Veras}, {Crepp}, \&
  {Ford}}]{2009ApJ...696.1600V}
{Veras}, D., {Crepp}, J.~R., \& {Ford}, E.~B. 2009, \apj, 696, 1600

\bibitem[{{Visscher} {et~al.}(2010){Visscher}, {Lodders}, \&
  {Fegley}}]{2010ApJ...716.1060V}
{Visscher}, C., {Lodders}, K., \& {Fegley}, B. 2010, \apj, 716, 1060

\bibitem[{{Voges} {et~al.}(2000){Voges}, {Aschenbach}, {Boller}, {Brauninger},
  {Briel}, {Burkert}, {Dennerl}, {Englhauser}, {Gruber}, {Haberl}, {Hartner},
  {Hasinger}, {Pfeffermann}, {Pietsch}, {Predehl}, {Schmitt}, {Trumper}, \&
  {Zimmermann}}]{2000IAUC.7432....3V}
{Voges}, W., {Aschenbach}, B., {Boller}, T., {Brauninger}, H., {Briel}, U.,
  {Burkert}, W., {Dennerl}, K., {Englhauser}, J., {Gruber}, R., {Haberl}, F.,
  {Hartner}, G., {Hasinger}, G., {Pfeffermann}, E., {Pietsch}, W., {Predehl},
  P., {Schmitt}, J., {Trumper}, J., \& {Zimmermann}, U. 2000, \iaucirc, 7432, 3

\bibitem[{{Vorobyov} \& {Basu}(2010)}]{2010ApJ...714L.133V}
{Vorobyov}, E.~I., \& {Basu}, S. 2010, \apjl, 714, L133

\bibitem[{{Vrba} {et~al.}(2004){Vrba}, {Henden}, {Luginbuhl}, {Guetter},
  {Munn}, {Canzian}, {Burgasser}, {Kirkpatrick}, {Fan}, {Geballe},
  {Golimowski}, {Knapp}, {Leggett}, {Schneider}, \&
  {Brinkmann}}]{2004AJ....127.2948V}
{Vrba}, F.~J., {Henden}, A.~A., {Luginbuhl}, C.~B., {Guetter}, H.~H., {Munn},
  J.~A., {Canzian}, B., {Burgasser}, A.~J., {Kirkpatrick}, J.~D., {Fan}, X.,
  {Geballe}, T.~R., {Golimowski}, D.~A., {Knapp}, G.~R., {Leggett}, S.~K.,
  {Schneider}, D.~P., \& {Brinkmann}, J. 2004, \aj, 127, 2948

\bibitem[{{Warren} {et~al.}(2007)}]{2007MNRAS.381.1400W}
{Warren}, S.~J., {et~al.} 2007, \mnras, 381, 1400

\bibitem[{{West} {et~al.}(2008){West}, {Hawley}, {Bochanski}, {Covey}, {Reid},
  {Dhital}, {Hilton}, \& {Masuda}}]{2008AJ....135..785W}
{West}, A.~A., {Hawley}, S.~L., {Bochanski}, J.~J., {Covey}, K.~R., {Reid},
  I.~N., {Dhital}, S., {Hilton}, E.~J., \& {Masuda}, M. 2008, \aj, 135, 785

\bibitem[{{Witte} {et~al.}(2009){Witte}, {Helling}, \&
  {Hauschildt}}]{2009A&A...506.1367W}
{Witte}, S., {Helling}, C., \& {Hauschildt}, P.~H. 2009, \aap, 506, 1367

\bibitem[{{Wright} {et~al.}(2008){Wright}, {Reike}, {Barella}, {Boeker},
  {Colina}, {van Dishoeck}, {Driggers}, {Goodson}, {Greene}, {Heske},
  {Henning}, {Lagage}, {Meixner}, {Norgaard-Nielsen}, {Olofsson}, {Ray},
  {Ressler}, {Thatcher}, {Waelkens}, {Wright}, \&
  {Zehnder}}]{2008SPIE.7010E..27W}
{Wright}, G.~S., {Reike}, G., {Barella}, P., {Boeker}, T., {Colina}, L., {van
  Dishoeck}, E., {Driggers}, P., {Goodson}, G., {Greene}, T., {Heske}, A.,
  {Henning}, T., {Lagage}, P., {Meixner}, M., {Norgaard-Nielsen}, H.,
  {Olofsson}, G., {Ray}, T., {Ressler}, M., {Thatcher}, J., {Waelkens}, C.,
  {Wright}, D., \& {Zehnder}, A. 2008, in Society of Photo-Optical
  Instrumentation Engineers (SPIE) Conference Series, Vol. 7010, Society of
  Photo-Optical Instrumentation Engineers (SPIE) Conference Series

\bibitem[{{Yamamura} {et~al.}(2010){Yamamura}, {Tsuji}, \&
  {Tanabe}}]{2010arXiv1008.3732Y}
{Yamamura}, I., {Tsuji}, T., \& {Tanabe}, T. 2010, ArXiv e-prints

\bibitem[{{Zapatero Osorio} {et~al.}(2002){Zapatero Osorio}, {B{\'e}jar},
  {Mart{\'{\i}}n}, {Rebolo}, {Barrado y Navascu{\'e}s}, {Mundt},
  {Eisl{\"o}ffel}, \& {Caballero}}]{2002ApJ...578..536Z}
{Zapatero Osorio}, M.~R., {B{\'e}jar}, V.~J.~S., {Mart{\'{\i}}n}, E.~L.,
  {Rebolo}, R., {Barrado y Navascu{\'e}s}, D., {Mundt}, R., {Eisl{\"o}ffel},
  J., \& {Caballero}, J.~A. 2002, \apj, 578, 536

\bibitem[{{Zapatero Osorio} {et~al.}(2004){Zapatero Osorio}, {Lane},
  {Pavlenko}, {Mart{\'{\i}}n}, {Britton}, \& {Kulkarni}}]{2004ApJ...615..958Z}
{Zapatero Osorio}, M.~R., {Lane}, B.~F., {Pavlenko}, Y., {Mart{\'{\i}}n},
  E.~L., {Britton}, M., \& {Kulkarni}, S.~R. 2004, \apj, 615, 958

\end{thebibliography}

\end{document}